**Several Action and Action Deformation Proposals in f(R) Variable G Modified Gravities and a Brief Exposé on Variable G Ontology**

Dara Faroughy

*Department of Sciences: Pima College West Campus, Tucson Az USA.*

*Contact email: darifaro@yahoo.com*

This note includes five sections. In section 2 a number of aesthetically motivated f(R) Lagrangians is proposed and frivolously explored in the metric formalism with variable Newton constant G(x). In section 3 the idea of deforming the terms in a given f(R) gravity-matter action S, and prior to extremizing it, is promoted with the aim of reducing S via an algebraic mixing between few selected gravity and matter terms without invoking new fields. The result of this is a number of constraint equations among the fields and often eccentric looking field equations derived from the reduced action. Section 4 has a similar upshot as section 3 but the difference is the action term mixing is carried out only after extremizing S and prior to any integration by part. Section 5 pertains to G(x) ontology where various personal views on G(x) gravity are expressed in the context of the Standard Model fermions (SMF). The prospect of mitigating gravity field effects for the SMF is investigated at the end via the variable speed of light (VSL) and new global Lorentz transformations fusing two upper speeds: the light speed c and a proper maximal speed given in the text assigned to all point mass $m_o$ and involving the Planck mass. The new LT in combination with VSL also houses a family of special noninertial frames in linear motions. The (spin sensitive) spacetime and momentum dependent VSL is modeled via the Hamiltonian formalism with the aim of canceling the effect of Newtonian gravity felt by a SMF test particle. I also model variable G for yielding finite gravity at all distances, including zero. Excising gravity from the SMF rank (via VSL plus G(x), or by gauging to zero the SMF gravitational masses) has epic brunt on quantum gravity, local SUSY, superstring models, leptoquarks (if any) and cosmology.

**Contents**

## 1. INTRODUCTION

Various proposals of diverse characters have been suggested by the physicists during the past decades for going beyond, or modifying, General Relativity (GR) and often for few viable reasons. In more modern times such reasons are mostly provoked by: the low energy quantum gravity (QG) limit, the string theory (the heterotic string in particular and its eventual connection to the Gauss-Bonnet quadratic terms), effective field theories, the early-time and the late-time acceleration of the universe, gravity alternative to dark energy, and so forth. The so-called f(R) modified gravities **[1]**, also one key theme of this note, is only one endeavor among others to go beyond GR.

Whereas Planck scale QG may entail alteration of GR (described by the Einstein-Hilbert action, linear in the curvature R) at high curvature limit, the discovery of an accelerating universe may inspire modifying the action at low-curvature cosmological scales. The dark energy (DE) scenario may require modifying the GR field equations at low curvature, while respecting Newtonian gravity locally, and dominating the field equations by the cosmological constant (CC), as the vacuum energy, asymptotically. However, not identifying the CC term with the vacuum energy (as normally done, leading to the so-called cosmological constant problem) but viewing it as a classical intrinsic Ricci curvature of a geometrical nature is another stance to DE, normally corresponding to the traditional ΛCDM model, which may also include a scalar field to model the non constituent DE. The latter model circumvents the orthodox GR+constituent DE of negative pressure and is a serious competitor to the many DE models based on the vacuum energy interpretation of the CC. It is quite challenging to find a single theory that can cope effectively with both curvature regimes at once. To my knowledge currently only the extra dimensional brane-world scenarios come close enough in this regard but only at the pre-quantized "effective" field level. In this paper we shall explore a number of f(R) gravity type actions in various guises as possible alternative models to the DE scenarios. But the issue of how these diverse models may score in cosmology at low or high curvature limits is not suitably studied here. Finally, the easy style of discussions adopted throughout the text is in large part to comfort our student, especially early student, readers.

In section 2 we shall formulate, and then lightly explore, a number of rather independent f(R) actions in combination with the variable Newton's constant G, which is taken seriously throughout this note as a quantity of great potentiality. All action formulations are carried out in the observationally relevant 4D Jordan frame, where spacetime and mass units remain constant. (The Jordan frame is favored by us, although this is a minor point, because it's metric yields geodesics normally coinciding with the free fall trajectories and has the appropriate Minkowski limit, and also because the matter $T_{\mu\nu}$ is divergence-free.) Few proposed f(R) models are found to exhibit a degree of novelty (e.g., limiting the scalar curvature range and avoiding singularities) not frequently seen in the majority of already published f(R) forms.

Most actions reported in the literature are justifiably designed for possible applications in quantum gravity, or more importantly for making predictions in cosmology. While this is imperative and definitely necessary, we shall adopt a more lenient attitude in section 2, and the following two sections, so to concentrate instead on the study of "reasonable" Lagrangian forms which yield aesthetically tempting actions and field equations following some term mixing procedures. Our lax attitude is prompted in part by the possibility that many tailored made f(R) models-unquestionably offering varieties of explanations on few astrophysics and cosmological observations and the late acceleration of the universe (which in the f(R) models is mainly due to gravity and not matter)-may be beset by serious stability problems and be already in conflict with the solar system measurements, and perhaps more seriously with the cosmological constraints! A number of authors have claimed lately that most existing f(R) models are to be ruled out despite their possible consistencies in local gravity and the supernovae, yet few others are more optimistic. In short, there are no assurances at this writing that the f(R) models can lead to a viable cosmology (at least for a class of models studied so far), or even score better than the simple (R-2Λ) Lagrangian model of GR in explaining the dark energy and the universe's acceleration, or even gravity at the micro scales. One perhaps testable peculiarity of most f(R) models is that whereas GR demands objects to move on geodesics, the same cannot be generally assured in the f(R) models. For an insight on hosts of possible inconsistencies in the f(R) models see [2].

In light of the current indecision on the overall merit of the f(R) "physical" models, we shall feel free to regard the f(R) models as basically phenomenological "math" tools and thus allow ourselves to toy around, in various guises, while remaining "reasonable", with the action/Lagrangian terms of these models for achieving an exploration purpose. On the whole, we want to explore what kind of unusual looking field equations we can generate if we cause some algebraic "mixing" (or combining) between the f(R), variable G, the cosmological constant and the matter field terms, thus resulting in few field type constraint equations. And we want to do that without introducing new fields. In carrying out the alleged "mixing" of the terms we shall be mostly guided by a dose of aesthetics and intuition, and naturally if things become too complicated for a particular mixing mode, which often can occur, we'll switch to a another mixing mode by choosing different combination of terms and constraints, and then take it from there! Most efforts in section 3 are about implementing this term mixing game prior to extremizing an action, meaning prior to "math" fields becoming physical fields. In section 4 the action term mixing game is carried out only after minimizing an action. I need to also mention this early on that once we derive the field equations, after minimizing the residual action for a particular mixing mode, our task is not always to jump into a serious study of these equations in cosmology-by, e.g., choosing a popular metric, like the FRW metric, for an accelerating universe-and see how they score in terms of predictions! In fact, often we shall be content with only finding the field equations or just seeing how variable G behaves under this or that mixing mode! Therefore, in order to carry out our aesthetic mixing mode with some degree of rational freedom we shall allow ourselves to fret less on certain standard issues like: an in-depth justification and interpretation of our mixing choices between the action terms, renormalizability if quantized, possible diffeomorphism-covariance defiance of the reduced actions, and, as I've just said, an overall predictability and viability of the derived field equations in the microworld or cosmology. One reason for this article is to spur few readers into looking at some of the many "field" equations and results presented here and use them as the starting point for studying gravity at very large and the solar system scales that may be omitted here.

Soon we shall elaborate in more details on what we have planned for sections 2 and 3, which for most part, but not always, have to do, as stressed, with gravity-matter action manipulations of a sort, regarding the individual action terms, prior to deriving the final field equations-a procedure that often results in few constraint equations involving the fields. Thus, the final field equations are only viable in the context of the resulting constraint equations. At first glance it seems the field equations so derived are of too limited values because of the constraint equations, that are normally absent in the traditional treatment of GR. We think this may not be so because of the shear number of fields involved in the action, counting also variable G. Variable G (in contrast to most scalar-tensor theories) is not treated by us as a genuine "physical field" (meaning no action variation with respect to G is ever performed, and no kinetic and potential terms are assigned to G), thus the constraint equations are essential to see how G varies in spacetime. And all that in addition to other "G-constraints", that we may use, and need to use from time to time, that have little to do with our term mixing procedure. (E.g., a frequently used constraint, having great simplifying effects, comes by viewing the product of G(x) by the square root of the metric determinant as a non dynamic quantity, and in this way variable G is brought closer to the geometrical sector rather than the matter sector.) For many models the final field equations that we shall obtain herald a great degree of generality, despite the constraint equations, and not limitations.

(It may be argued this early on that our peculiar view of variable G (i.e., in treating the dimensionless quantity $G_N/G(x)=\Phi(x)$ as only a scale function) in the context of f(R) theories is still a peculiar representation of the Brans-Dicke scalar-tensor gravity, an equivalency that is well established in the f(R) theories. I wouldn't be much concerned over this spat, because even if so and regarding G as physical then the equivalent BD action representation that we shall formulate is going to be the most reduced and sterile type of all! Because in the presumed BD action of ours the parameter ω and the BD potential V are simply set to zero and the matter action is independent of $\Phi(x)$! Thus, it is hard to see any utility for viewing $\Phi(x)$ as something physical instead of simply a scale function. In all actions $\Phi(x)$ enters the game only in form of $\Phi(x)f(R)$, and that's all we need for a start.)

Section 5 is an informal and an independent section. It contains among other things some of my own rough views on the variable G, the role of gravity (or the lack of it!) for the fermionic fields of the standard model, as well as few other topics. Among the discussed and pertinent topics notables are: the possibility of zero gravitational mass for the SM free leptons and quarks, and independently the variation of light speed in the context of a new form of special relativity dubbed DUSR. The latter needs two upper speeds in its formulation and has the additional potentiality of covering certain accelerated frame motions of a specific kind, and of particular interest is when the noninertialities are caused by gravity. Close to the end of the section two attempts are made to join both seemingly disconnected gravity and the VSL concepts together in the context of DUSR, and independently in the context of the Hamiltonian formulation. The latter formulation yields an expression for the SM fermionic VSL (by assuming the "canceling out-gravity" conjecture for such fermions) as a function of the coordinates and the fermionic test particle momentum. The heuristic menu leading to the fermionic VSL expression is thus the relativistic Hamiltonian for a given fermion with VSL in the presence of an externally prescribed variable G gravity potential "field" and then equating it to its (almost) free (Dirac-inspired) linearized Hamiltonian. We shall also propose an explicit form for $G(r)\sim e^{-r0/r}$ which I derived long time ago. The latter always leads to a finite and weak gravity, no matter the value of r (including r=0). Then we will discuss briefly the bosonic way of generating, as opposed to the fermionic way of canceling, finite gravity by means of VSL, or vice versa, and then conclude the section with a partial summary and few words on why our chosen means of tackling elementary particle gravity is variable G.

**Preliminaries:** The chief tool for starting our aesthetic explorations of sections 1-3 is of course the action formulation of gravity, and for our current purpose in the metric-formalism. In the forthcoming paper, if time permits, we shall undertake such explorations in the context of the (Palatini) metric-affine formalism, whereby the issue of whether they are alternative theories to GR is still been debated (see also **[2]**). The Palatini formalism is useful in many theoretical computations and may be overall easier to tackle than the metric formalism. For one thing it yields to a second-order system, in contrast to the fourth-order system obtained in the metric formalism, and there are indications (for at least a class of f(R) models) that it may have fewer theoretical (like the instabilities) and observational (like the solar system) inconsistencies than the metric formalism. (For completeness sake let me remind the reader that in the orthodox treatment of gravity there are three main and somewhat dynamically equivalent formulations of GR: the purely metric formulation of Einstein-Hilbert, the one used over here for varying the f(R) actions, the purely affine formulation of Einstein-Eddington, and, finally, the metric-affine formulation of Einstein-Palatini.)

Student readers should bare in mind that the standard menu that is closely followed in all action formulations of the classical fields is this: first propose a "suitable' action, by using a suitable Lagrangian that possesses certain local and possibly discrete symmetries of a kind, as well as other important features (such as diffeomorphism invariance in case of the GR-EH action), then use the functional variation of it (for our purpose with respect to the metric). And finally derive the field equations after some possible integrations by parts and employing the Stokes's theorem to bring forth the final surface terms that are normally discarded in the formulations of GR admitting a boundary (recall the famous Stokes's theorem relates the volume to surface boundary integrations of a vector V and its covariant derivative in an nD space as: $\int_m d^n x (\sqrt{|g|}) \nabla_\sigma V^\sigma = \int_{\partial m} d^{n-1} y (\sqrt{|\gamma|}) n_\mu V^\mu$). We also jog our memory on how the classical GR math does business by generating dynamically the continuous 4D geometry of spacetime under the banner of a topological delicacy called a manifold (more specifically, the smooth Riemannian manifold which is the spacetime tool for supporting tensors, like the metric tensor, and admitting a consistent differentiable structure). Loosely said, when gravity is switched off spacetime can assume a central role to serve as a fixed background stage for nongravitational field formulations and their quantizations, and for localizing relativistic physical objects. But this is no longer so when (relativistic) gravity is included in the grand picture. In that case the physical world (referring to all dynamical field theories and objects) is forced to behave in certain nontrivial way; the diffeomorphism invariant way, in that physical agents are localizable in a dynamically generated (curved) spacetime only with respect to each other, and not with respect to a fixed and dynamically sterile background spacetime, so goes the story. And "displacing" them all at once, informally said, would not alter the overall physics, thus the verbal meaning of diff-invariance which is indeed the symmetry of generalized spacetime transformations on suitable physical agents! For example, a suitable agent that leads to the Einstein field equations is the Einstein-Hilbert action, also used extensively in this note, which is diff-invariant. The diff-invariance is also closely tied to the principle of general covariance (which is tied to the equivalence principle) if you will, that says the physical meaning (or the lack of meaning) of "things" depends on whether the "things" are diff-invariant or not; basically "no diff invariance=no physical meaning"! Operationally, general covariance, holding only in regions small compared to the spacetime extension of gravity in a system (with negligible tidal effects), is identifiable as a pungent constraint on the diff-Inv of an action-functional, or the equations of motion. (Analogous of this invariance is also encountered in the string actions where they admit world sheet reparametrization invariance under infinitesimal shift in the world sheet coordinates ($\tau$, $\sigma$), a sheet created by the motion of the string in the background (target) spacetime.)

(Added note: The Einstein-Hilbert action ($\sim \int d^4 x \sqrt{g} R$) is used in several occasions in this paper for various purposes. This action is diffeomorphism invariant so let me say few brief words for benefiting early students on the aforementioned diffeomorphism invariance and few other points. First of all one should not confuse the diffeomorphism invariance with the

Lorentz invariance. In GR both the interval $ds^2=g_{\mu\nu}dy^\mu dy^\nu$ (which has a geometrical character) and the EH action are invariant with respect to the operation of diffeomorphisms on the field coordinates: $y^\mu \rightarrow \hat{y}^\mu(y^o,y^1,y^2,y^3)$. Whereas a gauge transformation transforms a field to another field at a point without affecting the physics, diffeomorphism, in diff-invariant theories like GR, moves, or rather maps, fields from one point on the manifold to another point and although in this way induces math changes in the fields the physics does not suffer observational changes and in this sense the field values like the metric and its by-products at a point are operationally pointless. (The issue of an analogy between the gauge group and the diffeomorphism group is however especially intriguing and still a topic of debate.) The Lorentz transformations are also spacetime (and generally linear) coordinate transformations, but the chief difference between both transformations is that the LT parameters and effects are observable, and thus physical in character, but this is not so for the diffeomorphism parameters! We also keep in mind that the general covariance of a theory does not automatically mean the theory is also Lorentz invariant (people in the late 1920s invented diff-inv gravity theories admitting local inertial frames and Galilean invariance but no LT invariance!). That is one reason why Einstein had to require of his diffeomorphism invariant gravity theory to be also locally Lorentz invariant so to have the right SR limit in the absence of gravity. Lorentz invariance is very restrictive in that physical equations, like the Maxwell's equations, must look exactly the same before and after a LT is performed and no extra terms must appear in the transformed equations. In general covariance though the transformed equations can acquire additional terms after the general transformation is performed so to reflect, loosely speaking, the effect of spacetime as being curved (i.e., the presence of gravity). Finally, we bare in mind that according to GR the *local* interaction Lagrangian density $L_{int}$ between the energy-momentum tensorial content $T^{\mu\nu}$ and the spacetime geometry (=metric=the gravitational tensor field $h_{\mu\nu}$) is given by $L_{int}=-\tfrac{1}{2}\kappa T^{\mu\nu}h_{\mu\nu}$. Here the gravitational field tensor $h_{\mu\nu}$ is defined in terms of the metric $g_{\mu\nu}$ via $g_{\mu\nu}=\eta_{\mu\nu}+\kappa h_{\mu\nu}$ while $\kappa$ is defined as a coupling constant and in terms of Newton's constant is given by $\kappa^2=32\pi G$ with c=1.)

As stressed earlier, besides looking at various "f(R)-variable G" action possibilities in section 2, we shall also investigate in section 3 the possibility of toying around, in ways that will become clearer later on, with the gravity+matter Lagrangians at the local level to introduce a degree of mixing between the gravity and matter terms. And we want to do that at the level of the "math" fields prior to extremizing the actions, so the constraint equations we'll be deriving pertain to the math fields. But what do we get out of this? There is no single answer to give, but generally the multifaceted purpose is to derive unorthodox (mixed) field equations, distinct from the plain GR's field equation format, which is of the unmixed form "geometry=physics". In addition, one aim is to also derive distinct field equations from the many existing f(R) models, but sometime we succeed in our derivation and other times we don't. Another purpose is to avoid surface terms, if at all possible, by calculating directly the variation of the scalar curvature. Yet another purpose is to generate few novel features in form of constraint field equations, etc. Basically, all this points to the exploratory nature of our approaches. The methodologies promoted here though are of general nature and not limited to only 4D gravity+matter actions; they can be applied to other forms of actions not discussed here, like the supergravity action in 4D, or generally in higher dimensional spacetime, the brane-world scenarios, and can be applied to diverse non-gravitational actions, the string actions, and what you will. In section 4 we shall explore a somewhat different method by beginning this time with few actions, perform variation on them and then toy around with the resulting infinitesimal terms, but all that game is to be played prior to performing certain integrations by part (if necessary) to derive the field equations! The purpose is to presumably avoid (if possible) the finalized surface terms that are ordinarily gotten by resorting to integrations by part. The aim is to render the orthodox integrations by part method, usually implemented after extremizing an action, unnecessary but this is sometimes possible and other times we must live with the surface terms. At the end of section 4 we shall toy around with the field equations in place of the action.

*An early warm up discussion:* The raison of being and motives behind our diverse ideas shall become more lucid only when we address these topics individually. But before I begin seriously in sect.1 let me offer a bit of an adhoc insight in below, and in the extra note to follow, on what got me motivated in the first place to eventually include the diverse topics we'll be discussing in Sections 2 and 3. Let me begin simplistically by considering GR at the field equations *level* in the presence of a constant cosmological term $\Lambda$: $G_{\mu\nu}+\Lambda g_{\mu\nu}=\kappa_N T_{\mu\nu}$, which is the natural outcome of the EH action involving the CC term, and set $\kappa_N=8\pi G_N$ (c=1). Now introduce a (dimensionless) generic scale field $\Phi(x)$ and decompose the field equations as $G_{\mu\nu}+\Lambda g_{\mu\nu}=\Phi\kappa_N T_{\mu\nu}+(1-\Phi)\kappa_N T_{\mu\nu}$. Next, and as an option, make the identification $(1-\Phi)\kappa_N T_{\mu\nu}=\Lambda g_{\mu\nu}$ so to knot $\Lambda$ to matter and the scale field $\Phi$

(implying also $\Phi=1-4\Lambda/\kappa_N T$). Obviously the latter equations represent the constraint equations and do not necessarily imply a single CC term. After doing so we end up with new field equations: $G_{\mu\nu}=\Phi\kappa_N T_{\mu\nu}$ that looks like $\Phi$ is coupled to matter and the "geometry" ($G_{\mu\nu}=R_{\mu\nu}-\frac{1}{2}Rg_{\mu\nu}$), and the effective gravitational coupling is now $\kappa=\Phi\kappa_N$ (also note $\partial^\mu\Phi T_{\mu\nu}=0$, thanks to the Bianchi identity and the conservation of $T_{\mu\nu}$, and $\Phi=-R/\kappa_N T$). This interesting example, which is rather trivial, is also an illustration of how one can manipulate (distort) the field equations, as in the above, without ever touching the original action. So, if one can do this with the field equations it is innate to ask whether one can play a similar game with an original action and modify, or reduce, it so to get different field equations, and so forth. The memo here is the inspiration for this paper was triggered with a simple observation of the field equations and their possible algebraic modification of a kind and then was extended to the action. After that one thing led to another, and that is basically the type of explorations we like to undertake in sections 2 and 3 working mostly with the actions (yet as stressed earlier toying with the field equations is also considered at the end of section 4).

(***Extra Note:*** To see better the level of possibilities presented in the above mixing approach let me digress a bit, but continue with the above discussions in the next paragraph, to present one interesting variant of the above field equation manipulation which basically amounts to imparting a degree of matter-geometry mixing by first distorting the field equations as: $G_{\mu\nu}+\Lambda g_{\mu\nu}=\kappa_N T_{\mu\nu}=\kappa_N T_{\mu\nu}\sin^2(\alpha R)+\kappa_N T_{\mu\nu}\cos^2(\alpha R)$ and then making the identification $G_{\mu\nu}=\kappa_N T_{\mu\nu}\sin^2(\alpha R)$ and $\Lambda g_{\mu\nu}=\kappa_N T_{\mu\nu}\cos^2(\alpha R)$, which vanishes if T=0. In this way we find the earlier $\Phi$ given by $\Phi=\sin^2(\alpha R)$. This in turn implies its vanishing in flat space, as well as at various quantized curvatures $R_n=n\pi/\alpha.$, and as noted the CC term survives in the flat spacetime limit. So $\Lambda$ can be related to the zero-point energy of the other three nongravitational quantum forces and can be very large as a VEV of these combined fields. But we must be a bit careful over here for the following reason. We can always enlarge the earlier identification we'd made to, e.g., this: $G_{\mu\nu}=\kappa_N T_{\mu\nu}\sin^2(\alpha R)+$"something" and $\Lambda g_{\mu\nu}=\kappa_N T_{\mu\nu}\cos^2(\alpha R)$-"the samething"! And then propose models for the "thing" as we please, of course subject to the Bianchi identity+energy conservation, and what have you, so care must be exercised! Also we note that throughout our arguments we made no assumption that the CC term has to be a constant, and for our current purpose we are not even concerned on what kind of action has led to $G_{\mu\nu}+\Lambda(x)g_{\mu\nu}=\kappa_N T_{\mu\nu}$ in the first place, because we are working at the level of the field equations (we further note that there are bunch of other expressions around and not shown due to the Bianchi identity and the energy conservation)! An exciting possibility that emerges out of these field equations is an explanation for why the observed $\Lambda$ (if the "thing" is set to zero) is so small. The explanation may be because we are currently living in a universe where $\cos^2(\alpha R)\sim 0$, and, thus, the effective Newtonian constant is near its observed value $G_N$! But, there are other quantized curvatures yielding $\Phi=1$ and a vanishing CC, and our present époque curvature may be one of those (Antrophic!) special curvatures $R_n=(n+\frac{1}{2})\pi/\alpha$ that resembles the quantized energy levels of a harmonic oscillator in quantum mechanics! One can now study the metric and the curvature fluctuations near these "eigen" curvatures perturbatively, where the nonlinear effects can be safely ignored. The real issue, of course, is what $R_o=1/\alpha$ is: is it a micro or a macro quantity? The latter may be favored, but with the universe being so old and so big one can envisage $R_o$ as a micro curvature, to relate it to field theory and the early time cosmology, and look at the continuum limit of very large integer n! We also notice that if the trace T is positive then R must be negative (as in de Sitter spaces with $\Lambda>0$), and for $\Lambda<0$ then R must be positive, note $R/4\Lambda=-\tan^2(\alpha R)$. Writing $\Phi=\sin^2(\alpha R)$ may also ring a bell from quantum mechanics and the possibility of interpreting $\Phi$ as the probability of finding the universe with curvature R, and then invoke some kind of superposition principle over the $R_n$ and mimic QM! Readers may want to also look at the possibility of decomposing the full nonlinear Einstein field equations in a manner to separate the linear and the nonlinear terms, like the nonlinear $\Gamma\Gamma$ terms, and then impose conditions on $\Phi$ to get rid of the nonlinear terms so that the "physical" equations are linear and friendly to quantization! In short, one scenario may go like this: write the Einstien equations symbolically as comprised of Linear+Nonlinear terms~$\Phi$T+(1-$\Phi$)T, and then make the identification: (1-$\Phi$)T= Nonlinear so to removes these terms, so the remaining "physical" equation is the linear part: Linear~$\Phi$T, which is presumably easier to quantize for the spin-2 gravitons, and the scale field $\Phi$ is not subject to quantization. Since $\Phi$~*1-(Nonlinear/T)* it is also easier to see, in a more compact form, how the perturbation apparatus, similar to linearized gravity $g_{\mu\nu}=\eta_{\mu\nu}+h_{\mu\nu}$, can be implemented thereafter.)

Let us now continue with of our earlier discussions. In the type of field equation manipulation presented previously no usage was ever made of either the underline GR action or the scale function $\Phi$ which was not even part of field equations initially! We, of course, know that the underline action is simply the textbook Einstein-Lemaitre action with constant gravitational coupling, so $\Phi=1$. To get the initial field equations, following the conventional procedures, one had to ultimately ignore (or eliminate by introducing counter terms) the surface terms ~$\delta R_{\mu\nu}$ in the variation (with fixed boundary hypersurfaces) at some stage to end up with the well-known field equations $G_{\mu\nu}+\Lambda g_{\mu\nu}=\kappa_N T_{\mu\nu}$. As noted, the above toying around of the field equations, and not the action, has not given us any proper field equation for the $\Phi$ scale field (although we have the important relation $\partial^\mu \Phi T_{\mu\nu}=0$). And that is desirable for our purpose pursued in this note; else the scale field $\Phi$ has to be treated as a "physical" field (like the dilaton) and that is not what we have in mind. So, the plan is to marginalize, to the extent possible, any dynamic interpretation for $\Phi$ and avoid the usage of, e.g., modified metrics that are conformally related to the physical metric $g_{\mu\nu}$ (yielding spatiotemporal distances between points) by means of $\Phi$ (the Einstein frame business)! (One tool for this dynamic "marginalization", that we shall often use, is to set the metric variation $\delta(\Phi\sqrt{-g})=0$, as stressed above.) Consequently, deprived of a full coupled wave equation for $\Phi$, we are free, at least in principle, to consider some robust possibilities for $\Phi$, like the vanishing of the kinetic energy term $\partial^\mu\Phi\partial_\mu\Phi=0$, or set $\Box\Phi=0$, etc. (In the latter case, e.g., one can envisage null-vectors $n^\mu\sim\partial^\mu\Phi$ such that $n^\mu n^\nu T_{\mu\nu}=0$-so that $n^\mu n^\nu G_{\mu\nu}=0$-and take it from there. Basically, the genuine game is to generate enough constraint equations to make the dimensionless $\Phi$ as distinct as possible from the real scalar fields encountered in the conventional scalar-tensor gravities (like the Brans-Dicke theory), where the dimensional scalars are treated as physical fields. This, however, is not always an easy task to complete and often interpretational dissonances arise. The ideal case, if realizable, is to prevent any variation of G from representing real gravitational energy and momentum and is a doxastic basicality, if you will, in all our treatments of $\Phi$ and is also in line with Einstein's view on relativistic gravity as a geometric theory, forbidding spacetime (but not necessarily spatial) topology changes, at least within the scales perhaps as small as $10^{-16}$ cm and as large as the universe!

At any rate, by taking the earlier finding $G_{\mu\nu}=\Phi\kappa_N T_{\mu\nu}$ at its face value, and as something esthetically palatable (and never mind now how it was derived!), one can inquire what kind of linear action (or actions), free of G-derivatives (e.g., a term like $\nabla^\mu\Phi\nabla_\mu R$), one can invent so that after extremizing it, and taking care of the surface terms, it yields the *exact* equations $G_{\mu\nu}=\Phi\kappa_N T_{\mu\nu}$ (evidently scalar-tensor actions would not do)? And if we invent such an action (see section 3), then how is the surface terms treated to arrive at the latter field equations without any extra terms? And what happens if one wants to expand the idea to the f(R) theories, and how stringently one should respect the exact general covariance of GR and the equality of the inertial and the gravitational masses in that case? And then there are always other issues to tackle. Can we manipulate the action integrand by combining certain terms together, or eliminate them algebraically by forcing them to eat each other (to borrow the Goldstone boson terminology!) prior to performing variations, so to end up with unorthodox field equations after performing the (metric) variation (Sect.2)? If so, then do we toy around with the gravity action only or with the matter action (or maybe both)? Or should we manipulate certain action terms (Sect.3) in the gravity sector, or the matter sector, after performing the metric variations and require of them to eat each other so to get field equations looking different from those derived by the action if left undisturbed? We have asked many questions, so for a further insight let us consider several prototype of the variable G action integrands to epitomize our motivation for wanting to include, e.g., Sect.2.

Take the following simple gravity-matter integrand$\sim(\sqrt{-g})(\Phi R+2\kappa\text{\pounds}_M)$ of the EH action with variable G and distort it as $\sqrt{-g}[R+(\Phi-1)R+2\kappa\text{\pounds}_M]$. Now, in order to proceed further we must decide what to do next and here is one option: do nothing to the matter Lagrangian $\text{\pounds}_M$ and toy around with only the gravity sector by, e.g., setting $(\Phi-1)R=-2\Lambda$ so that $\Phi=1-2\Lambda/R$, with $\Lambda$ the CC term. This option leads to the previous orthodox field equations $G_{\mu\nu}+\Lambda g_{\mu\nu}=\kappa_N T_{\mu\nu}$ and the relation $4\Lambda-R=\kappa T$. Other options yield other results and here is another example: rewrite $\sqrt{-g}(\Phi R+2\kappa\text{\pounds}_M)$ by absorbing the $(\Phi-1)R$ piece into the matter Lagrangian so that now the integrand becomes: $\sqrt{-g}[R+2\kappa(\text{\pounds}_M+(\Phi-1)R/2\kappa)]$. It is clear that if $(\Phi-1)R$ is assumed to be a constant then there is no effect on $T_{\mu\nu}$, and thus no effect on the equations of motion for the matter fields, but when taking gravity and matter terms in combination we end up with Einstein's equations in the presence of a CC term. But, there is even more to this. What if we do not require the term $(\Phi-1)R$ to be a pure constant but instead require the metric variation of it in form of $\delta(\sqrt{-g}(\Phi-1)R)$ to vanish, or it be a divergence term? Then the resulting picture is drastically altered. Obviously there is plenty more to say over here, and there are many variants to the above examples as well so it is best to stop for the time being until we get to sections 2 and 3, which are totally independent of the phenomenological approaches we'll be seeing in Sect.1.

There is a great interest these days in studying the f(R) gravity theories whose low-curvature predictions have the potentiality of being observed in isolated galaxy clusters, stellar interior and the solar system. There is a long history on f(R) gravity and the variable G. And at least for the f(R) modified gravity its tale begin only short years after the first attempt was made by Einstein and Hilbert to formulate GR in the language of the action principle. When Einstein first presented the finalized version of his field equations in Berlin on Nov 25 1915 he had not yet made a significant attempt to derive them from the principle of least action, rather he got them by certain amount of astute guessing. Hilbert, who was in close contact with Einstein before, had derived similar field equations about a week earlier by employing the techniques of the Euler-Lagrange equations and the meticulous methodology of the action principle, where the variation of the Lagrangian was made with respect to the metric. The outcome of such early efforts led (following some doze of historic controversies on who did what first and when) to what is known today as the Einstein-Hilbert (EH) action for matter-gravity coupling.

The EH action is essentially an invariant 4D volume integral over a manifold of a generally covariant Lagrangian density that is linear in the curvature scalar R. The EH action epitomizes the original tensorial formulation of general relativity (GR) using functional derivative variations that are rigorously well defined only when surface terms are absent (e.g., a vanishing divergence, that otherwise would have led to a surface term, is a suitable outcome to have after extremizing the action). The surface term present in the extremized EH action (stemming from the $g^{\mu\nu}\delta R_{\mu\nu}$ term) is usually discarded, or discarded by adding a counter surface term to the original action (closely related to the extrinsic curvature), and that is the end of this for the many common "engineering" applications of GR in, e.g. astrophysics and cosmology. But the interest in analyzing action surface terms more rigorously from the formal stand is still on the agenda, and even more so in the f(R) modified gravities, which bring forth extra complexities. In more modern provisos, the EH action is also Holographic in that the bulk and the surface action pieces, though not independent of each other, can be separated and studied apart. All this is naturally a classic textbook manner of conducting business regarding the 4D EH action. With the advent of the string dualities, 11D M-theory, holographic principle, etc., about a decade ago, some people like Horova (see [3]) suggested that the 4D EH action, including a cosmological term, may be the result of combining the holography principle and the compactification of the 11D supergravity theories (itself derived as the low energy limit of the M-theory, where the Riemann tensor is the Yang-Mills curvature of the spin connection) to 4D on a tori $T^7$. The tendency today is therefore to view the classic EH action as an effective action of some higher dimensional theory at low energy.

The action approach to gravity is also of prime importance for reconciling gravity with quantum mechanics. The textbook path integral formulation of quantum field theories-which main job is to compute probability amplitudes in form of path integrals over the factor $e^{iS/\hbar}$, with S standing for the action-has been also extended to quantum gravity. The ultimate goal is of course to unify all the basic quantum interactions of nature, and the path integral approach is one tool preferred by many theorists for doing the job. At the classical level the action formulation of gravity has also been useful for formulating GR in the Hamiltonian format.

Nowadays, of course there are ample motivations (for, e.g., explaining the cosmic acceleration, cosmological dark matter and energy, and the lack of fundamental gravitational vacuum state in GR, etc.) for wanting to modify the original EH action ($\sim$R) to higher order generic functions f(R) within the metric or the Palatini formalisms. It is often claimed that higher order curvature terms also emerge as the result of quantum corrections to GR from the matter sector, and are also seen by some to provide the possibility of avoiding spacetime singularity (e.g., in the isotropic Friedman universe). The nonlinear generalization of the EH action has become known as the f(R) theories of gravity, which is also a major subject matter of this paper. Inspired by the Mach principle, a well celebrated classic extension of the EH action in the context of spacetime variable G (known today as the BD scalar-tensor gravity) was formulated long ago by Brans and Dicke (see [4]) who essentially added a real scalar field $\Phi$ to replace G in the EH action, which source is the trace of the energy-momentum tensor T. As already stressed the BD scalar field $\Phi$, related to G as G$\sim$1/$\Phi$, is treated as a real physical field endowed with at least a kinetic energy term contributing to the overall Lagrangian and the action variations. In the string cosmology of nowadays there is an era, the dilatonic era, where the universe's dynamic evolution is mostly dominated by a number of massless fields that serve as the prime source of gravitational dynamics, and one of these massless fields (the dilaton) may be similar to the BD scalar. Consistent string theories, or the Kaluza-Klein theories, require higher dimensional spacetime for genuine theoretical consistency. Therefore, it is argued that the fundamental dimensionful constants of nature, such as G, are to be only true constants in those higher dimensions, and are to be viewed in our 4D world of familiarity as only effective variable constants-exhibiting, e.g., dependencies on the compactified extra dimension sizes and other factors, such as the dilaton that couples directly to matter, the string mass, and so forth. The dimensionless non-gravitational running coupling constants ($\alpha_{QCD}$, $\alpha_{QED}$ and $\alpha_W$), on the other hand, are sensitive to the probing energy scales, and the method for calculating their dependency on the energy is dictated by the renormalization group equations below some grand unification scale, where the three, otherwise distinct couplings, become identical in value.

Our view in this paper on variable G is a bit distinct. We shall regard the dimensionless $\Phi$ as a scale field obviously deprived of any kinetic energy and a potential term (so its possible relation to the real dilaton field is not clear) and then try to mingle it with the f(R) forms that we choose to eventually obtain the field equations for the action in various fashions. But as indicated earlier the lack of the kinetic $\Phi$-term in the Lagrangian usually means also a lack of a proper field equation for $\Phi$ (in contrast to the BD case with $\omega_{BD} \neq 0$). To compensate for this shortcoming and get information on $\Phi$ we shall introduce various simple techniques with diverse degrees of success. Let me stress also that the renormalization group equation (RGE) approach that is often employed, e.g., for the EH action, though very attractive, is not part of our present efforts! The RGE approach normally relies on some RG trajectory to determine G(x) and lacks, per se, a Lagrangian formulation. Yet even though the variable G in RGE approach has no functional dependency on the metric, it carries energy and momentum that in turn contributes to the curvature of spacetime-thus it is similar to the BD scalar field approach in that regard, and very dissimilar to what we have in mind.

One technique that we shall often make use of (though not exclusively) for designing the f(R) models of Sect.1 consists of tying variable G-viewed as an effective coupling due to perhaps the compactification of some higher dimensional theory to 4D-to the f(R) function according to the relation G=Go/($\partial$f/$\partial$R), where the bare Go is not necessarily what one measures in the Lab. One immediate benefit of this is reducing the number of the unknown gravity fields to only f(R). Our prime focus in Sect.1 is the calculation of the effective gravity coupling constant G, and how it may look for say this or that proposed form for f(R). Consequently, we shall discard, in most part, other details of our f(R) models that are unquestionably of great importance for undertaking any serious cosmological studies.

Most, but not all, proposed gravity actions in 4D (as we recall extension to higher dimensions, a la Kaluza-Klein (KK), and the intriguing Brane-world scenarios are discarded here) involve the variable Newtonian constant $G_N$ in combination with the nonlinear f(R) forms in the metric-formalism, and perhaps, more prominently, in the forthcoming paper the metric-affine formalism. Much has been said and written about these formalisms (see, e.g., [1]) so we shall not repeat them over here for the sake of brevity. Suffices to say that in the f(R) metric-formalism the main protagonist for performing variational calculations is the metric, while in the metric-affine formalism (also called the Palatini-formalism in some cases) the protagonists are the metric $g_{\mu\nu}$ and the connection $\Gamma_{\mu\nu}^{\lambda}$ that are treated as two independent geometrical quantities. We note that the familiar EH gravity action involving variable G=$G_o$/$\Phi$, i.e., (1/2$\kappa_o$)$\int$d$^4$x$\sqrt{}$-g$\Phi$R, can be also written as: (1/2$\kappa_o$)$\int$d$^4$x$\sqrt{}$-g$g^{\mu\nu}$R$_{\mu\nu}$=(1/2$\kappa_o$)$\int$d$^4$x$\sqrt{}$-g$g^{\mu\nu}$($\Gamma_{\mu\nu,\lambda}^{\lambda}$ - $\Gamma_{\lambda\nu,\nu}^{\lambda}$+$\Gamma_{\mu\nu}^{\lambda}\Gamma_{\sigma\lambda}^{\sigma}$-$\Gamma_{\sigma\mu}^{\lambda}\Gamma_{\lambda\nu}^{\sigma}$), where the affine-connection is $\Gamma_{\mu\nu}^{\lambda}$=½$g^{\lambda\sigma}$($g_{\nu\lambda,\mu}$+$g_{\mu\sigma,\nu}$-$g_{\mu\nu,\sigma}$). By just looking at this action one understands better why the Palatini-formalism may be so beneficial when considering $g^{\mu\nu}$ and $\Gamma_{\mu\nu}^{\lambda}$ as independent quantities! But either formalism, if formulated adequately, can be of benefit in explaining certain aspects of cosmology without necessarily invoking the dark energy (a suitable example is the f(R) dark energy models). In either formalism the Newtonian constant is treated by us as a scale field, and, as stressed earlier, for few of our nonlinear action models (but not all) we shall set G=$\mu_o$/f'(R), where $\mu_o$ is some reference value for G and f'(R)=$\partial$f(R)/$\partial$R.

Much has been said, beginning with Dirac, on the theoretical time variation of G and its current observational limits ($-\dot{G}/G$)<$10^{-12}$ (yr$^{-1}$). But the generic understanding of an effective and spatially variable G at the macroscopic level is a more involved tale, since it implies the possibility that G may assume different values in different parts of the universe. The observational limit on spatial variation of G is not encouraging and is indeed compatible with zero variation, for at least astrophysical scales in the range 0.01-5 AU. Nevertheless, the possibility of variable G($\mathbf{x}$) is tantalizing, e.g., for the theoretical modeling of the universe's acceleration, but is even more tantalizing at the microscopic level. The story is this: If one sticks to a constant G and does not modify the Einstein field equations and avoids the dark energy altogether then one gets a universe that is not accelerating and is dominated by matter (like in the old days of cosmology!). But since the acceleration is confirmed observationally then the inclusion of dark energy (constituting about 70% of the universe's total energy) becomes mandatory and this gives rise to a class of cosmological scenarios, among others, based on the cosmological constant viewed as vacuum energy. But why is this vacuum energy not gravitating is still an unresolved issue. For an extended review of our accelerating universe in the context of dark energy, see **[5]**.

In the late 90s the author surmised that the G variations (expectedly also modifying the Newtonian gravity at the low energy) may shed light on the observed Pioneer anomalous acceleration and was able to derive the experimental findings **[6]** by proposing an interesting form for G, **G(r)~e$^{-\mathbf{ro/r}}$**, which we shall use in section 5, while omitting further details over here because they are unnecessary for this paper. On the microscopic level involving the Standard Model particles, spacetime variation of G may have drastic effects at very high energies for quantum gravity, SUGRA and string theory applications. As already stated, the first "rigorous" treatment of variable G at the classical level was given by Brans-Dicke **[4]** in the early 1960s, and basically this is a theory that is still with us today, in spite its more modern modifications and interpretations as an effective field theory of a kind at low energies! The issue of variable G in the context of the KK theory or the string theory/SUGRA is obviously an interesting theme in its own right but it is not a discussable subject for this note.

Some of my own personal and rough views along with other pertinent topics on variable G are included briefly in section 5 at the end of this note, where I'll be discussing some possible reasons for G variations, and also discuss many other topics, including the possible lack of gravity for the fermionic standard model particles, to do away with quantum gravity, and also offer several reasons for why that can happen in connection to a number of scenarios, including variable light speed.

The last thing that we shall need to inject at this point, before beginning section 2, and explore its consequences, pertains to an embellishment of something I've discussed earlier on treating sometimes the determinant of the metric $\sqrt{-g}$, or $\sqrt{-g}$ in combination with some power of variable G (or $\Phi$), as a non-dynamic quantity. In practice, this approach has generally some noticeable simplifying effects for both deriving and reducing the number of terms in the field equations. (We note though some people object to this treatment because it contradicts the GR principle that all metric components, and including $\sqrt{-g}$, are to be regarded as dynamic fields.) It is worth mentioning that if $\sqrt{-g}$ alone is to be treated as a non dynamic field then the variation with respect to the metric $\delta\sqrt{-g}=0$ implies $g_{\mu\nu}\delta g^{\mu\nu}=0$. But by setting, e.g., $\delta(\Phi^\gamma\sqrt{-g})=0$ the above common objection can be mitigated, and at the same time allowing one to derive $\delta\Phi$ in terms of the nonvanishing term $g_{\mu\nu}\delta g^{\mu\nu}$. (For brevity sake I shall explore only the case of $\gamma=1$, and discard the more general possibility that comes about from factorizing, e.g., the gravity action $(1/2\kappa_o)\int d^4x\sqrt{-g}\Phi R$ as $(1/2\kappa_o)\int d^4x(\Phi^\gamma\sqrt{-g})(\Phi^{1-\gamma}R)$ and then setting $\delta(\Phi^\gamma\sqrt{-g})=0$ in the Jordan frame. The latter factorization though is found in few of my calculations to also impart some intriguing properties to the resulting field equations, which I must omit discussing here.) Imposing $\delta(\Phi\sqrt{-g})=0$, e.g., is not as naïve of an imposition as it may appear to be, it determines in part the metric in the presence of the scale field $\Phi$, of a kind, that itself has a functional dependency on the metric. Although we shall mostly limited ourselves to the case $\delta(\Phi\sqrt{-g})=0$, in Model 6 of sect 2 we shall extend the non-dynamicity to the case $\delta(\Phi\sqrt{-g}R_{\mu\nu})=0$ and explore its appealing consequences (one being a lack of a surface term for the EH action)!

Finally, for the purpose of this note (and regretfully) we shall omit any possible torsion and Gauss-Bonnet type Lagrangians in all our actions so that all the affine-connections $\Gamma_{\mu\nu}^{\lambda}$ are symmetrical in the lower indices $\mu$ and $\nu$.

Invoking the torsion may be of necessity to bring closer gauge theory and gravity and also for the geometrization of string theory. Relating variable G and Torsion, at least in part, is also of interest to this author who believes on the possibility of stripping, or mitigating, the role of gravity for the standard model fermions (see section 5). These lines of thoughts, however, are not pursued in the first four sections of this note and are only discussed casually in section 5. In case of fermions, e.g., the intrinsic spin plays an important role as the source of non propagating torsion. But the importance of torsion is mitigated in the macroscopic/cosmological treatment of matter, which is mostly unpolarized. Anyway, our variation convention under $g^{\mu\nu} \rightarrow g^{\mu\nu} + \delta g^{\mu\nu}$ for the determinant with respect to the inverse metric $g^{\mu\nu}$ is $\delta\sqrt{-g} = -\frac{1}{2}\sqrt{-g}\, g_{\mu\nu}\delta g^{\mu\nu}$, and the matter $T_{\mu\nu}$ is defined in the conventional way as $\delta(\sqrt{-g}\mathcal{L}_M) = -\frac{1}{2}\sqrt{-g}\, T_{\mu\nu}\delta g^{\mu\nu}$, and, finally, our metric sign convention is $(-,+,+,+)$.

## 2. Several f(R) Type Action Proposals (Metric-formalism)

***Model 1:*** The lack of a kinetic and a potential term in the action for the scale function $\Phi$ signifies, beyond what we've already discussed, that care must be exercised in regards to interpreting the way $\Phi$ combines with R, or with f(R), for a given action. It is easy, e.g., to look at the combination $\Phi(R)f(R)$ in the action, representing the overall gravity Lagrangian, and then assert that nothing should be fundamentally new in the way of treating the action and its minimization if one simply redefines the combined term $\Phi f(R)$ as another function of R, say $\Psi(R) = \Phi(R)f(R)$. And if so, then one expects to end up with the trivial field equations common to all f(R) theories with a constant gravitational constant (see next paragraph), namely: $\Psi' R_{\mu\nu} - \frac{1}{2}\Psi g_{\mu\nu} - (\nabla_\mu\nabla_\nu - g_{\mu\nu}\nabla_\sigma\nabla^\sigma)\Psi' = \kappa_o T_{\mu\nu}$. We like to avoid, prevent, or restrict, this reasonable assertion, by endowing $\Phi$ with certain proper character to make it special and not as trivial as a simple multiplicative function to f(R), so that the whole term can be redefined as a new function and dealt with as usual. Two attributes that make $\Phi$ somewhat special that we can think of are: (1) make the product $\Phi\sqrt{-g}$ not dynamic, or (2) fix $\Phi$ by requiring $\Phi = f'$ and then use the above field equations. With this said, let me begin with one of the simplest possible proposition for the f(R) gravity action following (2) in the presence of matter field action $S_M$ and involving variable G (which enters the action integrand normally as f(R)/G). Thus, I shall regard $\sqrt{-g}$ as a dynamic field, as usual, but assume $G \sim 1/f'$ (note also: $\kappa_o = 8\pi\mu_o$ and c=1). So the action is now written as:

$$S = (\tfrac{1}{4}\kappa_o)\int d^4x\sqrt{-g}\,(f^2)' + S_M \qquad\qquad (1)$$

The variation of S with respect to the metric yields the field equations 2, and upon contracting eq.3:

$$(f^2)''\, R_{\mu\nu} - \tfrac{1}{2}(f^2)'g_{\mu\nu} - (\nabla_\mu\nabla_\nu - g_{\mu\nu}\nabla_\sigma\nabla^\sigma)(f^2)'' = 2\kappa_o T_{\mu\nu} \qquad (2)$$

$$(f^2)''\, R - 2(f^2)' + 3\square f^{2\,''} = 2\kappa_o T \qquad\qquad (3)$$

Here $\nabla_\mu$ is the usual covariant derivative, $g_{\mu\nu}\nabla^\mu\nabla^\nu = \square$, and $\nabla_\mu\nabla_\nu\Phi = \partial_\mu\partial_\nu\Phi - \Gamma_{\mu\nu}{}^\lambda\,\partial_\lambda\Phi$.

Note that in the orthodox approach to f(R) modified gravity, where G is constant, equations 2-3 still apply provided $\frac{1}{2}(f^2)'$ is replaced by f(R). To be more precise, we have: $f'R_{\mu\nu} - \frac{1}{2}f g_{\mu\nu} - (\nabla_\mu\nabla_\nu - g_{\mu\nu}\nabla_\sigma\nabla^\sigma)f' = \kappa_o T_{\mu\nu}$, and R-$2(f/f') + (3/f')\square f' = (\kappa_o/f')T$. So, as we can see from the trace equation 3, we get an effective gravity coupling $\kappa = \kappa_o/f'$, or $G = G_o/f'$, in this theory, and this was one inspiration for us to set $G = G_o/f'$ in few of our models that are to be discussed in this section!

(Note: An immediate consequence of eq.3, e.g., is in cosmology during the era of radiation dominance (T=0) in an FWR homogeneous universe that gives $\partial_t^2(df^2/dR) = 0$, which in turn implies: $f = [\alpha\int t(dR/dt)dt + \beta R]^{1/2}$. If R is linear in time during that era, which is reasonable, then $f = (At^2 + Bt + C)^{1/2}$. Also note: In a more general case where we do not impose $G = G_o/f'$ but rather assume $\delta(\Phi^{-\gamma}\sqrt{-g}) = 0$ the resulting field equation for constant $\gamma$ is $f'R_{\mu\nu} - \frac{1}{2}[(1+\gamma)/\gamma]f g_{\mu\nu} - \Phi(\nabla_\mu\nabla_\nu - g_{\mu\nu}\nabla_\sigma\nabla^\sigma)(f'/\Phi) = \Phi\kappa_o T_{\mu\nu}$, and as observed the (G variable) conventional f(R) field equations are obtained in the limit $\gamma \rightarrow \infty$.)

To initiate our ***first model*** in this section we specialize next to the case where $f^{2\prime\prime}$ is taken to be a non-zero constant $2a_o$. The result is then the Einstein equation, provided one sets $\mu_o/a_o=G_N$. More precisely, the general solution for this particular case is $f^2(R)=(a_oR^2+bR+c_o^2)$. Setting b=0 gives the standard Einstein equation provided $\mu_o/a_o=G_N$. On the other hand, by also setting b=$-4a_o\Lambda$, where $\Lambda$ is identified with the cosmological constant, the resulting Einstein equation will involve the cosmological constant in the standard form. It is uncomplicated to also find few other possibilities, one viable possibility, e.g., is $f(R)\sim\pm(R^2-R^2_o)^{1/2}$. A more direct way of getting the latter is to set $f^2/G=R/G_N$, which is the standard Einstein-Hilbert action, in combination with $G_N/G=f^2$. This yields at once $f(R)=\pm(R^2-R^2_o)^{1/2}$. This model has been studied extensively in **[7]**. Another simple variant is to set $f^{2\prime\prime}=R/R_o$, this one leads to an effective G that can be tailored to vanish at R=$\infty$ (thus the possibility of preventing curvature singularities!) while converging to $G_N$ at R=0! The general expression for the effective G for any R is: $G(R)=2G_N(R^3/6R_o+\alpha R+\tfrac{1}{4}\alpha^2)^{1/2}/(\alpha+\tfrac{1}{2}R^2/R_o)$.

The study of maximally symmetric spaces for the vacuum (which often results also when the determinant is treated as non dynamic in many general cases) is also of interest (e.g., in the AdS case) in the context of the above field equations (details omitted).

We note that had we treated $\sqrt{-g}$ as non-dynamic in the action of eq.1, then the terms proportional to $(f^2)'$ would be absent in eqs.2-3 (and hence simplifying things a bit!), while also yielding to a "traceless" version of the Einstein field equation. The latter leads to some interesting consequences-like the emergence of the cosmological constant via the Bianchi identity (details omitted)

I shall next list in models 2-4 in below, a number of possible Lagrangians £ (among many others) but my purpose in doing so, at least for most part, is only limited to finding the effective G that result from these Lagrangians, so I'll be forcibly omitting other vital studies. Clearly, many important results like the critical curvature $R_c$ (if any), given by $(\delta£/\delta R)_{Rc}=f^{2\prime\prime}(R_c)/4\kappa o=0$, or the crucial ratio G'/G, for comparing with the existing data on the time variation of G, can be straightforwardly derived for each model once a metric is chosen for the cosmological studies.

-***Model 2:*** For this model we set £=$f^{2\prime}/4\kappa o$=(R-$\mu^4/R$)/2$\kappa o$ in eq.1 and obtain $f(R)=[R^2-2\mu^4Ln(R\sqrt{e}/\mu^2)]^{1/2}$ where $\sqrt{e}$ is included to assure the vanishing of f(R) at R=$\mu^2$. This model has some interesting predictions for the effective G, namely: $G(R)=G_oR[R^2-2\mu^4Ln(R\sqrt{e}/\mu^2)]^{1/2}/(R^2-\mu^4)$ that vanishes as R goes to zero (flat space) and goes to Go as R goes to $\infty$. But there is also a singularity at R=$\mu^2$, in fact for all R<$\mu^2$ we have G<0, whereas for all R>$\mu^2$ we get G>0! The form $f^{2\prime}/4\kappa o$=(R-$\mu^4/R$)/2$\kappa o$ is believed to give rise to a late time acceleration of the universe, see **[8]**.

-***Model 3:*** In this model, we set $f^{2\prime}/4\kappa_o=\alpha R^\gamma/2\kappa_o$ (for $\gamma\neq-1$). The desire for having $f^{2\prime}/4\kappa_o=\tfrac{1}{2}(R/\kappa_o)$ at small R, and $\gamma$=1, can serve as a guide for choosing the solution: $f(R)=(Ro/\sqrt{1+\gamma})[1-\gamma+2\gamma(R/Ro)^{1+\gamma}]^{1/2}$. The reality of f(R) imposes certain conditions on R/Ro for $\gamma$>1 (e.g., for $\gamma$=2, R>$(\tfrac{1}{4})^{1/3}$Ro). Setting G=$\mu_o/f^2$ provides the effective coupling at once: $G(R,\gamma)=(G_N/\gamma\sqrt{1+\gamma})(R/Ro)^{-\gamma}(1-\gamma+2\gamma(R/Ro)^{1+\gamma})^{1/2}$. And evidently, for $\gamma$=1 G(R) is independent of R and is identical to $G_N$. As perhaps noted, for all $\gamma$>1 the reality of G imposes a constraint R>Ro$((\gamma-1)/2\gamma)^{(1/1+\gamma)}$; the ratio G(R, $\gamma$)/$G_N$ also tends to zero as R=$\infty$!

For $\gamma$=-1 we find $f^{2\prime}=2\alpha/R$ which in turn gives $f(R)=(2\alpha Ln(R/Ro))^{1/2}$.The latter has a physical meaning for only R>Ro. Choosing $\alpha=\tfrac{1}{2}\Lambda^2$ yields $G(R)=G_N(R/\Lambda)$ and this implies G$\sim G_N$ for only R$\sim\Lambda$. The above Lagrangian, $f^{2\prime}/4\kappa o=\alpha R^\gamma/2\kappa o$, can also be extended to include a term having an arbitrary negative power of R (we have already alluded to a Lagrangian of the form $R^2-\mu^2 o/R$ earlier), which is certainly a more interesting case.

We may also seek **exponential** forms for the action by, e.g., setting $f^{2\prime}/4\kappa o\sim\exp(-\alpha R)$ and then calculate the effective G with ease, and then go on studying its cosmological consequences via eqs.1-3. Another interesting case is to select the gravity Lagrangian as $\Phi f(R)=\tfrac{1}{2}\Lambda(1-e^{-R/\Lambda})$ with $\Lambda$>0 and then set $\Phi=f^2$. Solving for f and then $\Phi$ gives $f(R)=\Lambda^{1/2}(R+\Lambda e^{-R/\Lambda})^{1/2}$ and $\Phi=\tfrac{1}{2}\Lambda^{1/2}(1-e^{-R/\Lambda})/(R+\Lambda e^{-R/\Lambda})^{1/2}$ respectively. As noted, $\Phi\sim 1/G$ tends to zero for strong curvatures, whereas it vanishes for R~0. Student readers may want to plot the curve of $\Phi$ versus (R/$\Lambda$). Yet, another exponential form of interest for the f(R), and assuming constant G (to simplify life!), has the form f(R)=Rexp(-$\alpha R^\gamma$), where $\gamma$ and $\alpha$ are constants. Such form can then be inserted into the well-known generic field equations for f(R) and then studied in detail. Of particular interest to this author is the special case **f(R)=Re$^{-Ro/R}$** which in both extreme limits gives f(R$\rightarrow\infty$)~R, f(0)=0 and for any R>>Ro: f(R)~R-Ro. By choosing Ro=2$\Lambda$ the latter gives f(R)~R-2$\Lambda$, which is basically the Einstein-Hilbert action in the presence of a cosmological term. This f(R) form deserves to be studied in detail within cosmology and some surprising results may emerge.

*-Model 4:* An interesting gravity Lagrangian model, that to my knowledge has not yet been reported in the literature, is: $f^2/4\kappa_o$=(1/2$\kappa_o$)Ln[(1+sin(R/Ro))/(1-sin(R/Ro)]/cosR/Ro. The latter choice converges to the standard GR form R/2$\kappa_o$ in the limit R<<Ro. The general solution for f(R) is f(R)=½RoLn[(1+sin(R/Ro))/(1-sin(R/Ro)] which tends to R at R<<Ro. What is appealing about this choice is a simple periodic (sinusoidal) form for the Newtonian coupling G, with possible alternating signs, that vanishes at R=0 and at many other curvature values given by $R_n$=n$\pi$Ro)! In fact we find G(R)=$G_N$sin(R/Ro)! Depending on the scale of Ro, the ramifications of this model to cosmology, or astrophysics, or even field theory may be most challenging, and thus deserves further studies.

*-Model 5:* The model to be discussed now is somewhat apart in terms of approach from the earlier models, that we've already presented, in that in the gravity integrand of eq.1, written as $\sqrt{-g}$(f'f/2$\kappa_o$), we shall first impose a specific constraint $\delta$(f'$\sqrt{-g}$)=0, which after some algebra leads to f''$\delta$R=½f'$g_{\mu\nu}\delta g^{\mu\nu}$, and only then see what ensues, thus the difference. The variation of the gravity action under the above constraint (which treats f'$\sqrt{-g}\sim\sqrt{-g}$/g as non-dynamic) gives $\delta S_G$=(1/2$\kappa_o$)$\int d^4x\sqrt{-g}f^2\delta$R which is also equal to (1/4$\kappa_o$)$\int d^4x\sqrt{-g}$(f³/f'')$g_{\mu\nu}\delta g^{\mu\nu}$. Combining the latter with the variation of the matter field action $S_M$ provides the field equations **(f³/f'')$g_{\mu\nu}$=2$\kappa_o T_{\mu\nu}$**, and upon contraction **2(f³/f'')=$\kappa_o$T** (here $\kappa_o$=8$\pi\mu_o$, c=1)). These equations, in turn, imply **$T_{\mu\nu}$=¼$g_{\mu\nu}$T**, which in a perfect fluid model scenario, and similar to dark energy (of mass density about $10^{-29}$ g/cm3), implies a negative pressure p=-$\rho$. Let us now look more closely at the relation f''$\delta$R=½f'$g_{\mu\nu}\delta g^{\mu\nu}$. By using the standard relation $\delta$R=($R_{\mu\nu}\delta g^{\mu\nu}$+$g^{\mu\nu}\delta R_{\mu\nu}$), the latter equation can be cast as: (f''$R_{\mu\nu}$-½f'$g_{\mu\nu}$)$\delta g^{\mu\nu}$=-f''$g^{\mu\nu}\delta R_{\mu\nu}$. The latter surface term is obviously an identity 0=0 for all constant f, while it is inconsistent with f~R! Standard tensorial calculations show $g^{\mu\nu}\delta R_{\mu\nu}=\nabla_\mu\nabla_\nu$(-$\delta g^{\mu\nu}$+$g^{\mu\nu}g_{\alpha\beta}\delta g^{\alpha\beta}$), which is a total divergent, and in the standard general relativity this is integrated over using the covariant Gauss' law to determine the flux of the commonly used infinitesimal vector given by $\zeta^\mu$=$\nabla_\nu$(-$\delta g^{\mu\nu}$+$g^{\mu\nu}g_{\alpha\beta}\delta g^{\alpha\beta}$). But because of the ambiguities involving the derivative of $\delta g^{\mu\nu}$ on the bounding hypersurfaces, it is customary in the metric-formalism (and unlike in the Palatini approach, where $\delta R_{\mu\nu}$/$\delta g^{\mu\nu}$=0) to introduce a counter boundary term in the Einstein-Hilbert action-related to the so-called extrinsic curvature-so to cancel the unwanted boundary flux. The extrinsic curvature has an important role to play in the Hamiltonian version of gravity where the role of "position" and "momentum" variables are now played by the metric on space (at a fixed time) and something related to the extrinsic curvature, and all these are quite handy for the canonical quantization of gravity. When it comes to the Hamiltonian description of GR in the finite (as opposed to infinite) universe, one chooses a definite inertial reference frame which can be taken as the cosmic microwave background (inertial) frame with a cosmic time and the observable space that is foliated, and thereafter proceed with the (long) canonical quantization program. Of course such a preferred (CMB) inertial frame represents a direct challenge to the notion of general coordinate transformation invariance practiced in GR, and so goes the story.

In the above even though thanks to our constraint, yielding $\delta$R=½(f'/f'')$g_{\mu\nu}\delta g^{\mu\nu}$, we do not have to introduce any counter term in the resulting action, which no longer includes $g^{\mu\nu}\delta R_{\mu\nu}$, it is amusing to choose a particular form for f(R) that says $g^{\mu\nu}\delta R_{\mu\nu}$=0! Doing that requires setting f''$R_{\mu\nu}$-½f'$g_{\mu\nu}$=0 (assuming f''$\neq$0) which, in turn, gives f(R)=K+R³/3Ro², where Ro and K are some constants, and the Lagrangian density in this case is £=(R/Ro)²f(R)/2$\kappa_o$. The curvature R is then related to T as R=Ro($\kappa_o$T/Ro)$^{1/5}$, and therefore is only weakly dependent on T. (Having R scaling as 1/5 power of the combined density and pressure in, say, the perfect fluid model is, the least to say, quite special and, moreover, it leads to R<0 as in the deSitter/antideSitter spaces!) The effective coupling in this case is G=$\mu_o$(Ro/R)² which diverges at the flat space limit R=0 (indicating strong gravity!) that can have a drastic impact on the very early universe geometry. And if we look at the condition on R for having $G \sim G_N$ we find $R \sim Ro\sqrt{(\mu_o/G_N)}$, implying $\mu_o=(G_N)^{5/3}(8\pi T(today)/Ro)^{2/3}$. To get an order of magnitude for $T(today)$ we set $\mu_o \sim G_N$ and $Ro \sim 10^{-56} cm^{-2}$ and obtain $T(today) \sim 5.10^{-30} g/cm^3$ and I leave it this way without commenting further. Despite deficiencies in this kind of argument, perhaps the most attractive feature in this model was the ability to derive an expression for $f(R)$. This is in contrast to most approaches that usually require plenty of guessing. We got $f(R)$ by simply setting $g^{\mu\nu}\delta R_{\mu\nu}=0$ and by disregarding its other type consequences (like introducing a counter boundary term in the action) that are not of importance in deriving the field equations which further in the above were shown in bold.

Yet, the restrictive condition $g^{\mu\nu}\delta R_{\mu\nu}=0$ implied by the latter derivation of $f(R)$ (which cannot be done in the linear action $\sim R$ approach, and must normally be eliminated by a counter term, or "else" type term) is also an indication that the infinitesimal vector $\zeta^\mu$ introduced earlier must be a covariantly conserved vector: $\nabla_\mu \zeta^\mu=0$. But this introduces an artificial constraint on the variation of the metric-that is not normally a Lie derivative to begin with (so to give us something to work with)-that we do not desire to have! So we shall leave $f(R)$ in the above field equations arbitrary, as it is conventionally done, and keep in mind that we can always introduce a counter boundary term, as in all the standard $f(R)$ and GR approaches, if ever needed, and be done with it. Armed with the above field equations and an arbitrary $f(R)$, we can proceed as usual with the study of the field equations and the effective coupling for diverse choices of $f(R)$. In section II I'll expand further the concepts outlined in this model.

*-Model 6:* This promising model is motivated by an extension of the previous non dynamic conjecture that was $\delta(\Phi\sqrt{-g})=0$. More precisely, the non dynamic conjecture we are following in this model is: $\delta(\Phi\sqrt{-g}R_{\mu\nu})=0$, leading to $\delta(R_{\mu\nu})=-\delta(\Phi\sqrt{-g})R_{\mu\nu}/\Phi\sqrt{-g}$, and $g^{\mu\nu}\delta(R_{\mu\nu})=-R\delta(Ln(\Phi\sqrt{-g}))$. Let us explore the consequences of the latter findings in the two cases of the EH and the $f(R)$ actions and see what kind of field equations (and other things) we can create.

First consider the EH action without a CC term. The action is: $(c^3/2\kappa_o)\int d^4x\sqrt{-g}[\Phi R+(2\kappa_o/c)\pounds_M]$. Next perform the metric variation and set $\delta(\Phi\sqrt{-g}R_{\mu\nu})=0$ and c=1. A worthy result, after doing some algebra and the use of $g^{\mu\nu}\delta(R_{\mu\nu})=-R\delta(Ln(\Phi\sqrt{-g}))$, is the gravity action variation $\delta(\Phi R\sqrt{-g})$ is completely free of the surface terms $\sim g^{\mu\nu}\delta(R_{\mu\nu})$ that are now cancelled out! In fact, we find: $\delta(\Phi R\sqrt{-g})=\Phi\sqrt{-g}R_{\mu\nu}\delta g^{\mu\nu}$! And by including the matter Lagrangian variation the simplistic field equations are $\boldsymbol{\Phi R_{\mu\nu}=\kappa_o T_{\mu\nu}}$, which upon contraction yields $\Phi R=\kappa_o T$, where $\Phi$=Go/G. (Notice also $R_{\mu\nu}-R(T_{\mu\nu}/T)=0$, and for a special case $T_{\mu\nu}/T=\frac{1}{2}g_{\mu\nu}$ the field equations mimic the Einstein vacuum field equations, and for the case $T_{\mu\nu}/T=\frac{1}{4}g_{\mu\nu}$, the resulting equations can impersonate the field equations for maximally symmetric spaces.)

At this point one can venture in many directions (including applying the above settings to the perfect fluid model, etc.). What comes next is only one plausible direction to venture. The above field equations can be written as $R_{\mu\nu}-\kappa T_{\mu\nu}=\Lambda g_{\mu\nu}+(\kappa-\kappa_o)T_{\mu\nu}$, where $\kappa=8\pi G$. Next we make the identification: $(\kappa-\kappa_o)T_{\mu\nu}=-\Lambda g_{\mu\nu}$, where $\Lambda$ is a dynamic cosmological constant. So the field equations are now cast as $R_{\mu\nu}+\Lambda g_{\mu\nu}=\kappa_o T_{\mu\nu}$, by contracting this and using $\Phi R=\kappa_o T$ we finally find the coupling $G=Go/(1+4\Lambda/R)$. Rewriting the latter field equation as $R_{\mu\nu}+\Lambda g_{\mu\nu}-\frac{1}{2}Rg_{\mu\nu}=\kappa_o T_{\mu\nu}-\frac{1}{2}Rg_{\mu\nu}$ and by taking the covariant derivative, using the Bianchi identity, and the conservation of $T_{\mu\nu}$ we find $\partial^\mu(\Lambda+\frac{1}{2}R)g_{\mu\nu}=0$. The simplest solution is $\Lambda+\frac{1}{2}R=\Lambda o$, where $\Lambda o$ is now the constant CC (and hence its origin!). Now, the original field equation $R_{\mu\nu}+\Lambda g_{\mu\nu}=\kappa_o T_{\mu\nu}$ can be finally cast as $R_{\mu\nu}-\frac{1}{2}g_{\mu\nu}+\Lambda g_{\mu\nu}=\kappa_o T_{\mu\nu}$ which is the familiar Einstein equations with a constant CC term and a constant gravity coupling Go! The effective coupling G is $G(R)=Go/(4\Lambda o/R-1)$, showing that flat space is deprived of gravity (i.e., $G(0)=0$). And for the case $R\rightarrow\infty$ $G(\infty)=-Go$ which is negative, and the interpretation of these findings is left to the readers! There is still some more we can do before concluding: reconsider our choice $(\kappa-\kappa_o)T_{\mu\nu}=-\Lambda g_{\mu\nu}$ and take the covariant derivative to find $\partial^\mu GT_{\mu\nu}=-\partial^\mu\Lambda g_{\mu\nu}$ and then combine this with $\partial^\mu G$ obtained from $G(R)=Go/(4\Lambda o/R-1)$. After some algebra we get: $\partial^\mu R[-\frac{1}{2}g_{\mu\nu}+4\Lambda o GoT_{\mu\nu}/(4\Lambda o-R)^2]=0$. And the simplest of possibilities is $4\Lambda o GoT_{\mu\nu}/(4\Lambda o-R)^2=\frac{1}{2}g_{\mu\nu}$, which also implies $T=(4\Lambda o-R)^2/2\Lambda o Go$. If this is so, then for flat space R=0, and restoring c, we find $T_{flat}=8\Lambda o c^4/Go$! If we set Go$\sim G_N$, then we can guesstimate the value of $\Lambda o$ in flat space by assuming the flat space to, e.g., contain one electron/$cm^3$! This crude guesstimate gives $\Lambda o \sim 0.74x10^{-56} cm^{-2}$ which is close to the observational estimates.

But naturally we are free to also invoke scalar, spinor or vector type matter fields, or invoke a perfect fluid model for expressing $T_{\mu\nu}$ and its trace and then find $\Lambda o$ in this way, which at this point does not concern us much!

To finish this model let me say few words on the case of f(R) gravity not yet discussed. For the latter case, short of invoking a counter term, we cannot in general get ride of the surface term as in the EH case, even by setting $\delta(\Phi\sqrt{-g}R_{\mu\nu})=0$. In fact, with the action given as $(1/2\kappa_o)\int d^4x\sqrt{-g}(f(R)+2\kappa_o f_M)$, we find after some work: $\delta(\Phi f\sqrt{-g})=\sqrt{-g}\Phi g^{\mu\nu}\delta R_{\mu\nu}(f'-f/R)+\Phi\sqrt{-g}f'R_{\mu\nu}\delta g^{\mu\nu}$. Since f(R) is no longer linear in R we have to now consider several phenomenological options. One reasonable option is to make the surface term as elemental as possible for ease of integration by, e.g., requiring $\sqrt{-g}\Phi g^{\mu\nu}\delta(R_{\mu\nu})(f'-f/R)=\sqrt{-g}g^{\mu\nu}\delta R_{\mu\nu}$, which is the type of surface term we find in ordinary EH action with constant G, implying also $\Phi=1/(f'-f/R)$. The latter relation can lead to a proper equation for f(R) if, as we did many times before, we set $\Phi=f'$, which is obviously an extra option. Then the equation for f is: $f'(f'-f/R)=1$, but as seen this is not a trivial DE, so I leave it this way.

**-*Model 7* (R-dependent periodic Lagrangians):** Another captivating class of models in the variable G category can be proposed by invoking in the gravity Lagrangian periodic sines or cosines in the variable R, or non periodic sinh and cosh, which are basically combinations of exponential functions. We shall concentrate here on the former combinations. Even more intriguingly, we may consider a linear superposition of sines and cosines in the variable R (or R-2$\Lambda$) for the gravity Lagrangian (like the EH Lagrangian~R-2$\Lambda$, with the right limit as R or R-2$\Lambda\rightarrow$0, in form of a Fourier series so that **£~$\sum[a_n\sin(nR/R_o)+b_n\cos(nR/R_o)$]**! This series represents the vacuum Lagrangians prior to invoking the matter fields. In the presence of matter, $\Phi$ may be also included in the sine angles, with the intriguing speculation that the Fourier coefficients may be interpreted as probabilities for the total action to be in this or that universal mode upon specifying an initial Lagrangian state. (This is an imitation of quantum mechanics with an eye on the Everett's many-world interpretation. In the latter, however, one has "generally" complex probability amplitudes as the main tool of the trade, whereas in the probabilistic interpretation of our gravity Lagrangian the Lagrangian is taken as a real function. It is however not inconceivable to complexify the Lagrangian terms to either find a real action, or a complex action which is not an unfamiliar concept to physicists and those familiar with the Hamilton-Jacobi treatment of classical physics!) A superposition possibility for the Lagrangian that may lead to the quantization of the curvature is obviously an area of great impact and deserves an entire paper by itself, so we shall thereafter limit ourselves to only a single periodic term.

A simple choice of a periodic function for f(R) may look like $f(R)=R_o(G_N/G(R))\sin(R/R_o)$ or like $f(R)=R_o\sin(G_N R/G(R)R_o)$, where $R_o$ is some constant curvature (e.g., it can be proportional to the CC term). The major difference between the two forms is that in the former the G and the R part factorize, but not so in the latter, and moreover in the former case we can invoke $G=G_N/f'$ with $f\sim\sin(R/R_o)$ but not so in the latter case. (Another obvious option is to include also a cosmological term $\Lambda$ and take, e.g., $f(R)=R_o\sin[G_N(R-2\Lambda)/G(R)R_o)$ and so on.) It is straightforward to see that at small angles the gravity action in either of the above forms is the standard Einstein-Hilbert action $\int d^4x\sqrt{-g}R/16\pi G$, plus an $R^3$ correction, but in company of variable G. In the former case by setting $G=G_N/f'$ we find $f=R_o\sin(f'R/R_o)$ from which f(R) can be obtained as the solution of the equation $f'=(R_o/R)\arcsin(f/R_o)$. Finally, with the total action (including matter) given by $S_{G+M}=\int d^4x\sqrt{-g}f(R)/16\pi G_N+S_M$, and by choosing $\sqrt{-g}$ to be *non dynamic,* we arrive at the field equations and its contracted version (details omitted).

Other possibilities for the gravity Lagrangians can also be envisaged. One may, e.g., envisage a sinc function $f(R)\sim\Phi R_o\text{sinc}(\Phi R/R_o)$ that I am particularly fund of, or try $f(R)\sim\Phi R\cos(\Phi R/R_o)$. The latter form, e.g., yields an $R^3$ correction term at small angles, but more than that both forms allow the option of getting rid of the $\delta R$ (and hence the surface term) altogether during the variation by relating $\Phi$ to R in a specific way. Let me contemplate the latter gravity Lagrangian $\sim\Phi R\cos(\Phi R/R_o)$ (which is an example of a more general Lagrangian of the form$\sim\Phi Rf(\Phi R/R_o)$) and include also the matter action $S_M$ but without setting $G\sim1/f'(R)$. After performing a metric variation on the overall action and setting $2\kappa_o\delta S_{G+M}=0$ (and choosing $\kappa_o=8\pi G_N$) the result is:

$\int d^4x\{\delta(\Phi\sqrt{-g})R\cos(\Phi R/R_o)+\Phi\sqrt{-g}[\delta R(\cos(\Phi R/R_o)-\sin(\Phi R/R_o)(\Phi R/R_o))-\delta\Phi(R^2/R_o)]-\kappa_o\sqrt{-g}T_{\mu\nu}\delta g^{\mu\nu}\}=0$.

At this point we have to make a choice. Basically we have to select one option among three possible options for the $\delta(\Phi\sqrt{-g})$ term appearing in the above integral, so to derive both the field equations and get the suitable information on the scale field $\Phi$, which by itself is deprived of a proper field equation. How the factor $\delta(\Phi\sqrt{-g})$ is viewed makes a genuine difference. Choice 1 is just the orthodox one in that $\Phi\sqrt{-g}$ is treated as dynamic. Choice 2 assumes $\Phi\sqrt{-g}$ is non dynamic and that means writing $\delta(\Phi\sqrt{-g})=0$. Finally, choice 3 consists in treating only the determinant as non-dynamic, meaning $\delta(\sqrt{-g})=0$. But this is only half of the story. The other half is a pivotal trick for getting rid of the $\delta R$ term in the above integral permanently (and that must be done for any of the three choices), to avoid surface terms. This boils down in allowing only specific forms of the $\Phi$ functions which are solutions to **$(\Phi R/R_o)\tan(\Phi R/R_o)=1$**, which I dub Eq.A. The many possible solutions of Eq.A in terms of the parameter X= $\Phi R/R_o$ are discrete values. To a good approximation, the quantized X are: *$\pm0.86033$, $\pm3.4256$, $\pm6.4373$, $\pm9.5293$, $\pm12.6453$, $\pm15.7713$,* and so on, showing a difference of roughly 3 (and $\to\pi$ at large values) between two adjacent solutions beyond the first fundamental solution. (Note: there is also another option, which is to leave $\delta R$ alone and go on deriving the field equations in the orthodox manner and then live with a surface term, this option however is not pursued over here.)

Having said all that, the resulting field equations for choice #1 are: $-\frac{1}{2}g_{\mu\nu}\Phi R\cos(\Phi R/R_o)=\kappa oT_{\mu\nu}$, and contracting it gives: $-2\Phi R\cos(\Phi R/R_o)=\kappa_o T$. Interestingly, it turns out that at the end the latter field equations of the choice#1 are also the field equations for choice#2, thanks to Eq.A! Thus, from the contracted trace equation, and by using the earlier numerical findings in proper order, we find a quantized relationship between T and $R_o$ with alternating signs, namely: $\kappa_o T/R_o=(-,+)1.2219$, $\pm6.5767$, $(-,+)12.7220$, $\pm18.9546$, etc. In the latter, the first value $(-,+)1.2219$ goes with the fundamental numerical solution of X, which is $X_0=\pm0.86033$, and has the opposite sign shown by the pair $(-,+)$, and so on. Looking at these numbers (which incidentally imply that $T_{\mu\nu}$ is piecewise covariantly conserved!) we realize that the real protagonist is the parameter $R_o$ and its numerical value (but is it microscopic or macroscopic?). The Trace T (energy/volume) is quantized in unit of $R_o c^4/\kappa_o$ and has a non zero "ground-state" value, per se, given by $T_0=(-,+)1.2219(R_o c^4/\kappa_o)$, so clearly the traceless scenarios are ruled out in such models. A reasonable first choice for the macro $R_o$ in cosmology is the cosmological constant~$1.5x10^{-56}$ cm$^{-2}$. Plugging in the numbers we find $T_0\sim(-,+)9.9x10^{-30}$g/cm$^3$ (which is identical to the WMAP spacecraft survey data found for the critical mass density), while the first "excited" state gives $T_1\sim\pm5.28x10^{-29}$ g/cm$^3$. As seen these numbers look quite reasonable. As noticed, there is also a sign ambiguity in these arguments. The $X_0$ sign is generally $\pm$, while the $T_0$ sign is always the opposite of that. If we wish a + sign for $T_0$ then $X_0$ must be negative, and that means the product $\Phi R/R_o$ must be negative. If the $R_o\sim\Lambda$ sign is chosen positive, and the gravity coupling $\Phi$ is also positive, then the curvature R must be a negative quantity, as in de Sitter spaces. Of course other sign options also exist for cosmological applications, and even more so if $R_o$ is microscopic, requiring the usage of the much larger numerical eigensolution range not covered in the above! Undoubtedly, the issues of how the quantized universes of diverse energies can communicate, or our earlier probabilistic conjecture for describing the universe, are interesting topics by themselves but regrettably are also beyond our present scope.

How about choice#3? This one implies $g_{\mu\nu}\delta g^{\mu\nu}=0$, and when combined with Eq.A, remaining valid even in this case, in the previous integral it yields $T_{\mu\nu}\delta g^{\mu\nu}=0$. Meaning, we either have a vacuum solution $T_{\mu\nu}=0$, or have $T_{\mu\nu}=hg_{\mu\nu}$, with h=¼T, and a covariantly conserved energy gives $\partial^\mu Tg_{\mu\nu}=0$. Clearly, a constant or zero T is always a solution which does not seem to be coupled to the findings of Eq.A!

Another promising class of Lagrangians, which is both appealing and intriguing, may look like $f(R)=R\sin(R_o/R)$ or $f(R)=R\cos(R_o/R)$, with interesting limits when $R\to0$ or $\infty$! Readers can easily treat these fellows for the case of constant G by simply using the conventional f(R) field equations that we recall were given by: $f'R_{\mu\nu}-\frac{1}{2}fg_{\mu\nu}-(\nabla_\mu\nabla_\nu-g_{\mu\nu}\nabla_\sigma\nabla^\sigma)(f')=\kappa_o T_{\mu\nu}$ (details omitted). Finally, one may ask whether variable G can have a vanishing value not only at say R=0 but also at some intermediary Ro which we denote by $2\Lambda$? The answer is affirmative, and here is one such a construction for $G(R)=Go[(R-2\Lambda)^2+2R\Lambda Ln(R/2\Lambda)]^{1/2}/(R+2\Lambda Ln(R/2\Lambda))$. As seen $G(\infty)=Go$, $G(0)=G(2\Lambda)=0$! By using the earlier ansatz $\Phi=\partial f/\partial R=Go/G(R)$ one can construct f(R) at once and the result is:

$f(R)=[(R-2\Lambda)^2+2R\Lambda Ln(R/2\Lambda)]^{1/2}$, thus the gravity Lagrangian in this case is $\sim\Phi f(R)=R+2\Lambda Ln(R/2\Lambda)$, and vanishing around $R/2\Lambda\sim0.567$!

## 3. Manipulating the Action.

**-Prelude**: The classical menu applying to all nongravitational matter-field couplings is always this: (1) how matter properties (e.g., charge) create a local field in some background spacetime (and that gives us the local field equations, like the Maxwell's equations), and (2) how a test matter piece behaves locally under said field when externally prescribed (and that gives us the special relativistic equations of motion for matter). With the background spacetime in place, the classical menu can be ultimately amenable to the quantization procedure to give us the relativistic renormalized quantum field theories, such as QCD and the electroweak theories, the Standard Model, GUT, and so forth-basically covering all the well-known nongravitational interactions. But the situation in the relativistic gravity formulation, even at the classical level, follows a very different menu. For one thing, and strictly said, there is no preexisting background spacetime in General Relativity (GR), as proposed by Einstein, to begin with, and that is already at odds with quantum physics, which needs, again loosely speaking, such a background for its formulation. And for another thing, the concept of local gravity field, and the action at distance criterion of classical physics, is removed from the GR menu in favor of the geometrization of gravity and the implementation of the equivalence principle (EP); basically the whole development of the GR theory, as we know it today, is the "math" realization of the EP idea. With the removal of the seemingly physical vector gravity field, a' la Newton, in favor of a (seemingly physical) geometric (and tensorial) character, like the Riemannian metric tensor $g^{\mu\nu}$, the quantization of now highly nonlinear gravity, in form of the metric, becomes beleaguered with all kinds of technical and nonrenormalizability issues. Yet, GR is highly successful at the classical level-aside providing the format for the big bang theory-in predicting hosts of solar and a number of astrophysical observations in (mostly) the weak field approximation. How well it will score at very large cosmic scales, or small microphysics scales, remains to be seen, but it is fair to say "so far so good"!

If one adheres to orthodox GR then the rough menu is this: there is no preexisting spacetime, spacetime's local metric, representing a curved non-Euclidean spacetime, is generated dynamically. It comes to existence locally due to local nongravitational field energies and the matter energy-momentum, if present, attributed to matter and its motion. As for the gravity field energy-momentum tensor, it is anybody's guess! The EP doesn't like it so we honor that. Finally, the theory has also plenty to say on the vacuum spacetime structure which can be Minkowskian, deSitter, and so forth.

On operational ground, the EP serves as the main ingredient for setting up the geodesic equations for the purpose of studying the *kinematics* of matter motion in curved spacetime. In GR the kinematics is dictated by the geodesic equations: $d^2x^\sigma/d\tau^2+\Gamma^\sigma_{\mu\nu}dx^\mu/d\tau dx^\nu/d\tau=0$ characterized by the affine-connection $\Gamma^\sigma_{\mu\nu}$ (itself derived from the metric) epitomizing the effect of gravity, but gravity is not treated as a force! In fact, the analogue of gravity 4-vector force, or rather gravitational force density, in GR is represented by the term $-\Gamma^\sigma_{\mu\nu}T^{\mu\nu}$ which does not even appear in the geodesic equations (providing the particle trajectories in curved spacetime). And this is indeed in sharp contrast to Newton's dynamic equation $md^2\mathbf{x}/dt^2=\mathbf{F}$(gravity)-even in 4-vector form-and its direct linking of the trajectories with the gravity force field! The *dynamics* (once more in contrast to Newton's dynamic law F=ma, yielding the equations of motion, and now defunct) is dictated by the tensorial, and generally covariant, type field equations fusing local geometry and energy-momentum locally. The (dynamic) field equations have the objective of yielding a metric in curved spacetime for a given $T_{\mu\nu}$ so to enable establishing a coordinate system. The field equations themselves are obtained by applying the action-principle to a suitable action integral (e.g., the EH or the f(R) actions) in curved spacetime over the total matter-gravity Lagrangian density. And the extremized action, yielding the field equations, gets it main contribution from the bulk Lagrangian variation, and the induced surface terms are either ignored or are eliminated by means of an added extrinsic curvature counter term.

Consider any decent looking "classical" gravity+matter action, having all the necessary inbuilt ingredients: general covariance, limited number of metric derivatives, and so forth. Now, go on to the quantum domain: the possibility that there can be a breaking of the general covariance for such decent looking classical actions, normally giving rise to also singular Hamiltonians, coupled with the lack of a canonical time evolution, so crucial to the quantization program, is a genuine concern in many quantum gravity (QG) purposes, even if the pre-quantum classical action-foundations are appropriate. The so-called time-problem (or problem of time) is a long debated thorny issue in QG and quantum cosmology. Unfortunately, a candidate "field" of a sort appears to be missing, in at least the orthodox approach, for coping with the time evolution in QG, which, according to GR, and unlike SR, is a dynamical quantity. (Incidentally, there are some indications that the problem of time may not be any longer a problem in a quantum cosmological setting if the light speed is allowed to vary!)

The GR time issue was a concern in the back of my mind, among perhaps other more pressing concerns, for wanting to include variable G in this text, with the inspiration that in some way it could be also related to the seemingly lacking time evolution in classic GR. But at the present I have not much to report on this speculation (unless invoking other fundamental constant variations along with variable G, requiring obviously another paper)! Invoking variable G has also other benefits for us and is especially relevant to the topic of this section, which has to do with the action manipulation that, again, has to do with the reduction, or deformation, of an action that in its initial form involves all the right covariance and other ingredients to be classified as a well behaved candidate for matter+gravity descriptions. Now, why one wants to reduce such actions in the first place if they are adequate? One answer is the actions are already known (e.g., the EH action, or the BD action) and there is not much to add, yet the overall Lagrangian can still be manipulated for various purposes that we shall discuss as we proceed. Another answer is technicalities, the available actions are quite rigid and the resulting field theories are somewhat predictable from the outset (modulo some subtleties). The f(R) actions, which go beyond the previously used actions, are somewhat more exciting than plain GR and open door to more possibilities and diverse field equations to explore. We have already seen few such possibilities in section 2 in conjunction with variable G and other assumptions. But in conventional f(R) approaches the field equation structure is already "prefabricated", per se, and all one has to do is to supply a specific f-form gravity Lagrangian and work out the details of the rigid field equations (recall the "rigid" field equations are given as $f'R_{\mu\nu}-\frac{1}{2}fg_{\mu\nu}-(\nabla_\mu\nabla_\nu-g_{\mu\nu}\nabla_\sigma\nabla^\sigma)f'=\kappa_o T_{\mu\nu}$) and the rest are technicalities and that is it.

There is no real "algebraic" interference, or constraints, per se, in traditional treatments between the matter sector and the gravity sector. Besides, the trivial covariant derivatives in the matter Lagrangian already "know" about the spacetime being curved prior to extremizing the action. We like to see more internal communiqué between the Lagrangian terms that originally look sterile and fixed, so to yield very different and exotic looking field equations in comparison to the f(R) equations in the above, or the GR field equations. We want to enhance communication between the Lagrangian terms by manipulating certain terms together to perhaps cancel each other or generate other terms, etc., for achieving our multifaceted purposes. And that means the resulting field equations, after extremizing the reduced action cannot be anticipated by simply looking at the original starting action, but of course there is also a price to pay, as we shall see soon! Anyhow, that is what we mean by deforming or manipulating the action without invoking new fields in addition to the ones already in existence in the original action. The idea is a bit analogous, at least in spirit, to the game we play in QFT toying around with the fields in a given Lagrangian prior to implementing spontaneous symmetry breaking, where the broken symmetry in most cases is a continuous one (thus the appearance of the massless Goldstone bosons). However, over there one wants to create a vacuum degeneracy (and some times at the classical level) that are no longer invariant under certain transformation group. And one gets there, by resorting to some aesthetic appeal, and by rotating the original vacuum that is invariant under the transformation group, and such rotation neither cost energy nor introduces any new parameter. The symmetry breaking in QFT is of course tailored for the quantum world usage and the reasons for wanting it, in say the standard model, is well documented and we have nothing more to add on this subject. As for us, and inspired by the symmetry breaking procedure, we like to play an aesthetic

Lagrangian game in this section, the game to be played in one or more steps, and as seen in below there are few ways to play this game, and variable G is certainly a crucial element in the game!

**-Part 1 Manipulating the gravity Lagrangian:** In this section and the next one the key word is action manipulation (or distortion) of an already given f(R)+matter action in the Jordan frame. More specifically, we shall promote here an approach for ultimately getting gravity field equations by directly manipulating the full Lagrangian density terms in a manner that sometimes by combining algebraically two or more of them we define a new term(s), while other times we may force the elimination of an existing term by cancellation with another term. The upshot of this is a residual (we call it physical) action, that may look quite different from the original undistorted action, plus a number of constraint equations involving the fields. And all this game is to be played prior to performing a metric variation (performed eventually on the physical action to yield the field equations) in this section, or after minimizing the original action in section 4. We will make a note that in almost all that we shall do the number of fields before and after the action reduction procedure remains unchanged and no new characters are to be introduced, yet it is entirely possible to invoke new fields and then play the game but that is not what we shall pursue over here. The practical aim for carrying out the Lagrangian distortion is to obtain new types of aesthetically appealing, or perhaps simply different, field equations than those normally gotten from the original action. The price to pay for this task is usually manifested in a number of "local" constraint equations between the fields that are basically external to the reduced action (which in turn is subjected to the action principle). It is clear that the field equations thus obtained is not to be viewed independently of the constraints, because both go hand to hand! The ultimate judge in this game is of course what kind of field equations, along with the constraint equations, we can produce and finally how they score in predicting this or that parameter in some real cosmological setting, or microphysics for this matter. No doubt, each field equation is also a statement about the allowable metrics, as in GR. (Again we note the same type of game shall be played in section 4 but only after performing a variation on the original action, in addition we end that section by toying around with the original field equations for a change and leave the original actions intact.) Needless to say that the methodology presented in this section can be easily implemented in the Einstein frame, if desired, which is related to the Jordan frame by a conformal transformation of the physical metric.

Let us now begin the game by commencing with an easy prototype model which is to serve to typify our methods. The aim in this model is to generate a cosmological term from the vacuum gravity action involving only the scale field $\Phi$=Go/G and the curvature scalar R (recall also the dual role of the CC term: (1) it is a QFT vacuum energy density, at least in some scenarios, and (2) it is a crucial factor for determining the large-scale behavior of the universe). We begin with the EH vacuum action in the presence of the scale function $\Phi$: $(1/2\kappa_o)\int d^4x\sqrt{-g}\Phi R$. To realize our aim we distort algebraically the Lagrangian prior to any action variation and cast it as $(1/2\kappa_o)\int d^4x\sqrt{-g}[R+(\Phi-1)R]$. Next, we identify the second term in the integrand as $(\Phi-1)R$=-2$\Lambda$, where $\Lambda$ is a relativistically invariant CC parameter that need not be a true constant (note: expressing the constant CC, or any dimensional parameter, as dynamic and expressible by a scalar field is also enthused by the string theory limit, even though in the CC case no variation of it with time is ever been observed!). This way of relating the CC to geometry (i.e., R) is also interesting because it says $\Lambda$ can be measured only through a gravitational type experiment! The "physical" action (the one to apply the variation on) is now the vacuum EH action in the presence of a (conditional!) CC term $\Lambda$ as determined by the constraint equation $(\Phi-1)R$ =-2$\Lambda$, and the Einstein vacuum field equations are derived by following the usual procedures (details omitted).

The above simplistic procedure for the vacuum action can be now extended to include the presence of the matter fields (i.e., $S_{G+M}$=$(1/2\kappa_o)\int d^4x\sqrt{-g}(\Phi R+2\kappa_o\mathcal{L}_M)$) with an added ingredient (to make it even more original!) that $\Lambda$ be T dependent, where T is the trace of the energy-momentum tensor (note $\Lambda$(T) is relativistically invariant so it is okay). By choosing the constraint $\Phi$=(1-2$\Lambda$(T)/R) we end up with a new (physical) action, prior to any minimization that looks like: $S_{G+M}$=$(1/2\kappa_o)\int d^4x\sqrt{-g}(R-2\Lambda(T)+2\kappa_o\mathcal{L}_M)$. Now we can perform the metric variation yielding: $2\kappa_o\delta S_{G+M}$=0=$\int d^4x\{\frac{1}{2}\sqrt{-g}Rg_{\mu\nu}\delta g^{\mu\nu}+\sqrt{-g}\delta R-2\delta(\Lambda\sqrt{-g})-\sqrt{-g}\kappa_o T_{\mu\nu}\delta g^{\mu\nu}\}$. At this stage we shall make certain (novel) choices to simplify life, as oppose to proceeding in the orthodox way, on the individual terms appearing in the action variation and evaluate each choice in terms of what ensues and what kind of matter condition we obtain. One possible collective choice is: (a) $\delta(\Phi\sqrt{-g})$=0, (b) $\delta(\Lambda\sqrt{-g})$ =0 and (c) $\delta(\sqrt{-g}T_{\mu\nu})$=0. Constraints (a) and

(b) taken together yield: $\delta\Lambda=\frac{1}{2}\Lambda g_{\mu\nu}\delta g^{\mu\nu}$ and $\delta R=\frac{1}{4}(R^2/\Lambda)g_{\mu\nu}\delta g^{\mu\nu}$. (We also find from these two relations another relation $\delta[(R-2\Lambda)/(\Lambda R)]=0$.) When a generic form $\delta\Lambda(T)=\Lambda'(T_{\mu\nu}\delta g^{\mu\nu}+g^{\mu\nu}\delta T_{\mu\nu})$, with $\Lambda'=d\Lambda/dT$, is combined with the above $\delta\Lambda$ we get: (d) $g^{\mu\nu}\delta T_{\mu\nu}=(\frac{1}{2}(\Lambda/\Lambda')g_{\mu\nu}-T_{\mu\nu})\delta g^{\mu\nu}$. Next, we make use of condition (c), which comes to the rescue. This one says $\delta T_{\mu\nu}=\frac{1}{2}g_{\sigma\rho}\delta g^{\sigma\rho}T_{\mu\nu}$ that in turn yields $g^{\mu\nu}\delta T_{\mu\nu}=\frac{1}{2}g_{\sigma\rho}T\delta g^{\sigma\rho}$. Clearly, this condition implies a very specific type of matter T but its nature (be it scalar, electromagnetic, perfect fluid, or what) is of no concern to us at this point! The last thing to do now is to equate the LHS of the latter equation with what we have found in (d) to finally get: $\frac{1}{2}(T-\Lambda/\Lambda')g_{\mu\nu}+T_{\mu\nu}=0$. Contracting the latter gives $\Lambda=\alpha T^{2/3}$, which in turn implies $T_{\mu\nu}=\frac{1}{4}g_{\mu\nu}T$. But we are not yet done, substituting $\delta\Lambda=\frac{1}{2}\Lambda g_{\mu\nu}\delta g^{\mu\nu}$ and $\delta R=\frac{1}{4}(R^2/\Lambda)g_{\mu\nu}\delta g^{\mu\nu}$ from the above integrand in $2\kappa_{o}\delta S_{G+M}=0$ (which also yields an expression for the surface term) gives the final field equations for this model: $\mathbf{g_{\mu\nu}(\Lambda+\frac{1}{2}R-\frac{1}{4}R^2/\Lambda)=-\kappa_{o}T_{\mu\nu}}$. Contracting this and substituting $\Lambda=\alpha T^{2/3}$ gives the relation between R and T: $R(T)=\alpha T^{2/3}(1\pm(5+\kappa_{o}/\alpha T^{1/3})^{\frac{1}{2}})$ which can easily be plotted. Clearly we must have $T\geq-(5\alpha/\kappa_{o})^3$ for having real R, and as also seen $R(0)=0$, but in addition the curvature also vanishes at $T_{o}=-(4\alpha/\kappa_{o})^3$ if the negative sign is adopted! The effective gravity coupling G (upon adopting the solution $(1-(5+\kappa_{o}/\alpha T^{1/3})^{\frac{1}{2}})$ is easily computed: $G(T)=\mu_{o}[-1+(5+\kappa_{o}/\alpha T^{1/3})^{\frac{1}{2}})/(1+(5+\kappa_{o}/\alpha T^{1/3})^{\frac{1}{2}}]$. The choice $G(0)=G_{N}$ fixes $\mu_{o}=2.61803399G_{N}$, and finally G is: $G(T)=2.61803399G_{N}[-1+(5+\kappa_{o}/\alpha T^{1/3})^{\frac{1}{2}})/(1+(5+\kappa_{o}/\alpha T^{1/3})^{\frac{1}{2}}]$. Readers may want to plot G(T) versus T in the interval $\infty\geq T\geq-(5\alpha/\kappa_{o})^3$ to see the effective coupling variation and its intriguing change of signs in this interval. Clearly for $T\rightarrow\infty$ $G(\infty)\sim2.6G_{N}$, and this is where we stop.

There is plenty more the reader can do in manipulating the gravity action by inventing different terms in the gravity Lagrangian and then begin playing the game. Here is a generic example: take the usual gravity+matter Lagrangian $(1/2\kappa_{o})(\Phi f(R)+2\kappa_{o}\mathcal{L}_{M})$ and invent a term $R_{o}f^2$ and then distort the expression $\Phi f(R)+2\kappa_{o}\mathcal{L}_{M}=\Phi f(R)+R_{o}f^2-R_{o}f^2+2\kappa_{o}\mathcal{L}_{M}$. Now, cancel the first term with the third term and use one of our earlier proposition $\Phi=f'$ to end up with $\Phi f'=f'f=R_{o}f^2$ providing the solution $f(R)=R_{o}e^{R/Ro}$, implying $\Phi=e^{R/Ro}$, and thus $G=G_{o}e^{-R/Ro}$ that goes to zero at infinite curvature if $R_{o}>0$! The reduce Lagrangian is $\frac{1}{2}(R_{o}/\kappa_{o})\Phi^2+\mathcal{L}_{M}=\frac{1}{2}(R_{o}/\kappa_{o})e^{2R/Ro}+\mathcal{L}_{M}$ and is ready for the minimization to yield the field equations (details omitted), moreover the term $\frac{1}{2}(R_{o}/\kappa_{o})\Phi^2$ is like a mass term in the action (you can also study the $R_{o}<0$ case and relate it to the CC term and see what you get).

**-Part 2 Manipulating the matter Lagrangian:** Part 1 had to do with distorting only the gravity Lagrangian terms in the full action to generate a CC term with the aid of variable G, subjected to some constraints. But, there is surely more to this than the above two limited examples. For one thing we can, e.g., begin with an f(R) Lagrangian in place of the previous EH Lagrangian. It is worth mentioning that the gravity action manipulation is not only limited to generating a CC term; it can also be extended to produce other terms! As anticipated, there is also another viable option to explore. One can envisage numerous scenarios whereby the matter Lagrangian is distorted instead, and the gravity Lagrangian is untouched (modifying both Lagrangians is only treated once, as seen shortly, even though it is a more intriguing case). Let us discuss briefly one such a matter distortion version as a warm up exercise.

We begin with the earlier EH gravity+matter action in the presence of the scale function $\Phi=Go/G$: $S_{G+M}=(1/2\kappa_{o})\int d^4x\sqrt{-g}(\Phi R+2\kappa_{o}\mathcal{L}_{M}))$. Next we make a constraint choice $\kappa_{o}\mathcal{L}_{M}=\Lambda\Phi(\Phi-1)$ for a class of matter field(s), which we don't care to specify at this point, and treat the CC as a constant so to eventually find $S_{G+M}=(1/2\kappa_{o})\int d^4x\sqrt{-g}\Phi[R+2\Lambda(\Phi-1)]$. Driven by the desire to simplify calculations we further specialize to the case where $\sqrt{-g}\Phi$ is treated as non-dynamic, and this fixes the metric variation of the scale field as: $\delta\Phi=\frac{1}{2}\Phi g_{\mu\nu}\delta g^{\mu\nu}$. Performing the variation with respect to the inverse metric, and after discarding the surface term, we end up with the simple field equations: $R_{\mu\nu}+\Lambda\Phi g_{\mu\nu}=0$, $R=-4\Lambda\Phi$, and also $T_{\mu\nu}=-(\Lambda/\kappa_{o})\Phi^2 g_{\mu\nu}$, and upon contracting the latter we find $T=-4(\Lambda/\kappa_{o})\Phi^2$. Suppose we had required only $\sqrt{-g}$ to be non-dynamic, then what are the field equations this time? The results are: $\kappa_{o}T_{\mu\nu}\{1+R/[2\Lambda(2\Phi-1)]\}=\Phi R_{\mu\nu}$, and $\kappa_{o}T\{1+R/[2\Lambda(2\Phi-1]\}=\Phi R$. A bit of algebra also gives $R=\kappa_{o}T/\{\Phi-\kappa_{o}T/[2\Lambda(2\Phi-1)]\}$ and the very interesting relation $RT_{\mu\nu}=TR_{\mu\nu}$. Having any of these basic field equations at hand, the next natural thing to do is to apply them to, e.g., cosmology and evaluate their overall performance.

Let us now generalize the above warm up example by starting again with the standard linear EH action in the presence of the scale field: $(1/2\kappa_o)\int d^4x\sqrt{-g}(\Phi R+2\kappa_o\pounds_M)$ and assume the matter Lagrangian is independent of $\Phi$, and that $T_{\mu\nu}$ is defined as usual:$-\frac{1}{2}\sqrt{-g}T_{\mu\nu}=\delta(\sqrt{-g}\pounds_M)/\delta g^{\mu\nu}$. As you know, the game now is to (1) break the above decent looking Lagrangian into few terms, while keeping its overall functional "value" unchanged. And (2) reduce it by identifying certain pieces for the purpose of canceling each other, and finally (3) find the field equations on the reduced action which now, and that is important, mixes matter and gravity terms together. The reduced action, prior to its minimization, may no longer have the same exact local symmetries of the original action, and may or may not be diffeomorphism invariant. As I've stressed much earlier, we do not heed much on these issues while carrying out our aesthetic game as long as we get some dandy looking field equations that are rich in message. And if it happens that the field equations, at least some of them, can be reduced to GR at certain limit then so much for the better! The Lagrangian manipulation game over here, as stressed before, is somewhat analogous to the symmetry breaking methods of a Lagrangian in the orthodox QFT, though the distinct purpose and differences between both procedures are to be also noted!

The breaking of the *matter* action and its further reduction (which is evidently a conditional constraint) is naturally not going to be a unique procedure. In part 1 no attempt was made to break the matter Lagrangian, however this case is by no means deprived of real interest. In this part the emphasis is on the matter Lagrangian distortion. Let us now contemplate one version, among others, of matter Lagrangian breaking. Write $\pounds_{G+M}=(\Phi R+2\kappa_o\pounds_M)/2\kappa_o$ and break it as $(\Phi R+2\kappa_o\pounds_M(\sin^2\Phi+\cos^2\Phi))/2\kappa_o$. As easily observed, we could have chosen a different way to break the Lagrangian, e.g., as $\Phi R+2\kappa_o\pounds_M(\cosh^2\Phi-\sinh^2\Phi)$ which we shall not pursue here-even though the latter seems somewhat more appropriate to use because it does not lead to singularities of the type $1/\sin\Phi$ and $1/\cos\Phi$ at some $\Phi$. Of course, other less symmetrical periodic possibilities also exist for breaking the matter Lagrangian. Generally, though one can envisage any type of functions $P(\Phi,R)$ and $Q(\Phi,R)$ (and even include the trace T variable) satisfying $P+Q=1$ and then commence the matter breaking by writing $\pounds_M=\pounds_M(P(\Phi,R)+Q(\Phi,R))$. Here, though, we shall limit our scope to only few elementary periodic functions, or consider the plausible non periodic case $\pounds_M=\Phi\pounds_M+(1-\Phi)\pounds_M$. Next on the menu is the reduction of the above expressions. Here are few revealing examples of our breaking method:

(i) We set $\pounds(\text{reduced})=\pounds_M\cos^2\Phi$ upon choosing $\Phi R=-2\kappa_o\pounds_M\sin^2\Phi$, which in turn determines $\delta\Phi$ in terms of $\delta R$ "locally", and externally to the integration containing the reduced action, which as before is labeled the "physical-action" that always remains in form of an integrand. The local expression relating $\delta\Phi$ and $\delta R$ can be then used to extremize the physical action $(\int d^4x\delta[\sqrt{-g}\pounds(\text{reduced})]=0)$ for obtaining the (physical) field equations. In our terminology the "local" Lagrangian piece $(\Phi R+2\kappa_o\pounds_M\sin^2\Phi)/2\kappa_o$ upon multiplication by $\sqrt{-g}$ and integration is identified as an unphysical action that we then set equal to zero! We also note the identification $\Phi R=-2\kappa_o\pounds_M\sin^2\Phi$ implies negative curvature R<0 if $\Phi$>0.

(A comment is now in order: given the additivity of the action (as physical+unphysical) nothing prevents us from, e.g., multiplying the above constraint $\Phi R+2\kappa_o\pounds_M\sin^2\Phi=0$, which is a "local setting", by $\sqrt{-g}$ and then divide both sides by say $\Phi$ and then perform an integration over it. And again nothing prevents us to extremize the latter with respect to the metric, so to single out the unwanted surface term $g^{\mu\nu}\delta R_{\mu\nu}$ that is then disregarded. Meaning, we can write from the above setting: $g^{\mu\nu}\delta R_{\mu\nu}=2\kappa_o\delta(\pounds_M\sin^2\Phi/\Phi)-R_{\mu\nu}\delta g^{\mu\nu}$, which is now singled out, and then integrate over upon multiplication by $\sqrt{-g}$, and finally perform variation to get some relations. It is interesting to note that we could have also "locally" set $\Phi R+2\kappa_o\pounds_M\sin^2\Phi$ equal to a perfect divergence, instead of zero.)

Anyhow, back to our reduced Lagrangian $\pounds(\text{reduced})=\pounds_M\cos^2\Phi$, we see that the shortcoming for this particular setting is obviously the lack of vacuum field equations in the absence of matter and a cosmological constant term! Thus in models in below we shall henceforth include also a cosmological term $\Lambda$ in the original action. Thus, the new Lagrangian is now: $\pounds=(\Phi(R-2\Lambda)+2\kappa_o\pounds_M)/2\kappa_o$.

(ii) The latest Lagrangian can be broken up as $£=(\Phi R-2\Phi\Lambda+2\kappa_o£_M\cos^2\Phi+2\kappa_o£_M\sin^2\Phi)/2\kappa_o$, and by setting $\Phi\Lambda=\kappa_o£_M\sin^2\Phi$ (which relates directly $\Phi$ to $\Lambda$ and non-gravitational matter, but not to R) we obtain the reduced Lagrangian $£(\text{reduced})=(\Phi R+2\kappa_o£_M\cos^2\Phi)/2\kappa_o=(R+2\Lambda\cot an^2\Phi)/2\kappa_o$. Here again $\delta\Phi$ is related to $\delta(£_M)$ that is external to the action integral involving only the reduced Lagrangian. In more detail, we have: $\delta[\sqrt{-g}\Phi/\sin^2\Phi]=(\kappa_o/\Lambda)\delta[\sqrt{-g}\;£_M]$ and we must now decide again how to treat $\sqrt{-g}$, or $\sqrt{-g}$ in combination with any ''$\Phi$-stuff'', as dynamical or non-dynamical. Suppose we begin in the orthodox way and treat $\sqrt{-g}$ dynamical, then after some algebra we find: $\delta\Phi=(\delta\Phi/\delta g_{\mu\nu})\delta g^{\mu\nu}=\frac{1}{2}\delta g^{\mu\nu}[\Phi g_{\mu\nu}-(\kappa_o/\Lambda)\sin^2\Phi T_{\mu\nu}]/(1-2\Phi\cot an\Phi)$ and this, in turn, can be used for minimizing the reduced action, forcibly having a surface term from the $g^{\mu\nu}\delta R_{\mu\nu}$ piece that is then disregarded. The above finding yields also the metric variation of $\Phi$: $(\delta\Phi/\delta g_{\mu\nu})=\frac{1}{2}[\Phi g_{\mu\nu}-(\kappa_o/\Lambda)\sin^2\Phi T_{\mu\nu}]/(1-2\Phi\cot an\Phi)$.

The field equation is: $G_{\mu\nu}-\Lambda\cot an^2\Phi(1+4\Phi/[\sin2\Phi(1-2\Phi\cot an\Phi)])g_{\mu\nu}=-2\kappa o\cot an\Phi T_{\mu\nu}/(1-2\Phi\cot an\Phi)$ and as seen it admits also a vacuum solution. The Bianchi identity can be used for further insight (details omitted). But, as seen, the drawback here is that the overall equations although rich are quite complicated, and for $G\sim0.8576Go$, which is the solution to $1-2\Phi\cot an\Phi=0$ (recall $\Phi=Go/G$), we encounter a singularity. Whereas, the value $G\sim0.7342Go$, e.g., yields $G_{\mu\nu}-1.471\Lambda g_{\mu\nu}=-8\pi GoT_{\mu\nu}$. So, the choosing $Go=G_N$ and $\Lambda o=1.471\Lambda$, leads to the standard Einstein equation with a CC term $\Lambda o$!

(**Supplementary Note:** The idea of having field equations admitting a singularity at some finite value of $G<Go$ as we'd just encountered in the above seems rather synthetic (but note that the usage of $\cosh\Phi$ or $\sinh\Phi$ instead of the sinusoidal functions we are using for breaking the matter Lagrangian may have been more appropriate to prevent such singularities!). Anyhow, now that we encounter singularities we must do something about it! One possible option for avoiding this issue is to regard the main field equation in the above as two separate equations: (1) $G_{\mu\nu}-\Lambda\cot an^2\Phi g_{\mu\nu}=0$, and (2) $\Lambda\Phi/\sin^2\Phi g_{\mu\nu}=\kappa o T_{\mu\nu}$ (so $\Lambda$ here arises from matter). As seen, now each equation's singularity arises only in the limit $G\rightarrow\infty$, which is more satisfactory than the previous situation. In this case the trace of the first equation yields the curvature R as $R=-4\Lambda\cot an^2\Phi$, while the second one yields $4\Phi\Lambda/\sin^2\Phi=\kappa o T$. For the empty space with $T_{\mu\nu}=0$ we find $\Lambda=0$, which in turn implies $G_{\mu\nu}=0$, and then the second trace equation requires $R=0$ and consequently $R_{\mu\nu}=0$, so the picture is consistent! Also we note other findings: $T_{\mu\nu}=\frac{1}{4}g_{\mu\nu}T$ and $R_{\mu\nu}=\frac{1}{4}g_{\mu\nu}R$, and in case R is assumed constant then the possibility of constructing a maximally symmetric space can be attempted.)

Up to now we have treated $\sqrt{-g}$ dynamical, which in turn led to complicated, albeit interesting, field equations. This is somewhat unattractive, though perhaps right! What we want to do next is to explore what happens if some combination of $\sqrt{-g}$ with $\Phi$ can be treated as non-dynamical so to obtain more simplistic field equations, and more importantly to allow an equation for the scale field $\Phi$ which in the above treatment is unspecified due to the lack of a kinetic term in the action! The starting point is our earlier equation: $\delta[\sqrt{-g}\Phi/\sin^2\Phi]=(\kappa o/\Lambda)\delta[\sqrt{-g}\;£_M]=\frac{1}{2}(-\kappa o/\Lambda)\sqrt{-g}T_{\mu\nu}\delta g^{\mu\nu}$. As before, we are now facing a choice of either setting $\delta[\sqrt{-g}\Phi]=0$, or setting $\delta[\sqrt{-g}/\sin^2\Phi]=0$. The former implies $\delta\Phi=-\frac{1}{2}\Phi g_{\mu\nu}\delta g^{\mu\nu}$, while the latter yields $\delta\Phi=-\frac{1}{4}\tan\Phi g_{\mu\nu}\delta g^{\mu\nu}$. These settings in turn yield at once the field equations. For the first one, e.g., the field equations are: $-2\Phi^2(\cos\Phi/\sin^3\Phi)g_{\mu\nu}=(\kappa_o/\Lambda)T_{\mu\nu}$, while the second setting gives: $\Lambda g_{\mu\nu}/\sin2\Phi\cos\Phi=\kappa_o T_{\mu\nu}$ and consequently $4\Lambda/\sin2\Phi\cos\Phi=\kappa_o T$. Thus, the second choice implies more simplicity in the equations and will be our choice. On the other hand, after performing the variation of the reduced action and by using the second setting for $\delta\Phi$ we find: $G_{\mu\nu}=\Lambda g_{\mu\nu}(1+\cos^2\Phi)/\sin^2\Phi=2\cos^2\Phi(1+\cos^2\Phi)\kappa_o T_{\mu\nu}/\sin\Phi$. Taking the trace of the latter equation and using the T expression just derived from the earlier trace we find the curvature $R=-4\Lambda(1+\cos^2\Phi)/\sin^2\Phi$, which is also an equation for relating $\Phi$ to R.

So far the matter Lagrangian decomposition was implemented only via $\Phi$ by writing $£_{M=}£_M(\sin^2\Phi+\cos^2\Phi)$. Now we want, as another option, to use the same breaking-method that we've outlined in Part 1 but employ instead the sines and the cosines of the curvature scalar R, meaning: $£_M=£_M(\sin^2R/Ro+\cos^2R/Ro)$ where Ro is some reference curvature. (Recall, in Part 1 we left the matter Lagrangian intact and broke the gravity action instead.) One can ask at this point whether there is a deep reason behind matter Lagrangian decomposition. My answer is going to be conjectural. Our conjecture here is that the presence of gravity or variable G (or perhaps more importantly, the presence of both factors) may be the cause of decomposing the matter Lagrangian as $£_M=£_M(\sin^2\chi+\cos^2\chi)$, with $\chi$ (if linear) equal to any of the quantities: $\Phi$, R/Ro, $\Phi$R/Ro, $\alpha\Phi$, $\alpha/\Phi$, $\beta/R$, or powers of these, or related to both $\Phi$ and R in some way (recall again that we do not wish to introduce additional fields besides those, here $\Phi$ and R, included in an original $\Phi$-f(R) action; at least this is the policy in this note).

At any rate, as in part 1, we shall concentrate here only on the EH action, while further in below we shall deal with the Brans-Dicke action instead. Start with the EH action involving a cosmological constant term $\Lambda$: £=[$\Phi$(R-2$\Lambda$)+2$\kappa_o$£$_M$]/2$\kappa_o$=($\Phi$R-2$\Phi\Lambda$+2$\kappa_o$£$_M$cos$^2$R/Ro+2$\kappa_o$£$_M$sin$^2$R/Ro)/2$\kappa_o$, and then set 2$\Phi\Lambda$=2$\kappa_o$£$_M$cos$^2$(R/Ro) to see what type "action" consequences results. The reduced action is now: £(reduced)=$\Phi$(R+2$\Lambda$tan$^2$R/Ro)/2$\kappa_o$, and the matter Lagrangian being equivalent to $\Phi\Lambda$/$\kappa_o$cos$^2$(R/Ro) allows the determination of the energy-momentum tensor at once, provided one knows what to do with the metric variation of $\Phi$. The reduced Lagrangian appears as the "$\Phi$f(R)" theory for the vacuum, so it is tempting, but not followed here, to use our earlier ansatz $\Phi$=$\partial$f/$\partial$R to eventually find G=Go/[1+(4$\Lambda$/Ro)(tanR/Ro/cos$^2$R/Ro)]! Anyhow, from the above setting we find the variation for $\Phi$ but still things remain rather undetermined unless we decide to simplify life considerably and make a decision on whether to treat $\sqrt{}$-g, or it in combination with $\Phi$, as dynamic or not. Such choices simplify the calculations considerably, I choose $\delta$[$\sqrt{}$-g$\Phi$]=0 and then calculate the matter $T_{\mu\nu}$. What we finally find is an expression for the surface term: g$^{\mu\nu}\delta$R$_{\mu\nu}$=-[R$_{\mu\nu}$+¼(Ro/$\Phi$)(cos$^3$(R/Ro)/sin(R/Ro)$T_{\mu\nu}$]$\delta$g$^{\mu\nu}$. On the other hand, the reduced Lagrangian density, which is still under the spacetime integration as an integrand, is suppose to give us the physical field equations upon its metric variation. So the variation result, and upon assuming $\delta$[$\sqrt{}$-g$\Phi$]=0, is finally: $\delta$S$_{reduced}$=0$\rightarrow$$\int$d$^4$x$\sqrt{}$-g$\Phi$[1+(4$\Lambda$/Ro)(sinR/Ro/cos$^3$R/Ro]$\delta$R=0. The latter exhibits piecewise constant curvatures in the vacuum as solutions to (4$\Lambda$/Ro)sinR/Ro=-cos$^3$R/Ro. Obviously, had we chosen say $\delta$[$\sqrt{}$-g]=0 instead, the results would have been very different, and even more so when $\sqrt{}$-g is not restricted! Readers can try another option by using this time $\delta$[$\sqrt{}$-g]=0 and see what ensues.

Let me now study a useful variant of the above method with *constant* G for the purpose of getting an effective f(R) Lagrangian that may not need an evaluation of the surface terms! Start with the orthodox EH action in the presence of a cosmological term and then distort it in the same manner as we did in the above. Namely, write £= (R-2$\Lambda$+2$\kappa_o$£$_M$cos$^2$R/Ro+2$\kappa_o$£$_M$sin$^2$R/Ro)/2$\kappa_o$, and then set c$^4\Lambda$/$\kappa_o$=£$_M$sin$^2$R/Ro (I've restored c momentarily). This yields at once the reduced action: S$_{reduced}$=(c$^3$/2$\kappa_o$)$\int$d$^4$x$\sqrt{}$-g(R+2$\Lambda$cot$^2$R/Ro). As seen, in this appealing scenario f(R)=R+2$\Lambda$cot$^2$R/Ro. The latter for small R<<Ro behaves as ~R+2$\Lambda$(Ro/R)$^2$, meaning a drastic change for small-curvature cosmology if (R,$\Lambda$)>0! It has also poles at various quantized values of R, and reduces to pure EH gravity, at least piecewise, for other quantized values of R in unit of Ro (hence the tantalizing case where matter+CC can mimic pure gravity and also quantize the curvature!). Normally, with the above f(R) at hand we can find, upon dropping the surface term, the field equations at once by using the familiar menu: f'R$_{\mu\nu}$-½fg$_{\mu\nu}$-($\nabla_\mu\nabla_\nu$-g$_{\mu\nu}\nabla_\sigma\nabla^\sigma$)f'=$\kappa_o$T$_{\mu\nu}$. And this would have been the normal process hadn't been for our constraint equation c$^4\Lambda$/$\kappa_o$=£$_M$sin$^2$R/Ro. The latter constraint gives us the opportunity of doing things that we could have not been able to do with a generic f(R), e.g., calculating the exact variation of R! So next we do the metric variation on the latter constraint equation to find $\delta$R. After some lengthy algebra and using the orthodox definition of T$_{\mu\nu}$ we finally find: $\delta$R=[-¼Rotan(R/Ro)g$_{\mu\nu}$+¼($\kappa_o$/c$^4\Lambda$)RoT$_{\mu\nu}$tan(R/Ro)sin$^2$R/Ro]$\delta$g$^{\mu\nu}$. To get the physical field equations all is needed is to perform the (inverse) metric variation to extremize the reduced action, which will surely contain $\delta$R that pleasingly we've already calculated. The field equations are derived without any need to perform a surface integral and that means retaining probably more amount of action "information" in comparison to the standard treatment of GR which normally dispenses with the surface term that may contain, shall we say, some holographic" information). Well, here are the field equations:

g$_{\mu\nu}$$\{$-$\Lambda$+(2$\Lambda$/Ro)cotR/Ro(R+2$\Lambda$cot$^2$R/Ro)/[1+(4$\Lambda$/Ro)cotR/Ro/sin$^2$R/Ro]$\}$+($\kappa_o$/c$^4$)sin$^2$R/RoT$_{\mu\nu}$=0.

As seen, these equations are involved because we have no $\Phi$ to play around with and simplify life by, e.g., setting $\delta$[$\sqrt{}$-g$\Phi$]=0. The field equations are also of the type T$_{\mu\nu}$=¼g$_{\mu\nu}$T, were the trace T(R, $\Lambda$) is easily derivable. Clearly, at R=½$\pi$Ro (or in general at R$_n$=½(1+4n)$\pi$Ro) the world is dominated by the CC term with matter density $\rho$=4$\Lambda$c$^2$/$\kappa_o$, which for $\rho$~10$^{-29}$ g/cm$^3$ yields $\Lambda$~10$^{-56}$ cm$^{-2}$! We also note that if Ro is to be microscopic (?), instead of being of cosmological scale, then the integer n has to be huge. The next obvious step is the application of this model to cosmology for a given $T_{\mu\nu}$ (e.g., the pressureless dust model may serve the purpose), and that's the end of this.

Before exploring the BD theory in below, let me provide briefly an example that we have not seen so far regarding the full Lagrangian distortion when both the gravity and the matter Lagrangians are manipulated at the same time. Take the EH action including a CC term: $£=(\Phi(R-2\Lambda)+2\kappa_o£_M)/2\kappa_o$, and decompose it as $£=[(R-2\Lambda)+2\kappa_o\Phi£_M+(\Phi-1)(R-2\Lambda)+2\kappa_o(1-\Phi)£_M)]/2\kappa_o$. For simplicity we have not invoked periodic forms over here. Next, cancel off the first and the second terms by setting the local constraint as $(R-2\Lambda)=-2\kappa_o\Phi£_M$ and calling it Eq.A. The reduced action is now $S_{red}=\int d^4x\sqrt{-g}(1-\Phi^2)£_M$ and upon minimizing it, making use of Eq.A and setting $\delta[\sqrt{-g}\Phi]$ $=0$, we finally obtain the following field equations: $\frac{1}{2}(R-2\Lambda)\Phi g_{\mu\nu}=(1-\Phi^2)\kappa_oT_{\mu\nu}$, and contracting it we get $\Phi=U[1\pm(1+U^{-2})^{1/2}]$ with $U=(R-2\Lambda)/\kappa_oT$. There is plenty more to say over here, especially in the context of exploring further the constraint Eq.A, exploring the dust model, etc, that is left as exercises to our student readers.

Let us now investigate what we can do with the BD scalar-tensor gravity in the context already divulged in this part. According to BD there is a "real "physical scalar field out there defined as $\Phi=1/G(x)$. The total BD action in the Jordan frame is: $S_{tot}=(1/16\pi)\int d^4x\sqrt{-g}(\Phi R-\omega\partial\Phi\partial\Phi/\Phi^2+2\kappa_o£_M)$, where $\partial\Phi\partial\Phi=\nabla_\sigma\Phi\nabla^\sigma\Phi$ and where $\kappa_o=16\pi$. (For brevity sake I shall not bother with the more general case where the BD action can include a CC term, and where the dimensionless coupling $\omega$ can be $\Phi$ dependent.) Normally, the variation business on the action is to be carried out independently for both the metric and the scalar field to get the pertinent field equations. Here we say yes to $\Phi$ as a physical field, but say no to its independent variation! We want to get its ups and downs through constraint equations and our earlier non dynamic conjecture. So here is the menu. Decompose the matter Lagrangian in some chosen way, like the two most common decompositions we'd employed earlier $£_M(\sin^2R/Ro+\cos^2R/Ro)$, or $£_M(\sin^2\Phi+\cos^2\Phi)$. Let us choose the former decomposition and impose the constraint: $\omega\partial\Phi\partial\Phi/\Phi^2=£_M\sin^2R/Ro$, so the remaining action to be subjected to the metric variation is now: $S=(1/16\pi)\int d^4x\sqrt{-g}(\Phi R+2\kappa_o£_M\cos^2R/Ro)$ looking like few actions we had before! Next, regard the product $\sqrt{-g}\Phi$ as non dynamic (i.e., set $\delta(\sqrt{-g}\Phi)=0$) so to get closer to the Einsteinian mentality that says gravity influence is only through the metric (the strong equivalence principle). Anyway, performing the metric variation of the action S gives $\int d^4x\sqrt{-g}[(\Phi-2\kappa_o/Ro£_M\sin2R/Ro)\delta R-\kappa_oT_{\mu\nu}\cos^2R/Ro\delta g^{\mu\nu}]=0$, and after eliminating $£_M$, and by using the above constraint, we finally get: $\int d^4x\sqrt{-g}[(\Phi-2(\omega/Ro)\cot(R/Ro)\partial\Phi\partial\Phi/\Phi^2)\delta R-\kappa_oT_{\mu\nu}\cos^2R/Ro\delta g^{\mu\nu}]=0$. As seen, one must expect a rather complicated surface term over here. It is, however, possible to reduce the would-be complicated surface term to a more trivial surface term as "uncomplicated" as the one encountered in the orthodox EH action (with $G=G_N$) by simply imposing: $\Phi-2(\omega/Ro)\cot(R/Ro)\partial\Phi\partial\Phi/\Phi^2=\Phi o$, where $\Phi o$ is a constant. So, the "field" equation for $\Phi$ is now: $\mathbf{\Phi^2(\Phi-\Phi o)=2(\omega/Ro)\cot(R/Ro)\partial\Phi\partial\Phi}$. And, upon neglecting the trivial surface term the final field equations read: $\mathbf{R_{\mu\nu}=(16\pi/\Phi o)T_{\mu\nu}\cos^2R/Ro}$ (implying also $T=(\Phi o/16\pi)R/\cos^2R/Ro$). Here, like in few models encountered before, interesting consequences can emerge for the "R-quantized" case when $\cos R/Ro=0$, for which $\Phi\rightarrow\Phi o$, and $R_{\mu\nu}\rightarrow 0$! We also note that no action variation was ever performed with respect to the $\Phi$ field, in contrast to the BD theory, to derive the field equation for $\Phi$, yet we have already an equation for the latter scalar field which reads: $\Phi^2(\Phi-\Phi o)=2(\omega/Ro)\cot(R/Ro)\partial\Phi\partial\Phi$. And by choosing a suitable matter Lagrangian of a kind (e.g., an electromagnetic field) our initial choice relationship $\omega\partial\Phi\partial\Phi/\Phi^2=£_M\sin^2R/Ro$ can become quite handy in evaluating $\Phi$. Interested readers can also try other types of matter Lagrangian decompositions (like $£_M(\sin^2\Phi+\cos^2\Phi)$, or $\Phi£_M+(1-\Phi)£_M$, among others) and explore their consequences (Note: the Bianchi identity also yields $\partial^\mu R[R_{\mu\nu}\tan(R/Ro)+\frac{1}{4}Rog_{\mu\nu}]=0)$.

Thus far we manipulated the matter Lagrangian while considering the BD theory, but it is also possible to manipulate the gravity Lagrangian instead. Here is an example: write the total BD Lagrangian as: $\Phi R-\omega\partial\Phi\partial\Phi/\Phi^2+2\kappa_o£_M=\Phi R(\sin^2R/Ro+\cos^2R/Ro)-\omega\partial\Phi\partial\Phi/\Phi^2+2\kappa_o£_M$ and then impose $R\sin^2R/Ro=\omega\partial\Phi\partial\Phi/\Phi^3$, so the remaining "physical" integrand is $\Phi R\cos^2R/Ro+2\kappa_o£_M$, which becomes the EH action for $R/Ro<<1$ (or $\sin R/Ro=0$) where $\Phi\rightarrow\Phi o$ is an acceptable solution. After performing the metric variation (and also assuming as earlier $\sqrt{-g}\Phi$ as non dynamic) the resulting Field equations are found at once. Naturally, we can say more but I think I've already said enough in relation to this tail and our overall methodology so I'll stop here.

Another class of gravity+matter Lagrangian modification idea that I have on my menu is of somewhat different nature. What we do next is to explore the possibility of getting rid of the inherent nonlinearity that exist in the EH action (with no CC term) that normally arises from the Ricci tensor. To do that first take the EH action: $(1/2\kappa_o)\int d^4x\sqrt{-g}\Phi g^{\mu\nu}R_{\mu\nu}+S_M$ and write it as $(1/2\kappa_o)\int d^4x\sqrt{-g}\{\Phi g^{\mu\nu}(\Gamma^\lambda_{\mu\nu,\lambda}-\Gamma^\lambda_{\lambda\mu,\nu}+\Gamma^\lambda_{\mu\nu}\Gamma^\sigma_{\sigma\lambda}-\Gamma^\lambda_{\sigma\mu}\Gamma^\sigma_{\lambda\nu})-2\kappa_o\pounds_M\}$, and then set $\pounds_M=\Phi\pounds_M+(1-\Phi)\pounds_M$, and finally impose $-2\kappa_o(1-\Phi)\pounds_M=\Phi g^{\mu\nu}(\Gamma^\lambda_{\mu\nu}\Gamma^\sigma_{\sigma\lambda}-\Gamma^\lambda_{\sigma\mu}\Gamma^\sigma_{\lambda\nu})$ as the constraint, which has an exciting interpretation in the limit $\Phi\to0$, 1 and $\infty$! The remaining integrand terms in the action are therefore: $S_{physical}=(1/2\kappa_o)\int d^4x\sqrt{-g}\{\Phi g^{\mu\nu}(\Gamma^\lambda_{\mu\nu,\lambda}-\Gamma^\lambda_{\lambda\mu,\nu})+2\kappa_o\Phi\pounds_M\}$. Next, extremize the action: $\delta S=0=\int d^4x[\delta(\sqrt{-g}\Phi g^{\mu\nu})(\Gamma^\lambda_{\mu\nu,\lambda}-\Gamma^\lambda_{\lambda\mu,\nu})+\sqrt{-g}\Phi g^{\mu\nu}\delta(\Gamma^\lambda_{\mu\nu,\lambda}-\Gamma^\lambda_{\lambda\mu,\nu})+2\kappa_o\sqrt{-g}\delta\Phi\pounds_M-\sqrt{-g}\kappa_o\sqrt{-g}\Phi T_{\mu\nu}\delta g^{\mu\nu}]$, and then take it from there. (*It is noteworthy to mention that in the standard treatment one usually writes $\delta(\Phi R)=(\Phi R_{\mu\nu}+g_{\mu\nu}g^{\alpha\beta}\Phi_{;\alpha\beta}-\Phi_{;\mu\nu})\delta g^{\mu\nu}$, this nice result however is obtained after performing an integration by part and using the Stokes's theorem, but in our case it cannot be used because all our effort here is to toy around with the terms prior to performing any integration by part, if it becomes necessary at all!*) We note that the limit $\Phi\sim1$ is also the weak field limit around the Minkowski spacetime background (i.e., we are now within the linearized gravity limit) because $(\Gamma^\lambda_{\mu\nu}\Gamma^\sigma_{\sigma\lambda}-\Gamma^\lambda_{\sigma\mu}\Gamma^\sigma_{\lambda\nu})$ is small when $\Phi\sim1$. In this limit we can write $g^{\mu\nu}=\eta^{\mu\nu}+\phi^{\mu\nu}$, where the small tensor field $\phi^{\mu\nu}$ represents a (spin-2) perturbation around the Minkowski metric $\eta^{\mu\nu}$, and the metric sign convention is "mostly-plus". Let us now assume $\phi^{\mu\nu}=\Lambda\partial^\mu\partial^\nu\Phi$ so that $\phi^\mu_\mu=\Lambda\square\Phi$, where $\Lambda$ is the cosmological constant and is perceived as the source for the variable G field, in contrast to the "physical" BD scalar field whose source is the trace of the matter energy-momentum tensor. To $O(\phi^2)$ approximation we have $g_{\mu\nu}=\eta_{\mu\nu}+\phi_{\mu\nu}$ and we can then compute $\Gamma^\lambda_{\mu\nu,\lambda}-\Gamma^\lambda_{\lambda\mu,\nu}=\frac{1}{2}(\partial_\mu\partial_\lambda\Phi^\lambda_{\ \nu}-\partial_\nu\partial_\mu\Phi^\lambda_\lambda+\partial_\nu\partial^\lambda\Phi_{\lambda\mu}-\square\Phi_{\mu\nu})$, which in turn can be expressed in terms of $\Phi$ (lengthy details omitted).

To proceed, we shall introduce another simplification that we've used in a more reduced form before. We shall assume that $\delta(\sqrt{-g}\Phi g^{\mu\nu})=0$ which leads to $\sqrt{-g}g^{\mu\nu}\delta\Phi=-\Phi\delta(g^{\mu\nu}\sqrt{-g})$, and by using the identity $\delta(g^{\mu\nu})=-g^{\mu\sigma}g^{\nu\rho}\delta(g_{\sigma\rho})$ we arrive at $\delta\Phi=-\frac{1}{4}g_{\mu\nu}\delta g^{\mu\nu}\Phi$, so the earlier $\delta S_{phy}=0$ reads: $\int d^4x[\sqrt{-g}\Phi g^{\mu\nu}\delta(\Gamma^\lambda_{\mu\nu,\lambda}-\Gamma^\lambda_{\lambda\mu,\nu})-\frac{1}{2}\kappa_o\sqrt{-g}g_{\mu\nu}\delta g^{\mu\nu}\Phi\pounds_M-\frac{1}{2}\kappa_o\Phi T_{\mu\nu}\delta g^{\mu\nu}]=0$. Next comes the elimination of the nonlinear terms in the integrand-due to our earlier setting $\pounds_M=(-\Phi/2\kappa_o(1-\Phi))g^{\mu\nu}(\Gamma^\lambda_{\mu\nu}\Gamma^\sigma_{\sigma\lambda}-\Gamma^\lambda_{\sigma\mu}\Gamma^\sigma_{\lambda\nu})$-upon imposing $g_{\mu\nu}\Phi\pounds_M=-T_{\mu\nu}$, which also gives $4\Phi\pounds_M=-T$ (and $T_{\mu\nu}=\frac{1}{4}g_{\mu\nu}T$). With the last setting the above vanishing integrand yields the field equations: $g^{\sigma\tau}\delta(\Gamma^\lambda_{\sigma\tau,\lambda}-\Gamma^\lambda_{\lambda\sigma,\tau})/\delta g^{\mu\nu}-\frac{1}{2}(1-\Phi)\kappa_o T_{\mu\nu}=0$. Also by using the earlier relation $\pounds_M=-\frac{1}{4}T\Phi$ and the definition $-\frac{1}{2}\kappa_o\Phi T_{\mu\nu}\delta g^{\mu\nu}=\delta(\sqrt{-g}\pounds_M)$ and some algebra we find $2\Phi T_{\mu\nu}\delta g^{\mu\nu}=g^{\mu\nu}\delta T_{\mu\nu}$ which is of importance once we start choosing a $T_{\mu\nu}$ of a kind that must forcibly verify the later relation (also note $\delta T=(2\Phi-1)T_{\mu\nu}\delta g^{\mu\nu}$). Now, by using the earlier relation $\pounds_M=(-\Phi/2\kappa_o(1-\Phi))g^{\mu\nu}(\Gamma^\lambda_{\mu\nu}\Gamma^\sigma_{\sigma\lambda}-\Gamma^\lambda_{\sigma\mu}\Gamma^\sigma_{\lambda\nu})$ and after performing some algebra we find the field equations: $g^{\sigma\tau}\delta(\Gamma^\lambda_{\sigma\tau,\lambda}-\Gamma^\lambda_{\lambda\sigma,\tau})/\delta g^{\mu\nu}+\frac{1}{4}g_{\mu\nu}(\Phi/(1-\Phi))g^{\alpha\beta}(\Gamma^\lambda_{\alpha\beta}\Gamma^\sigma_{\sigma\lambda}-\Gamma^\lambda_{\sigma\alpha}\Gamma^\sigma_{\lambda\beta})-\frac{1}{2}\kappa_o T_{\mu\nu}=0$, and by using the relation $g^{\mu\nu}(\Gamma^\lambda_{\mu\nu}\Gamma^\sigma_{\sigma\lambda}-\Gamma^\lambda_{\sigma\mu}\Gamma^\sigma_{\lambda\nu})=R_{\mu\nu}-g^{\mu\nu}(\Gamma^\lambda_{\mu\nu,\lambda}-\Gamma^\lambda_{\lambda\mu,\nu})$ the field equations can be rewritten as: $g^{\sigma\tau}\delta(\Gamma^\lambda_{\sigma\tau,\lambda}-\Gamma^\lambda_{\lambda\sigma,\tau})/\delta g^{\mu\nu}+\frac{1}{4}g_{\mu\nu}(\Phi/(1-\Phi))[R_{\alpha\beta}-g^{\alpha\beta}(\Gamma^\lambda_{\alpha\beta,\lambda}-\Gamma^\lambda_{\lambda\alpha,\beta})]-\frac{1}{2}\kappa_o T_{\mu\nu}=0$. The latter equation can be now used by the readers to explore the weak field limit of the above work and see what results.

-*Why the CC term is so small*? At this end let us use the methods outlined in this section to show why the cosmological constant $\Lambda$ is so small, normally considered a tricky and still an unresolved issue. Start with the constant G, and c restored, EH Lagrangian in the absence of any CC term and then distort it as: $(c^3R/2\kappa_o+1/c\pounds_M)=(c^3R/2\kappa_o+1/c\pounds_M+\nabla_\sigma V^\sigma-\nabla_\sigma V^\sigma)=[(c^3R/2\kappa_o-\nabla_\sigma V^\sigma)+1/c\pounds_M+\nabla_\sigma V^\sigma]$. Subsequently, place $\nabla_\sigma V^\sigma$ next to the gravity (and not the matter) Lagrangian, and set $\nabla_\sigma V^\sigma=c^3\Lambda/\kappa_o$ to get an overall matter-gravity action $S=(c^3/2\kappa_o)\int d^4x\sqrt{-g}(R-2\Lambda+2\kappa_o\pounds_M/c^4)+\int d^4x\sqrt{-g}\nabla_\sigma V^\sigma$. The last integral is simply $(c^3/\kappa_o)\Lambda x\Omega$, where $\Omega$ is the invariant volume of the 4D system. On the other hand, it is also well known that the last integral is $\int d^4x\partial_\sigma(\sqrt{-g}V^\sigma)$, and in case V (itself a solution to $(c^3/\kappa_o)\Lambda\sqrt{-g}=\partial_\sigma(\sqrt{-g}V^\sigma))$ goes to zero sufficiently rapidly at infinity that integral also goes to zero. Now, for today's cosmology $\Omega$ is proportional to the visible size of the universe, meaning $\Omega$ is certainly large enough for our purpose to set $\int d^4x\partial_\sigma(\sqrt{-g}V^\sigma)\sim0$. This in turn motivates taking $(c^3/\kappa_o)\Lambda x\Omega<<1$, where "1" has a unit of action, and since $\Omega$ is assumed very large then forcibly the CC term must be indeed very small! But, we can also relax the constancy of $\Lambda$ and assume it to be cosmic time dependent in an expanding universe and choose the reasonable estimate $\Lambda(t)x\Omega(t)\sim\Omega^{1/2}(t)$, to make $\Lambda$ proportional to the area of the universe's expanding 3-manifold volume, so to end up with today's reasonable value $\Lambda \sim \Omega^{-1/2} \sim R^{-2} \sim 10^{-56}$ cm$^{-2}$, where R is the radius of the present universe $\sim 10^{28}$ cm. And if so, then $\Lambda$ must have been very large during the pre big bang era where the vacuum energy was equally huge! We also note that by using the Stokes's theorem we have $\int_m d^4x(\sqrt{|g|})\nabla_\sigma V^\sigma = \int_{\partial m} d^3y(\sqrt{|\gamma|})n_\mu V^\mu) = (c^3/\kappa_o)\Lambda x\Omega$, so an interpretation of $\Lambda$ via the 3-volume is also possible as equivalent to some averaged 3D energy density because $V^\sigma$ has the unit momentum/area.

Attempting to model $V^\sigma$ is an intriguing theoretical process, especially if it is set proportional to the curvature R, which is a covariant scalar, or the Ricci tensor to make the geometrization of $\Lambda$ even more manifest. Proposing something like $V^\sigma \sim R\xi^\sigma$ (with $\xi^\sigma$ having unit of momentum, treated as a constant vector or $\partial_\sigma \xi^\sigma = 0$!) is crucial for pinning down the origin of the CC term (now in par with the dynamical fields) in terms of the curvature so to put an end as to why it must exist in cosmology in the first place.

Finally, let me state that in our earlier sinusoidal decomposition of the matter or the gravity Lagrangians, e.g., generically shown as $\pounds_M = \pounds_M(\sin^2\chi + \cos^2\chi)$, we got obviously only two positive sub-Lagrangian terms. Students can study the case of breaking, say, $\pounds_M$ into three sub-Lagrangians with two positives and the last one negative. An example is $\pounds_M = \pounds_M[2(\sin^4\frac{1}{2}\chi + \cos^4\frac{1}{2}\chi) - \cos^2\chi]$. So, begin with the latter by "inventing" your $\chi$ (e.g., why not invoke an extra d.o.f in place of R or $\Phi$?) and then combine terms of your choosing with the gravity Lagrangian terms $(\Phi f(R) - 2\Lambda)/2\kappa_o$ and see what surprising constraints and field equations you can come up with (details omitted). To make things even more exciting you may want to also include the radiation Lagrangian $\pounds_{Rad}$, in addition to gravity and matter, and apply our overall term mixing game to this extended action. Possibilities are indeed abundant. Curious students may want to consider a combination of $\Phi$ and R in the total action in form of sines and cosines or a mixture of sines and cosines and sinch and cosh and try their luck!

## 4. Manipulating the Extremized Actions

In contrast to section 3, where we dealt uniquely with the manipulation of the Lagrangian terms prior to action minimization, in this short section we shall allow variation of a generic action to take place at first and then toy around with the infinitesimal terms to see how we can distort and combining them, and all that prior to performing integrations by part. The objective, if it can be fully implemented, is to step-by-step get rid of those potential term(s), and again prior to performing integrations by part, which would normally lead to the surface term(s). To make the point let us begin with the ***first model***, which basically contains a generic f(R) action and the variable G in the form: $S_{G+M} = (1/2\kappa_o)\int d^4x\sqrt{-g}[\Phi f(R) + 2\kappa_o \pounds_M]$. The next step is extremizing this action with respect to the metric variation in the orthodox way and equating it to zero: $\delta S_{G+M} = 0 = (1/2\kappa_o)\int d^4x[\delta\sqrt{-g}\Phi f(R) + \sqrt{-g}\delta\Phi f(R) + \sqrt{-g}\Phi f'R_{\mu\nu}\delta g^{\mu\nu} + \sqrt{-g}\Phi f'g^{\mu\nu}\delta R_{\mu\nu} - \sqrt{-g}\kappa_o T_{\mu\nu}\delta g^{\mu\nu}]$. Next, we decompose the $T_{\mu\nu}$ tensor as: $T_{\mu\nu} = \Phi T_{\mu\nu} + (1-\Phi)T_{\mu\nu}$ and insert that into the above integrand to end up with six terms. Following that, we set the sum of the first and the fifth term equal to zero, this gives us the field equations (1) $\kappa_o T_{\mu\nu} = -\frac{1}{2}fg_{\mu\nu}$ and therefore $f = -\frac{1}{2}\kappa_o T$, and we are now left with four terms in the integrand! Next, we set the sum of the second and the fourth terms equal to zero and this gives another field equation (2) $\kappa_o T_{\mu\nu} = -\Phi f'R_{\mu\nu}/(1-\Phi)$ and contracting it yields $\kappa_o T = -\Phi f'R/(1-\Phi)$. As the final step we demand the sum of the two remaining infinitesimal terms in the integrand to also vanish, and therefore by following this step-by-step reduction procedure we basically get rid of any would be surface term while getting our field equations. This reduction method however is not necessarily unique and also comes at a heavy price. We have imposed many constraints to get all these equations which at face value may look okay and in line with our exploration desire to see what kind of (hopefully) consistent field equations and rich in content we can fabricate. Equations 1 and 2 in the above, e.g., appear consistent. As for the remaining procedure to be viable we must require (3) $g^{\mu\nu}\delta R_{\mu\nu} = -(f'/f)\delta(Ln\Phi)$. Now by combining (1) and (2) we get $R_{\mu\nu}$ and by using it in (3) and doing some algebra we get (4) $\frac{1}{2}g^{\mu\nu}\delta[g_{\mu\nu}(1-\Phi)f/\Phi f'] = -(f'/f)\delta\Phi$. (We also note that a metric $h_{\mu\nu}$ conformally related to $g_{\mu\nu}$ as $h_{\mu\nu} = (f/f')g_{\mu\nu}$ can be used to write (4) as $\frac{1}{2}h^{\mu\nu}\delta[h_{\mu\nu}(1-\Phi)/\Phi] = -\delta\Phi$.) There is a drastic oversimplification of these equations if in addition to the previous four constraints one imposes another one: $(1-\Phi)f/\Phi f' = a_o$, where $a_o$ is a constant. Then (4) becomes: $\frac{1}{2}a_o g^{\mu\nu}\delta g_{\mu\nu} = (\Phi/(1-\Phi))\delta\Phi$ and the condition for this synopsis to be viable is $\delta Ln(\sqrt{-g}) =$

$(\Phi/(1-\Phi))\delta\Phi=\delta(-\Phi-\text{Ln}(1-\Phi))$, implying finally $\sqrt{-g}=e^{-\Phi}/(1-\Phi)$. There is visibly more that one can do over here by, e.g., utilizing the above trace equations, but I think I've already said enough on this issue!

**-Model 2(Manipulating the field equations):** As discussed much earlier, one can adopt another view that is somewhat different from toying around directly with the terms in the extremized action, and is this: let the action and its variation to be as in the conventional approaches (For example, take any suitable f(R) Lagrangian in the metric formalism with variable G which we know obeys the familiar general field equations: $f'R_{\mu\nu}-\frac{1}{2}[(1+\gamma)/\gamma]fg_{\mu\nu}-\Phi(\nabla_\mu\nabla_\nu-g_{\mu\nu}\nabla_\sigma\nabla^\sigma)(f'/\Phi)=\Phi\kappa_0 T_{\mu\nu}$) and then choose to toy around with the resulting field equations instead. In the early part of this note we saw few examples of this procedure in the context of the Einstein field equations, so we will not discuss those types here and concentrate on the f(R) type field equations instead. An illustration of this is to begin with the special case of the above "f(R)" field equations and write $f'R_{\mu\nu}-\Phi(\nabla_\mu\nabla_\nu-g_{\mu\nu}\nabla_\sigma\nabla^\sigma)(f'/\Phi)=\Phi\kappa_0 T_{\mu\nu}=(\Phi-1)\kappa_0 T_{\mu\nu}+\kappa_0 T_{\mu\nu}$. The latter is gotten from the general field equations by choosing $\gamma=-1$, along with the non dynamic constraint $\delta(\Phi\sqrt{-g})=0$. At this time we can begin our manipulation procedure by identifying the first term in the RHS of that equation with the second term on the LHS of it, namely: $\Phi(\nabla_\mu\nabla_\nu-g_{\mu\nu}\nabla_\sigma\nabla^\sigma)(f'/\Phi)=(-\Phi+1)\kappa_0 T_{\mu\nu}$. Upon contracting the latter equation we arrive at the field equations given by: $-3\Phi\Box(f'/\Phi)=(1-\Phi)\kappa_0 T$. On the other hand, we also have $f'R_{\mu\nu}=\kappa_0 T_{\mu\nu}$ and by contracting it we get $f'R=\kappa_0 T$. Thus, combining both findings yields: $\Box(f'/\Phi)=(-1+\Phi)R(f'/\Phi)$. And the special case of constant T, or R, simplifies things even further. Armed with the above bunch of equations one can go ahead and explore their ramifications in cosmology, and perhaps microphysics.

Expectedly, the idea of toying around with the field equations, and leaving the action to be eventually sorted out, is obviously a much broader method than simply toying around with the actions as we did previously. The only basic requirement is to make a reasonable guessing on the choice of the terms to be manipulated. This guessing can be further guided by, e.g., respecting the Bianchi identity and possibly the well-known conservation laws, but as for the rest the limit is the limit of your own imagination and your sense of aesthetics! One can even distort the field equations to the point that they look quite distinct from any of the orthodox ones. Here is an example: Imagine, or decree, if you will, that the field equations pertinent to "real" gravity must look like: $\mathbf{G_{\mu\nu}n^\mu n^\nu=\chi sin(\kappa_0 T_{\mu\nu}n^\mu n^\nu/\chi)}$, where $\chi$ is a dimensionful constant or a function (e.g., $\chi$ can be the CC term, or be $\kappa_0 T$, as two possibilities among many others) and $n^\mu$ is a constant vector (e.g., a null vector, or a unit vector of a kind). At this stage the issue of whether this vector sets a preferred frame, or what, is of no immediate interest. The proposed field equations for large $\chi$ values become $(G_{\mu\nu}-\kappa_0 T_{\mu\nu})n^\mu n^\nu=0$. And as seen, the limit of large $\chi$ is at least compatible with the traditional Einstein equations in that $G_{\mu\nu}-\kappa_0 T_{\mu\nu}$ can be either zero or be of a more general superposition of terms like $-\Lambda g_{\mu\nu}+An_\mu\varepsilon_\nu+Bn_\mu n_\nu$. Then Einstein equations in the presence of a CC term is obtained when one sets A=B=0. (Perhaps an, easier to digest, equation to propose-easier in the sense of ultimately getting the underline f(R) action more easily, and maybe invoking at the same time some Lie derivatives in the argument to help us move ahead-may be an equation of the form $\kappa_0 T_{\mu\nu}n^\mu n^\nu=\chi sin(G_{\mu\nu}n^\mu n^\nu/\chi)$. This one also gives the same "Einstein" results as the one in the above when $\chi$ is large, and presumably $\chi$ may be large in regions of the spacetime if the present universe is Einsteinian!)

In general, it is not hard to find many variants to the above field equations, which look even more distinct from the Einstein equations when $\chi$ is not necessarily small. Few examples involving either the sinusoidal or the exponential superposition of terms are: (1) $(G_{\mu\nu}-\kappa_0 T_{\mu\nu})n^\mu n^\nu=\chi sin[(G_{\mu\nu}-\kappa_0 T_{\mu\nu})n^\mu n^\nu/\chi]$, which is a sinc type function yielding solutions as functions of $\chi$. And (2) the exponential ones chosen as $(G_{\mu\nu}-\kappa_0 T_{\mu\nu})n^\mu n^\nu=\chi\{exp[(G_{\mu\nu}-\kappa_0 T_{\mu\nu})n^\mu n^\nu/\chi]-1\}$. And perhaps the more interesting of these types is (3) $G_{\mu\nu}n^\mu n^\nu=\chi sin[\kappa_0 T_{\mu\nu}n^\mu n^\nu/\chi]+\psi sin[G_{\mu\nu}n^\mu n^\nu/\psi]$, where the vacuum geometry is constraint by the sinc function, becoming specified when $\psi^{-1}$ is specified, representing the superposition of both energy and geometry. One can also draw other types of conclusions for the latter field equations (like in the limit of large $\psi$ the quantization of matter $T_{\mu\nu}$ may take place independently of the geometry!) but I shall have to stop here. Before leaving this tour of frenzy, though, let me go back to the very first field equations we'd proposed, $G_{\mu\nu}n^\mu n^\nu=\chi sin(\kappa_0 T_{\mu\nu}n^\mu n^\nu/\chi)$, and explore few of their features.

First of all we note that the field equations are invariant under shifting either $G_{\mu\nu}$ or $\kappa_o T_{\mu\nu}$, or even both, by an amount proportional to $-\Lambda g_{\mu\nu} + An_\mu\varepsilon_\nu + Bn_\mu n_\nu$. This observation can have many consequences for gravity, and few of them may be related to the mainstream research in GR! This observation however is a trivial one and there is indeed a less trivial one around due to the sine function itself: Let us ask what happens to the geometry if we change $T_{\mu\nu} \rightarrow T_{\mu\nu} + \tau_{\mu\nu}$? It is clear that if the added $\tau_{\mu\nu}$ term is chosen as antisymmetric then the answer is nothing happens to the overall $G_{\mu\nu} n^\mu n^\nu$ term (this is obviously in sharp contradiction with the geometrical conclusions drawn from the Einstein field equations!). But interesting results can arise when $\tau_{\mu\nu}$ is chosen to be symmetric in that if we insist in the invariance of $G_{\mu\nu} n^\mu n^\nu$ then the following quantized rule is observed: $\kappa_o \tau_{\mu\nu} n^\mu n^\nu = 2n\pi\chi$! Now, a shift in $T_{\mu\nu}$ is obviously related to a shift of a kind in the matter Lagrangian, $\pounds_M \rightarrow \pounds_M + \Omega(x)$ subject to the condition $\delta(\sqrt{-g}\Omega) = -\frac{1}{2}\sqrt{-g}\tau_{\mu\nu}\delta g^{\mu\nu}$. And in case the term $\sqrt{-g}\Omega$ is treated as non-dynamic then there is constraint $\tau_{\mu\nu}\delta g^{\mu\nu} = 0$! (Clearly, any choice of the form $\tau_{\mu\nu} \sim \alpha g_{\mu\nu} f(x)$ will do the job, but it implies also the vanishing of the integer n.) Anyway, the spacetime geometry $G_{\mu\nu} n^\mu n^\nu$ is compatible with the vacuum solutions $G_{\mu\nu} = 0$ for $T_{\mu\nu} = 0$, but moreover we have also $G_{\mu\nu} n^\mu n^\nu = 0$ every time $\kappa_o T_{\mu\nu} n^\mu n^\nu = n\pi\chi$, and that may be of some theoretical interest.

Let me conclude this section by saying few words on toying around with the traditional Einstein field equations to complement the procedure we followed way before. If we take the most simplistic Einstein equations with constant $G_o$ and no CC term then we can show that this can lead to the Einstein equations in the presence of both a CC term and variable $G = G_o/\Phi$ which are also derivable under some constraining conditions. To see that, begin with $G_{\mu\nu} = \kappa_o T_{\mu\nu}$ (yielding $R = -\kappa_o T$) and rewrite it as $G_{\mu\nu} = \Phi\kappa_o T_{\mu\nu} + \kappa_o(1-\Phi)T_{\mu\nu}$ and then make the simple identification: $\kappa_o(1-\Phi)T_{\mu\nu} = -\Lambda g_{\mu\nu}$ (implying $\kappa_o(1-\Phi)T = -4\Lambda$). This now leads you, as advertised, to the Einstein field equations that include both variable $G(x)$ and the CC term, namely $G_{\mu\nu} + \Lambda g_{\mu\nu} = \Phi\kappa_o T_{\mu\nu}$ and moreover $-R + 4\Lambda = \Phi\kappa_o T$! The so derived CC term and variable G are not of any arbitrary nature, but are rather generated in accordance with the above constraints that serve to define them. Now you can combine the above three trace relations to find these quantities explicitly (details omitted).

## 5. Several Reflections on Variable G Ontology

This section is left open for raising and asking many questions of diverse nature and speculating freely on matters of gravity in general. However, we shall spend more time on two peculiar and rather broad issues: The first one is the possibility of fermionic matter fields, as classified in the standard model (SM), being either deprived of orthodox gravity coupling or coupling in an unusual manner, and the other issue is exploring the phenomenology of variable G in a broad range of topics with the goal of yielding always finite and weak gravity at all scales, including zero. I have said enough on variable G throughout this text in the context of generic $f(R)$ and the EH actions, so I will not repeat those over here. I shall begin instead with a detour to advance several new and personal reflections on variable G.

Whatever the variation of G may mean theoretically, the existing cosmological data on its variation with the cosmic time scale are quite dismal, and even more so when it comes to its spatial variations. But on the small scales things can be quite different, and in particular we desire to have a vanishing G at scales much smaller than some fundamental scale that is needed to design G. Most of our attention in this section is therefore focused on the spatial variation of G with respect to a fundamental length scale. The subject matter of why G, or other fundamental constants, which I'll omit discussing here for brevity sake, may be a variable is not much of an issue. After all, even the non-Lagrangian renormalization group equations (RGE) approach, e.g., implies such a variation, albeit as a function of momentum. (We also remark that the gravity coupling G, responsible for the nonrenormalizability of gravity, is somewhat peculiar as compared to the other three dimensionless fundamental coupling constants, yielding renormalizable theories, because it is a dimensional entity. Yet, another peculiarity is that divergences can appear, according to Larsen and Wilczek, in renormalizing G in the perturbative quantum gravity approach due to the nonminimal matter-curvature couplings, see **[9].**) Rather, what is at issue is what really causes G to vary in the first place, from, say, the field-theoretic perspectives? And by that we mean what is the underline dynamic basis for variable G in the physics sense, perhaps interrelating local geometry of spacetime with the local energy of the underline bosonic and fermionic particles that after all constitute our universe.

From the modeling stand, there are hosts of scalar-tensor field theories around, the least to say, and all can model $G(x)=G(x^\mu)$ in one way or another. But even in these theories the correct epistemic norms for such a variation (modulo the Mach principle argument!) is ambiguous, and that is not meant as a criticism because we were also facing the same ambiguity in the previous sections. And then there is the string theory, endowed with a dilaton "physical" field, and few other none stringy proposals, around that demand variation of not only G but all dimensional constants. But, despite all our current faith in string theory we don't know how far this theory can stretch in the future, or possibly mutate, to shed enough light on the G variation at all scales. In short: presently nobody has any concrete answer on the G variation ontology, including this author. My own sense is that ultimately the variation of G must be studied in the context of microphysics in combination with the "nontrivial" spacetime variation of the other fundamental constants, notably ℏ. Variable ℏ is certainly the one having the most impact in microphysics and I have done a great deal of work and given seminars on this wide topic but alas have not yet concluded the multifaceted study! With the exception of the light speed variation, discussed at the end of this section, and its possible relation to variable G, we shall not elaborate further on the variation of the other fundamental constants like ℏ in this section, for an extensive review on variation of the fundamental constants the readers may want to consult reference [10] by J. P. Uzan.

Plainly speaking, it is not unreasonable to anticipate the G-variation cause(s), whatever it may be, has to be related to the microscopic property of matter, or topology (!), or be related to the macroscopic nature of the universe, as epitomized by its dark matter/energy and its cosmic time scale. But there are also recent conjectures that variable G, apart from all the undergoing RGE activities, may be the result of the compactification to 4D of an internal space of a gravity theory of a kind (e.g., supergravity) in higher dimensions. There are also recent proposals to tie the time variation of G to perhaps Lorentz-violating dark energy (see [11]). Dark energy evolution may also cause variation of other fundamental constants and a variation in the proton to the electron mass ratio (see [12]). While all these activities proceed their evolutionary course, one may also speculate, and on very different grounds, that perhaps variable G may be also an indication that the weak version of the equivalence principle (WEP) in form of m(inertial)=m(gravitational)-so crucial for formulating the GR kinematics-although an excellent approximation (as verified in so many experiments involving all types of test objects) may not be a strict equality for everything we know in this universe, and even more so in the quantum world of the building-block particles as appearing in the standard model. (Note: henceforth we shall bare in mind that throughout this section the abbreviation WEP shall always mean for us the equality $m_i=m_g$ and must not be confused with other definitions.) Once we are ready to "reasonably" challenge the exact equality $m_i=m_g$, for say a category of particles, then we are opening doors to loads of speculations. And talking about speculation here is the very first one in this section: can variation of G for a class of fundamental particles be related to the variation of their gravity masses $m_g$ acting as sources? This may sound novel, but how to formulate it is a different matter! Towards the end of this section I will contemplate that "scalar" gravity for the bosonic SM particles (or the lack of for the fermionic SM particles!) may be a direct result of light speed variation of a kind. And, moreover, our desire for wanting always a weak and finite gravity at all separations r (including zero) may demand a variable G of the form $G(r)\sim e^{-r_0/r}$ in the static limit. The latter though is only one viable form among others, but it is also my favorite one because of its simplicity and sufficiency for doing the job. (It is noteworthy to stress that there have been attempts in the literature to modify gravity's coupling to matter at high energies. One prominent model is the so-called "fat gravity" which basically says gravity is suppressed at short-distances below some 100 microns.)

It is worth mentioning that ample articles exist in the literature dealing with the important topic of VSL (variable speed of light). Many such articles attribute the VSL as a generic quantum phenomenon in nontrivial vacua (e.g., E&M or gravitational fields) stemming from one or more loop corrections, or at classical level as due to the non commutativity of space, where in general the Lorentz symmetry is lost [13]. There are also indications of VSL in Minkowski spacetime in the presence of a constant background field. But there is much more to this (older and newer) VSL story, that has been around ever since the advent of SR (and even earlier!). The modern approach to VSL was initiated in the seminal paper by J.W. Moffat on the spontaneous breaking of the Lorentz symmetry under inflation. Interested readers can learn about these issues and much more by consulting [14]. As we shall see later on, for us the emergence of VSL arises for different reasons; one may be traced to the peculiarity of the modified LT forms that we shall propose to formulate a new form of relativity involving not one but two maximal speed limits per particle: one for light and the other(s) for massive elementary particles.

Whatever the viable theoretical reasons are (and there are ample) for wanting to impose an exact $m_i=m_g$-forcibly entailing an equal treatment of all sorts of matter-energy, regardless of the spin statistics-the experimental fact is this. At this writing we have no concrete evidence that, e.g., free electrons, or, say, muons really fall under gravity! (More precisely, I know of only one existing free-fall experimental technique performed 40 years ago by Witteborn and Fairbank on electrons in the vacuum, enclosed by a copper tube, with an almost zero result. The no fall result for the electrons was due presumably to the cancellation of an induced electric field outside the tube with gravity force! In the authors own words: "The force was shown to be less than 0.09mg…This supports the contention that gravity induces an electric field outside a metal surface, of magnitude and direction such that the gravitational force on electrons is cancelled", see [15]. This experiment is at best inconclusive, yet we shall keep in mind the negative conclusion!)

Currently, we do not know, e.g., whether the time-of-flight of an electron is mass and spin independent. Nor we can claim to understand the origin of inertia-gravity coupling with the spin! And all that despite the fact that atoms containing electrons do fall, and even there are indications that individual quantum neutrons feel the gravity, thus indicating that gravity is felt at least at the hadronic level (an obvious conclusion!). Since we have no concrete experimental proof currently on whether an electron falls under gravity, then a speculation on the electron mass like $m_i \neq m_g$ cannot be rules out, provided it can benefit us in some way. But if somehow the suggestion $m_i \neq m_g$ is tied to the universe at large then the technical issues are how the universe of certain scalar curvature R breaks the exact equality between the inertial and the gravitational mass of, e.g., the quantum electron? (After all, the only quantum effect on the curvature in cosmology is thought to be related to the cosmological constant.) It is well known that in the context of the Schwarzschild metric in GR even though the equality of the inertial and the gravitational mass is established at rest, both masses are not necessarily equal in motion, and the moving gravitational mass $m_g=m_i\sqrt{(1-v^2/c^2)}$ may vanish at the v=c limit. Anyway, there may be various road maps to explore this involved issue within the framework of the Newtonian scheme of gravity, or even within the Machian doctrine, but not so if we adhere to the stringent classical GR, which is so special of a theory, based on the requirement of general covariance. General covariance implies strong locality, and the covariant continuity of the energy-momentum tensor and down playing that principle is like playing with fire! As for the quantum gravity world, we can only speculate on whether the WEP is preserved, but if there is a limit on locality, in form of say an elemental quantum time, then the interpretation of any WEP breaking becomes even more exigent. And one suggestion to make in QG is to appeal to variable G in the microworld to be an element, among others, for inducing the violation of the WEP. Obviously, this is a suggestion that runs contrary to, say, the classic scalar-tensor theories respecting the WEP. G variation can occur without violating the WEP, this is, e.g., the case of the BD theories, which respect the WEP but not the stricter strong EP.

In what follows our discussions on variable G are going to be rather informal and are mostly carried out with the flavor of physics, rather than mathematics, in mind. The math interest may be focused on, e.g., possible fusing of gauge invariance with the diffeomorphism invariance of gravity, and the likes. We shall conduct ourselves as inquisitive physicists and stay away from any rigorous modeling of the G variation based on a detailed mainstream mathematical formalism, and the reason is simple: we have little to present beyond those we've already presented in sections1-3! We shall instead talk and talk, and then talk and act a bit and then act a bit more, and then stop acting and that should take care of variable G! But sometime we also mean business, so let us begin the talking.

Suppose we desire to model the WEP violation of an electron classically, and blame it on the universe's curvature R, then the question is how do we do it? This is obviously an ill defined question; for one thing it does not specify what the scenario for the flat space is. (E.g., do we want a zero gravitational mass at zero curvature, or what?) Anyway, a more clear-cut answer to the above question is nobody knows how to proceed! An electron being a quantum particle requires, so says quantum physics, a background space to exhibit its "quantum-ness". In the absence of any exact menu on how to violate the WEP the only course of action, or talking, is speculation. And talking again about speculation here is a (twisted) Machian toy model, with a fictitious short story behind it to make it more interesting, or perhaps duller: A contemporary tenure track professor working at an elite university announces to his peers his latest finding in a rather loud manner: "behold friends, the gravitational mass of an electron feels the background spacetime curvature R around it and is determined by: $m_g(R)=m_i e^{-Ro/R}$, implying no gravitational mass in flat space, and $m_i=m_g$ at the curvature singularities, and this relation, my valued colleagues, says the obstinate professor, is non negotiable, even if it spells trouble for the action=reaction dogma between two causally separated masses"! Others at school react more cautiously about the news, but finally decide to take a deeper look at the professor's finding. And on the basis of the Newtonian version of the gravity force, $F\sim-G_N m_g{}^2$, they conclude (in a secret meeting, of course) that in order to avoid nonlocalities it is more prudent to absorb the professor's exponential factor into the redefinition of the local gravity coupling- $G(R(x))\sim G_N e^{-Ro/R(x)}$ and assigning a universal Ro to the formula independent of the particles feeling the potential-while leaving the equality $m_i=m_g$ intact, and, yes, dismiss also the professor, and this dismissal was also non negotiable! Learning about his dismissal, the professor vowed to never reveal the secret behind his magic formula to anybody, and thus deprived us from learning more about his monumental discovery!

We, on the other hand, think the professor's formula for $m_g$ does not follow from any first principle of a kind, but taken at face value implies variable G is geometrical, and thus far not implicitly related to the matter fields, so in this sense it becomes easier to, e.g., construct local potentials and mass densities. The moral of the story is that in the relativistic version of gravity one is free to cook up modified Einstein field equations, involving, e.g., the above variable G, and propose an alternative to Einstein's field equations, maybe something like $G_{\mu\nu}\sim\kappa_N T_{\mu\nu}e^{-Ro/R}$, implying also $\partial^\mu R T_{\mu\nu}=0$ and $\kappa_N T=-Re^{Ro/R}$, or equivalently $G\sim-T/R$ in the absence of a CC term. Such modified field equations, or the likes, however present a challenge in terms of finding the corresponding actions! (Readers may want to begin their investigation more appropriately by choosing their gravity action as $(1/2\kappa_o)\int d^4x\sqrt{-g}e^{-Ro/R}R$ and then make use of the field equations $f'R_{\mu\nu}-\frac{1}{2}fg_{\mu\nu}-(\nabla_\mu\nabla_\nu-g_{\mu\nu}\nabla_\sigma\nabla)f'=\kappa_o T_{\mu\nu}$, and $R-2(f/f')+(3/f')\Box f'=(\kappa_o/f')T$ with $f(R)=e^{-Ro/R}R$ and then see what can they do afterwards!)

(It is also possible to adopt an opposite view and say unlike in the above case not only strong gravity must exist but it must be also indicative of large curvature, and then go on proposing a model for variation of G as, e.g., $G\sim G_N e^{R/Ro}$ and, finally, honor the WEP! Having done that, one can follow the orthodox line of including Yukawa type corrections to the 1/r gravity potential (that is anticipated anyway in the gauge theories all the way to the string theory) and lump what these theories say on the corrections into the redefinition of the variable G as $G(r)\sim G_N(1+\alpha e^{-r/\beta})$. Having done that one can then equate both expressions in a spherically symmetric space (ignoring subtleties) to get an expression for the curvature as a function of separation r: $R(r)\sim RoLn(1+\alpha e^{-r/\beta})$, so that R vanishes at large separations, and lastly look at the most recent observational data to determine $\alpha$ and $\beta$! But this kind of analysis, which is closer to the orthodox approach, is not of interest to us because we are seeking a permanently weak and not strong gravity, even at r=0.)

It is evident that the kinds of speculations made so far demand also inventing different types of actions, which regulate the dynamics of GR, beyond the naïve EH action with a constant G! This is fine, but the real problem over here, though, does not stem much from the dynamic field equations, which is still an issue; rather the difficulty comes from the way GR prescribes the other half of the story. That is, the "kinetic" equations for the geodesics of any particle, which are (thanks to the EP) utterly insensitive to the mass parameter, in fact no mass parameter appears at all. So it is not straightforward to see how in a variable G modified GR, based on geometry, these geodesics are to be affected by variable G, if G is not a functional of the metric in the first place. In the original BD theory variable G was not assumed to affect the geodesics, and its source was due purely to the trace of the energy momentum tensor T, while the matter Lagrangian is not to include variable G by virtue of the EP. Apparently the BD prescription is not that clear-cut in the formulas we've proposed earlier in the above. A phenomenological example of variable G, that I can think of, can be realized in the Schwarzchild metric upon replacing the standard line element and using a variable G of the form $G(r)=G_N e^{-r_0/r}$: $ds^2=c^2(1-2MG_N e^{-r_0/r}/c^2r)dt^2-(1-2MG_N e^{-r_0/r}/c^2r)^{-1}dr^2-r^2(\sin^2\theta d\phi^2+d\theta^2)$. This modified metric has the same metric behavior as the orthodox one for $r>>r_0$ plus some small corrections, yet it does not have the usual naked singularity at $r=0$, and moreover the spacetime at $r=0$ is just the gravity free Minkowski spacetime in spherical coordinates! The existence of an event horizon here depends on how $r_0$ compares to the Schwarzchild radius $2MG_N/c^2$, and I leave it as an exercise to our student readers to find the nonvanishing affine connections and the curvature things and then see what they can articulate about the energy-momentum tensor! Students can also envisage an internal world of hadrons by using the above-distorted metric and fix $r_0$ to stand for say the nucleon radius such that an event horizon is possible. To make the event horizon construction realizable around $r_0$ one may set the gravity constant below the event horizon to be many orders of magnitude larger than the Newton's constant. Outside the hadron you may want to keep the usual Newton's constant. This kind of scenario may provide a dandy example of quark confinement via strong gravity (and its possible relation to QCD?). Later on we shall say few more words on the general issue of strong gravity inside the hadrons and the variable G therein. It is interesting to note that the above modified Schwarzchild metric leads automatically to the Reissner-Nordstrom (RN) type metric in the limit $r>>r_0$ provided M is charged and $r_0$ is taken to be about the charge radius of M: $r_0=Q^2/2Mc^2$! It is noteworthy that in the traditional RN metric the charge contribution serves to diminish and not augment the metric, and this observation is in contradiction to what GR says, our proposed metric though begins with a single analytic exponential function that integrates both charge (if any) and gravity in $r_0$ and $2MG_N/c^2$!

(**Important note:** In the Newtonian version of gravity one can phenomenologically force the gravity coupling **F** between a point electron and, say, a much heavier mass M (consider WEP is holding here) with variable G to, e.g., look like this: $\mathbf{F(r)}=-m_e M[\mathbf{r}\hat{\ }G/r^2-\nabla(G/l_0)]$, where $l_0$ is some characteristic length scale (I leave it open whether $l_0$ is microscopic or macroscopic). In this form we can at least get rid of the radial part of **F** for the electron by choosing $G(r,\theta,\phi)=G_N e^{-l_0/r}g(\theta,\phi)$ (and this is the least of what I'll imagine in this paragraph in place of setting **F**=0 for the fermions at all r!). Consequently, the overall "non radial" gravity force now reads as follows: $\mathbf{F(r)}=-(G_N m_e M/l_0)(e^{-l_0/r}/r)[\hat{\theta}\,\partial g/\partial\theta+\hat{\phi}\,(1/\sin\theta)\partial g/\partial\phi]$. It is clear that this residual (rotational) gravity force vanishes at zero separation and for any $r>>l_0$, while reaching its maximum value at $r=l_0$. The kinematics of the electron can be readily studied under such a residual force (defying the orthodox action-at-distance gravity of Newton) in the conventional NR sense to derive the equations of motion with possible intriguing results-like the electron angular momentum-depending on the choice of $g(\theta,\phi)$. If one desires to measure the Newton's constant $G_N$ based on tabletop type experiments, by using, say, electrons, then that is the case of $r>>l_0$, and setting g=1 is certainly the simplest option for getting F=0 and a (presumably) null experimental result! One intricacy though is to get the gravitational potential from the above force expression, which Laplacien is supposed to yield the local mass density. The above force formula is surely not unique, other variants are also possible, but F=0 for the electron is definitely the simplest of all. The moral of this story is one has no problem accommodating the Newtonian gravity, invoking variable G and its first derivative, for our weird electron behaving in a peculiar way under gravity, or rather the lack of it. Spin undoubtedly must play a dominant role to justify the emergence of the above force formula in the first place, but how this can be justified from any first principle is not clear. Finally, the above phenomenological force formula is clearly not the orthodox Newtonian limit of any known GR model at weak field limit and that has surely dire consequences for a modified formulation of the relativistic gravity.)

On the microscopic scales things are generally more involved in case one wants to design an alternative local theory of gravity to explain why, e.g., $m_g=0$ can happen for an electron. One usually runs into an immediate trouble here because of the EP in connection to the local laws of quantum field interactions, pertaining to non-gravitational forces, and basically the lack of a full quantum theory for gravity. The latter schemes are all traditional theories and are understandably not equipped to deal with cases like $m_g=0$ when $m_i\neq0$. Even in the most revered string theory one still adheres to the EP (modulo subtleties), and the same is true for the WEP respecting scalar-tensor theories, and basically the more modern likes!

But what happens, in theory, if, say, free leptons do not feel gravity, or say curve the spacetime differently from the GR mathematical expectations? Or even better, let me speculate (as I did few years ago in several seminars and unpublished technical essays) that our conjecture of "no-gravity", or "less-gravity" be extended to all matter fields of the Standard Model (including possibly to the SUSYSM extension, if SUSY is ever confirmed), and that means free leptons/antileptons and (the unobserved) quarks/antiquarks and possibly the leptoquarks as well. Then what? One reply may be that GR does not apply to the fundamental quantum matter fields of the SM, at least in its current form. But any reason for that? Nobody can provide a clear answer to the extreme oddities of these kinds, be it in string theory, induced gravity, loop quantum gravity or what have you, which all respect the EP! A partial reason, if gravity is nonexistent may be that gravity fades because it is somehow sensitive to the leptonic and the fractional baryonic numbers, or simply sensitive to the spin ½ nature of these matter fields (that happen to be mostly charged, and that is important for the QED-mass role, with the exception of neutrinos) in a manner that defies conventional understanding of building block fermions in the curved spacetime, where the Poincaré invariance is replaced by the EP, let alone the full quantum gravity. By conventional here we mean the Dirac equation expressed in terms of the vierbein and the spin connection. Another speculation for possible gravity demise may be that whatever the underline reason, the SM fermions are simply deprived of any gravitational mass (even at the Newtonian level!), or in a milder version of this one may say for the SM matter fields $m_i \neq m_g$. A variant to this scenario may be that only leptons do not feel gravity but quarks do, etc. All these speculations obviously challenge our common understanding of how the (inertial) mass generation takes place for such fermions (as well as the massive vector bosons of the weak interaction) by means of the (massive!) Higgs particles in the gravity-blind SM, if the matter fields are to have $m_g = 0$ in, e.g., 4D. And this may be even a more serious concern in supergravity theories, which are the low energy limits of the compactified Strings and the M-theory! (For completeness sake we add that even within the SM there are still vague issues remaining regarding the Higgs, like the interaction of the Higgs boson with itself, or whether they are elementary particles, or do they really provide mass to fermions or only the weak gauge bosons, and so forth.) As for the Higgs and its tangoing with gravity, the Higgs is somewhat of a nuisance (asides the possibility that for gravity it may induce violation of the Lorentz invariance) and tend to create extra complexities when it comes to unifying GR and the SM in $R^4$ under a "fat" action, which normally would have been the algebraic sum of the individual GR and the SM actions. The Higgs field $\Phi_h$ alters the latter recipe in the sense that it contributes a multiplicative factor, multiplying the curvature R in the EH gravity action, having the form $(1-\Phi_h^2/\Phi_o^2)$, thus restricting regions subject to $\Phi_h^2 < \Phi_o^2$ and forcing us to modify the metric in some intricate way! The Higgs-gravity affinity, however, may be deeper than we think is currently.

(Note: One may argue that a different approach to the geodesics departing from GR is needed for fermionic motion in $R^4$. E.g., we may surmise that in the context of variable G, the "total" affine connection in $R^4$ is the sum of two parts. One is derived by the *metric* $g_{\mu\nu}$ in the orthodox-GR manner and is not a tensor, and the other is a tensor and directly related to variable G, which is non zero even at the flat spacetime limit where R=0, and the field equations are the ones for the vacuum. An example of the latter extra connection can be proposed by setting $\Delta\Gamma^\sigma_{\mu\nu} = \delta\nabla^\sigma\Phi/\delta g_{\mu\nu} = \frac{1}{2}g_{\mu\nu}\nabla^\sigma\Phi$ as a constraint. So that the equations for the geodesics in $R^4$ are $d^2x^\sigma/d\tau^2 + \Gamma^\sigma_{\mu\nu}(dx^\mu/d\tau dx^\nu/d\tau) - \frac{1}{2}\nabla^\sigma\Phi(ds/d\tau)^2 = 0$, with $ds^2 = g_{\mu\nu}dx^\mu dx^\nu$, and obviously in the case of flat spacetime they are not straight worldlines in $M^4$, which we desire to have, unless G is just a constant, which is one reasonable outcome! The latter geodesic equations in $R^4$ are a vestige of an extra force due to the $\nabla\Phi$ piece. If our principal conjecture is to hold in curved spacetime then for fermions we should get $d^2x^\sigma/d\tau^2 = 0$ implying $\Gamma^\sigma_{\mu\nu}(U^\mu U^\nu) = \frac{1}{2}\nabla^\sigma\Phi(ds/d\tau)^2$. In the Newtonian limit $g_{\mu\nu} = \eta_{\mu\nu} + h_{\mu\nu}$ and $\Gamma^\sigma_{00} \sim \frac{1}{2}\eta^{\sigma\nu}\partial h_{00}/\partial x^\nu$, we find $\nabla h_{00} \sim -\nabla\Phi(u/c)^2$ where $h_{00} \sim -2\varphi$ with $\varphi$ the gravitational static potential (recall how $\Phi$ is related to the variable G). Thus, it appears that variable G can restore straight geodesics in $R^4$ by canceling gravity for the fermions! At the very end of this section we shall see a similar type result by invoking variable light speed canceling gravity, actually this is only for the fermions because of the linearization of the Full Hamiltonian in the spirit of Dirac!)

Classical GR defies any a priori fixed background spacetime. Here geometry is both dynamic and nonlinear and this runs contrary to the orthodox and linear QM, which is in need of a fixed spacetime background (like the

Minkowski spacetime) to formulate its local wave-particle duality business. Quantum gravity, on the other hand, comes in various forms. And one is the induced gravity conjecture where it tries to build the background spacetime of GR as a sort of mean field approximation to an underline quantum space and its quantum degrees of freedom. And our no-gravity conjecture, as one can imagine, is obviously a fatal blow to all the spacetime geometry and the mean field conjectures.

Demanding $m_g=0$ but $m_i \neq 0$ for, say, an electron requires from the modeling stand the intervention of a field agent, or agents, of a kind for the job, and the possible role of the variable G cannot be underrated in this regard! Consider the following witty toy model for the variable SM fermionic gravitational masses: $m_g(x)=m_i(1-l^2_o \square \Phi)$, where $l_o$ is some characteristic length scale depending on $m_i$, the spin, and what you will, and $\Phi$ is the dimensionless scale function (e.g., $\Phi=G_o/G(x)$, but it can be also tied to other variable fundamental constants). The game now is to set $m_g=0$ every time $m_i$ hits the observed discrete masses for say the leptons and the free quarks (though for free quark masses we have no knowledge of their exact values). If you are a student reader, you may want to invoke other fancy things for describing the $m_g$ variation, like the Bessel functions of a kind, while promoting the above equation to a more general K.G type equation: $m_g(x)=m_i(1\pm(l^2_o/\Phi)\square\Phi)$. In this case $m_g(x)=0$ occurs at the discrete zeros of the Bessel functions incorporating $m_i$ where the above KG equation also vanishes. The interesting thing is that these zeros are also endowed with two labels, and your next task can be to provide an interpretation for the labels as quantum labels of a sort, like perhaps the spin and the isospin and so forth!. And in this way you kill two birds with one stone: you get the quantized inertial masses that make the $m_g=0$, and have a field equation for $\Phi$ in flat spacetime. So let your fun begin and see how far you can go with this toy model. As seen, there is plenty that we can do phenomenologically, but there is no denial that we lack a fundamental underline principle for doing all that, especially in the absence of any experimental proof telling us whether an electron falls under gravity or what. So all we can do is to propose scenarios and keep an open mind on the possibility $m_g=0$ or $m_g \neq m_i$!

The idea of stripping gravity, or mitigating it beyond the usual norm (e.g., by conjecturing it to be asymptotically free and always weak, no matter the separation between the particles), is not as senseless as it may appear at first! It may provide a way out for avoiding the necessity of wanting full quantum gravity at high energies that is normally a nonrenormalizable theory, for at least the matter sector of the SM particles. But if hadrons composed of confined quarks are to feel external gravity but not so for the free quarks, wandering around at any separation then how are we to resolve this apparent inconsistency? It may be that QCD, a nonabelian gauge field theory itself, is the source of inner hadronic strong gravity between the confined quarks. And an effective type Riemannian metric may arise in that color environment proportional to the dressed gluon-gluon configuration $g_{\mu\nu}\sim G^a_\mu G^b_\nu \eta_{ab}$, as was suggested long ago by A. Salam and C. Sivaram [16]. Actually, what the latter authors proposed was the possible emergence of a QCD type confining linear potential from a Weyl type strong gravity action quadratic in the curvature. The role of the gravity-like spin-2 particle may be played here by an effective di-gluon field $G_{\mu\nu}$ acting as a Riemannian metric (with $D_\sigma G_{\mu\nu}=0$) and as exchange graviton-like particles in the IR region (a more modern terminology for the above scenario is chromogravity). Now, it is a well-known fact that quantized gravity, unlike QCD, is perturbatively nonrenormalizable. So, does this mean no analogy can be drawn between Einstein type quantum gravity (EQG), whose underline d.o.f are those of the spacetime metric, and QCD (e.g., in the IR region) inside the hadrons because of the EQG nonrenormalizability? I don't know the answer to this, but a recent paper by Reuter and Weyer, also rich in references, seem to suggest that recent developments in the nonperturbative treatment of the EQG have shown that the theory is renormalizable and in their words "asymptotically-safe" (see [17]). Consequently, there seem to be an analogy between QCD and EQG after all, where the asymptotic free scales $\Lambda_{QCD}$ and the Planck mass for EQG play a similar role. The latter reference also claims a linear rise for G(r) that in turn causes the antiscreening behavior of quantum gravity and induces an extra pulling force between the test masses. (Of interest to student may be the issue of also an analogy with QED in that even though graviton Feynman diagrams can include triple graviton couplings there is a common trend for calculating graviton couplings to matter in analogy to photon coupling using the helicity formalism, where the amplitudes are decomposed in definite helicities.)

From our stand, we may now want to search for a new theory of inner hadronic gravity whereby the effective gravity coupling between say two confined quarks goes to zero at vanishing quark separations (call it gravitational asymptotic freedom!) while the interaction increases as the separation is increased, and it is reasonable to consider variable G playing a crucial role over here! Beyond some critical quark separation of about the size of a hadron we may anticipate gravity to become "normal" and follow the orthodox scenario of GR. (Of course we are not saying anything here about the tensorial nature of gravity within the hadron! By assuming an eternally weak gravity then a scalar or spin-2 quanta makes no real difference inside the hadrons.) So the battle tested GR is to emerge in this universe only at the *hadronic* level and beyond, while avoiding the classical singularity in the inner world of hadrons which for our purpose can be considered nonrelativistic (but this need not be an absolute rule). In this way GR cosmology is finally linked to hadronic matter at the Fermi scale. In our modified gravity picture of the hadronic interior, gravity is to remain always weak between the quarks no matter the separation, and even more so at very small separations where the gravitational constant is allowed to go to zero. At separations of a Fermi, or so, gravity is to reach its maximum yet finite value, meaning we do not require of it to go to infinity for confinement purposes. And the reason for this is we still want QCD to dominate and be responsible for the quark confinement and not a strong gravity of a kind. Thus, in this picture the gluonic role in producing gravity is mitigated if free leptons, which are deprived of strong interaction, are to follow also the same gravity pattern, or the lack of it, as their quark cousins.

So, the idea is to avoid, to the extent possible, a phase transition type occurrence for variable G as the function of the quark or the leptonic separations. G has to be continuous so to apply to both free quarks and leptons, and ultimately the early universe. By eliminating any QCD role in generating variable G gravity we are then back again to the spin, and maybe some other quantum number and degrees of freedom to explain the odd behavior of our variable G gravity, or the lack of it, for the SM class of fermions and antifermions. A trivial observation though is light quarks and leptons have small inertial masses (or effective masses, in case of confined quarks) relative to the observed hadrons and that may also be a factor, a part from the spin and the other d.o.f, in depleting the gravitational mass while $m_i$ is left alone (e.g., writing the gravitational mass as $m_g \sim m_i e^{-\mu/m_i}$ may do the job phenomenologically, where the mass scale $\mu$ may be in the MeV-GeV range!). (Note: Small fermionic mass also means a small coupling to the Higgs, and one speculation may be whether the scalar Higgs can play also the role of an internal strong scalar gravity?) But as indicated before the fact remains that there are no viable A-Z scenarios around to explain our conjecture for a gross violation of the weak EP regarding the SM fermionic sector (aside from saying this is the way nature is, especially if confirmed experimentally!).

If what we are saying has some validity, then perhaps something more fundamental must be at work, and be common between quarks and leptons, beyond just spin, mass, etc. to defy ordinary gravity. That thing may be related to the possible compositeness of these particular spinors, perhaps something along the line of the preons of the yesteryears, but in some modified version of course, as common constituents of quarks and leptons that can (somehow!) be the cause of the gravity demise or screening! If so then that obviously opens doors to a whole new type of investigation on why preons are to defy gravity. If free SM fermions don't fall that may also mean they have to be insensitive to any spacetime warping around them (because of whatever cause there may be: spin, mass, variable $\Phi$, and what have you, effects). And that in turn means they must ignore motion along the bosonic geodesics, if present. They must be somehow shielded from gravity (or curvature) at all time and locations, and must be living in the Minkowski spacetime $M^4$ (as one logical flat spacetime background) and drag that flat spacetime along their geodesics, now free worldlines, and all that is to presumably occur in a Riemannian curved spacetime background due to the bosons or the macro matter, if present! But how is this realized formally? And what if at the end of the day the SM bosons, in contrast to macro matter, live in a non Riemannian kind of a spacetime? No easy answer here! (Few years ago I cooked up a scenario, with mixed degrees of success, for canceling the $1/r$ (Newtonian) gravity potential generated around a spin ½ particle, but the price to pay was to introduce a variable Planck constant protagonist for the job. The idea was to see whether gravity potential can be cancelled on classical grounds by treating the spin "vector" classically, and if the effort failed then teat the spin quantum mechanically, which undoubtedly is an added complexity, given the aim of the problem. In short, the idea is still around and of some value too, but I'm still hoping for a real break!)

(**Note**: The pivotal question is of course how a flat spacetime background (e.g., a Minkowskian one) for the fermions can coexist simultaneously with a curved spacetime due to the bosons, or macro matter, and in addition allowing diverse coupling between these particles as in the real world! This is undoubtedly an issue of complicated interface between quantum mechanics and GR and thus very murky by nature. It is perhaps central to distinguish two distinct cases, i.e., Feynman diagrammatically speaking, when fermions and bosons are allowed to couple at vertices. One when the coupling is mediated by virtual particles and two when real bosons are involved as mediators of the coupling. We may envisage, e.g., for the former case diverse fermion-fermion scattering mediated by virtual weak gauge bosons, gluons or photon exchanges, not to be omitted are also self-coupling of single fermions or bosons, likewise the latter case may be exemplified by the Rutherford scattering of an electron off a nucleus, or the Compton scattering mediated by real bosonic (photon) exchanges. According to GR, each virtual or real boson mediating a coupling must gravitate, forcibly yielding a non flat metric in the intermediary regions between, e.g., the fermion-fermion vertices! (Recall, in flat spacetime QED the bare fermionic vertex "things", like the charge and the mass, get corrected by the vacuum polarization loops to yield observational "things", while modifying also the photon leading propagator.) According to QED the intermediary regions where virtual bosons can propagate cannot be observed, thus any GR induced curvature must be observationally doomed, so it seems! On the other extreme the mediating real photon causing spacetime curvature in say the Rutherford scattering is not subject to the observational restriction as seen in the virtual particles, so in principle its induced curvature may be observed. But the intriguing "quantum-diagrammatic" fact is this: the real exchanged photon in the Rutherford scattering has various probability amplitudes to materialize into $e^-e^+$ loops-normally yielding infinities due to integrals over unrestricted internal moments but ultimately managed by the QED renormalization procedures-during its flight before returning to a photon and hitting the nucleus (similar story applies to a much more complex case of, e.g., photon-photon coupling, and other more familiar cases). The same fermionic loop-story applies also to the virtual photon exchanges. Now consider our conjecture. The order-by-order perturbative fermionic loops, which are permitted during the mediating photon flight, being fermions (we are assuming SM fermions) cannot cause spacetime curvature during the times that the exchanged photon spends as any virtual $e^-e^+$ pair. In applications though things are much more involved as we go from one loop to two and beyond because of considering couplings between the fermions within the loops and other graphs. In dealing with QED we learn that loop corrections are much smaller in comparison to the leading graph contribution to the scattering, where the exchanged photon remains a photon during its flight (care must be exercised for other gauge forces). Yet we must keep in mind that we have in principle an infinite number of allowed fermionic loops (we are discarding bosonic loops, more on this in the next few paragraphs!) and the idea is that the infinite sum of all higher than say four loops remains still small. For us, therefore, the curvature picture associated to a real photon exchange during its flight is a probabilistic series of intermediary Minkowski metric, assigned to the fermion loops, along with a series of presumably curved metrics assigned to the cases when the photon remains a photon! Quantum gravity allows the coupling of an external graviton to the fermionic loops, but not so in our picture! The same curvature picture applies to the virtual photons but with the understanding that virtual photons in the Feynman diagrams are not observationally meaningfull when not disturbed by external probes. Most Feynman diagram calculations are done using the momentum space wavefunctions and momentum space loop integrations for many practical reasons (one being the ill defined notion of photon localization in the configuration space). Thus, assigning a series of *deterministic* spacetime metrics at this or that definite spacetime point along the trajectory of the exchanged boson is operationally futile, or the least a nightmare. Incidentally, let me digress a bit by saying imagine incorporating our conjecture of no fermionic SM gravity to the superstring fermionic and bosonic vibration modes and then realize its dire impact on the (a la Einstein) desire of generating the very fabric of spacetime dynamically from such strings, free at last from the notion of any pre-existing background spacetime! And to complicate things even further recall in string interactions the well defined notion of a pointlike vertex, where the worldlines meet and interaction takes place, normally also an observer independent point, becomes totally distorted in that now we are dealing with world sheets, endowed with supersymmetry, joining and there is no well defined and observer independent point for the interaction, depicting the strings merging into one! So much more to say but I better stop! The exchanged virtual photon in case of $e^-e^-$ scattering moves in a dynamically self generated curvy spacetime between the two non gravitating leptonic locations (vertices), while the electrons move, according to us, in flat spacetime. Thus, at the vertex of say photon emission by an electron, the electron must still remain an electron (albeit renormalized) and not undergo any spacetime geometrical changes (obviously changing to what?) while upholding the meaning of the flat spacetime coordinates! At ultra high energies when the two electrons are almost touching the tiny intermediary exchanged photon spacetime (assuming c and h remaining constants) must be nearly flat so to avoid contradiction. Hence in this extreme case the very high frequency, or short wavelength, virtual photon exchange should lead to Einstein vacuum solution in the photonic infinitesimal region, which runs contrary to GR which says the opposite should happen! It is evident how hard it is to conceive a single (amalgamated) metric that can accommodate at once both bosonic and fermionic species-given the fact that leptonic positions or spacetime must be singled out. Of course the question is who is doing the singling out? By an observer who knows how to distinguish instantly between the F (fermion) and the B particles at a vertex in the sense of a physical measurement and the

"curvature" probing? This Fermi Bose bimetric conjecture with a "flat-curvy" duality character promoted by us must fit into some geometrical structure of a kind. (Maybe an 8D superspace of a kind, or else?) In the GR sense our free fermionic conjecture can appear at best as the vacuum solutions of the field equations (ignoring E&M contributions). But the notion of a flat Minkowskian "background" spacetime for the fermions with locally non-Minkowskian metric transcends GR (though keep the merit of a vierbein formulation for the case of fermions in mind). Maybe one approach to follow, injected with some dose of physics and the distinguishably of fermions versus bosons regarding local experiments is the concept of superspace endowed with both bosonic (orthodox) and fermionic (Grassmann) coordinates. Yet another approach may come from how to reconcile pure supergravity with its coupling to matter (fermionic) fields, should we invoke a unified view and a unique metric tensor in higher dimensions which upon "breaking" lead to two distinct (flat-fermionic and Riemannian-bosonic) metrics in 4D? Naturally, it is always much easier to speculate than to design such a metric; the mathematical consistency must be outlandish over here! Even the pure supergravity by itself is irreconcilable with our conjecture, I mean bringing a bosonic graviton with a fermionic gravitino in the Einsteinian spirit of dynamically generating a single Riemannian spacetime. It would have been many folds easier to simply declare that all the SM particles (including all fermionic and bosonic particles) are deprived of any gravity coupling, if it was not for the photon that bends under gravity! Can anybody dare to say the SM "photon" is not necessarily the quantum of light generated in the lab, or measured by observing the heaven? How to relate both flat and curved metrics locally and still remain in 4D (and the issue of local supersymmetry transformations and the coupling between the gravity supermultiplet and the chiral supermultiplet) is simply beyond our current reach. Even within the GR framework there are always surprises popping out here and there, an example, that may not be relevant to us but it is good know, may be the finding by Ashtekar and collaborators a decade ago that the 4D vacuum gravity is equivalent to 3D gravity coupled to a scalar matter field. So, can the scalar be related to variable G? Another advance that I've recently became aware of is the topological geometrodynamics (TGD) which grew out of a the desire to build a Poincare' invariant gravity theory in 8D space H with the neat feature that the 4D Minkowski spacetime $M^4$ are surfaces in H. And the other 4D space, fixed by the symmetry of the standard model, is the complex projective space $CP_2$ (i.e., $H=M^4 \times CP_2$), which spinor structure encodes the known particle quantum numbers and couplings in its geometry (thus the geometrization of these quantum numbers). The $CP_2$ isometries is the color group, whereas the isometry group of $M^4$ is chosen as the Poincare' group. There is much more to this story, which may be pertinent to our desperate discussions trying to connect our novel conjecture to something more mainstream, but I doubt this can be done at this writing. Readers interested in learning more about TGD may consult the internet which contains several articles by Matti Pitkanen at the University of Helsinki. To end, I am confident that there is a genuine "math" solution to the issues we've raised thus far, given the arsenal of knowledge already accumulated in the field of topology and differential geometry, if tackled by the right experts in the field. A unique case that comes to mind is the matrix model formulation of microphysics that is becoming especially popular these days, leading eventually to a noncommutative (NC) version of the gauge theory. Who knows, maybe gravity can be studied or induced within the NC gauge theory framework, deviating obviously from GR at high Planckian energies?)

Returning back to earth (or rather inside the hadron) once again there is no doubt that orthodox GR can no longer apply to the SM matter fields, even in the orthodox curved spacetime formulation, using the tetrad and the spin connection, if the latter are either deprived of gravitational masses, or in a milder version, are gravitationally very distinct from say the photon field that is GR friendly. Our demand that the gravity coupling G is to vanish at small quark separations, while reaching a finite maximum at another separation, and beyond that converge to either a constant or a lower constant value $G_N$, as the separation is increased further, is not, technically speaking, a trivial issue-even at the level of assuming for simplicity a maximally symmetric 4D geometry for the interior spacetime of the hadrons with easily derivable Killing vectors.

All the eccentric G-behavior thus far alluded is to be happening presumably because of some kind of yet unspecified reason, be it in form of an unusual curvature coupling to the spin, variable G, QCD, or what have you, even without requiring the quark gravitational masses to vanish, because now the real protagonist is variable G and not $m_g$! Despite all these concerns, one attractive feature, I think, of desiring always a weak gravity via variable G is in some way equivalent to disregarding "nonlinear" GR contributions. Because one major difference between GR and the Newtonian counterpart is at high energies and thus high curvatures where according to GR gravity becomes very strong. But if in nature gravity is always weak then an immediate question is who needs the nonlinear GR, which is a captivating theory by itself and hard to let go, but perhaps not in the eye of Mother Nature, or in the eyes of those adhering to the quantum principles? A weak limit of GR as the whole story of gravity with variable G may serve the purpose of quantization and provide us the necessary corrections to the

Newtonian gravity we need to confirm to observations. It is noteworthy that most experimental tests of GR are made at nonrelativistic and almost Newtonian limit, the observation of blackholes (strong curvature) or gravitational waves (small curvatures), perhaps slowing down the binary star motions, are not at this writing sufficiently conclusive and direct. Though few well-known (solar) observations favor GR predictions over the Newton's gravity involving constant G, it is hard not to entertain the possibility of devising a variable G type Newtonian gravity (perhaps promoted by special relativity) making similar observational predictions to those that are claimed to support (weak) GR with constant G.

In GR the equations of motion for the geodesics in curved spacetime, which give the kinematics, have no dependency, thanks to the EP, on the mass of the particles. And making those geodesics "Minkowskian" straight worldlines for free quarks and leptons of the SM, if deprived of gravitational masses (as in our proposed conjecture) means presumably killing the affine connection effect in shaping up the geodesics and imposing a zero curvature. This is basically the end of the Riemannian geometry in the applied sense, and in essence the end of the quantum geometrical unification programs for these fermionic particles! (In this scenario, e.g., the SM matter fields, regardless of nongravitational quantum corrections to the inertial mass, cannot tell spacetime how to curve, and spacetime cannot tell them particles how to move!). So, if there is something peculiar about the SM fermions regarding gravity, then the three main scenarios (but there could be few more) are: (1) there is no gravity/curvature effect on fermions to play around with (e.g., $m_g$=0), or (2) such particles feel the gravity effect but the effect must be canceled by some yet unknown agent (e.g., by means of variable c as entertained at the end of this section), and (3) they feel gravity but this force is asymptotically free and is neither in full accord with the Newtonian dogma nor with the Einstein's geometrical prescription. For (2) and (3) variable G can, and indeed must, play a crucial role to lead a finite gravity at all scales. To formulate GR, Einstein had to rely on $m_g$=$m_i$ at rest and also assume the same relativistic moving mass formula for $m_g$ as for $m_i$, as already derived in SR. An inequality between $m_g$ (without vanishing it) and $m_i$ creates an ambiguity when it comes to their moving mass values. Setting $m_g$=0 at rest for all the SM fermions certainly removes any necessity to say anything further about their moving mass behavior under the Lorentz transformations, thus the advantage! Often I tried to dispel the thought of setting $m_g$=0, but it keeps returning, and I guess this will remain so until somebody finds, once and for all, the closing answer in the Lab: "to be fall or not to be fall"!

At this point we may argue that perhaps our "no-gravity" fermionic conjecture is simply too far fetched, and it may be best to change the past stringent scenario that SM fermions do not feel spacetime curvature to a more GR friendly statement that the fermionic spacetime is to be only "spatially" flat so that the metric in this universe without spatial curvature can look something like $ds^2=-c^2e^{f(t)}dt^2+\sum(dx^i)^2$ in Cartesian coordinates (this is a kind of VSL metric). The latter is worth exploring because of perhaps two reasons: firstly I got my inspiration here from the canonical quantum gravity that presupposes a spatially fixed three-manifold and is austere to any topology changes, and secondly because there are already cosmological scenarios that are spatially flat, although to my knowledge in those scenarios the f function in the above metric depends only on the spatial coordinates and not time, but there are other metrics (e.g., Bianchi types) that fit the spatially flat criterion. With a "fermionic" metric, like the one in the above (or simply a different one), at hand one can proceed to find the connection and the curvature things and then apply it to cosmology to see what ensues.

However, if we keep on insisting and holding on to our earlier $m_g$=0 conjecture for the SM fermions then we shall need to do some hard and critical rethinking over here. Either the conjecture can be ultimately explained by the currently available mainstream physics/math, or we shall need to invent an explanation for it, and I don't see any other way except to say this is the way things are! Let us explore a bit the mainstream physics and then do some speculation. Consider, e.g., an electron. According to QED prescription for calculating quantum corrections to bare charge one can envisage a "cloud" of electron-positron virtual pairs surrounding it to screen its (infinite or whatever) bare charge. Now according to item (1) of the summary at the end of this section the spacetime around the bare source should still behave as the vacuum regarding gravity, if the e+e- virtual pairs are deprived of gravitational masses (obviously this runs contrary to GR where the vacuum energy, or virtual particles obey also the equivalence principle, fall along geodesics, and obey mechanical laws of motion modulo QM, and never mind if GR is not even a quantum theory!). The situation seems to be manageable regarding our conjecture provided the only allowed virtual pairs in the cloud are fermionic-antifermionic SM matter fields and not, say gauge boson pairs of a kind that can curve the surrounding spacetime?! Yet, QFT prescriptions do not necessarily preclude bosonic virtual pair production (albeit having smaller probabilities). Does this mean we should tone down (1) and, e.g., demand that the spacetime surrounding an electron (presumably having an infinite inertial bare mass!) to be flat only on an average (and perhaps in a stochastic) sense? Or, do we envisage a local scenario for canceling the effective curvature around a "physical" electron by demanding a kind of fermionic-bosonic "virtual" curvature cancellation (most likely nonsensical, or a technical nightmarish) because in GR both sign curvatures are allowed (e.g., assigning R<0 for de Sitter spacetime of the fermions). Or should we invoke the need for a new gravity theory for the SM matter fields where the curvature is also dependent on the charge sign, or perhaps the spin directions? Or how about making a decree, since we are free to frenzy, that "all" virtual particles are deprived of gravitational masses, as I indicated earlier?

Setting the bare gravitational mass to zero is only useful to us in part, because the electron can still feel external gravity if the virtual particles in the cloud are capable of gravitating. The EH action or its modified f(R) versions involve both bulk and surface effects. But in GR geometry is dynamic, so one can ask whether the surface dynamics cannot induce a gravity version of a Casimir type negative energy to cancel out the allowed bosonic virtual pair production effects, inducing curvature within an electron cloud? Certainly a far fetched idea of nonlocal nature! How about the GR-friendly photon field, curving the spacetime, and also capable of real and virtual leptonic pair producing? The photon must then yield a flat spacetime after say the real leptonic pair creation, and all that independently of the energy content of these fermions. There is also the issue of early cosmology in that which form of energy came first, the bosonic or the fermionic energy? In the latter case gravity could have not existed at t=0, whereas in the former case fermions could not have existed at t=0 and it took a short while for them to be created, once nature knew what the number π meant for spin! (Recall spin is expressed in units of ℏ=h/2π, so no π at t=0, where spacetime continuum was just in the making, means also no spin at t=0!).

(Query: Do the above arguments necessitate an *inherent* distinction between the fermionic energy $T_{\mu\nu}$ and the bosonic one regarding curving spacetime-similar in spirit to the fact that nature recognizes leptonic and baryonic numbers and their conservations that are quantum labels associated only to the fermions? Hence can energy at this very fundamental level be also fermionized (so that we may jokingly speak of 1 joule of Fermi energy or I joule of bosonic energy at the SM scales)? A brute way out of this flight of imagination is to cook up a conjecture that says that the final SM-fermionic field theory cannot be constructed while respecting simultaneously the invariance under the group of diffeomorphism, as required by GR, and the whole gauge transformation invariance related to the nongraviational couplings. So, this would mean electron energy is nongravitating, and in this case fermions would have been better suited as carriers of nongravitational forces than the usual gauge bosons in the first place!! Another quandary comes from *unbroken* supersymmetry. In the latter case an electron and its scalar selectron are supposed to be identical in all aspects except the spin. Now, if an electron is nongravitating and a selectron is gravitating then the spin is to be the prime agent for the lack of gravity in electrons. But in a realistic world SUSY (assuming to be valid) is broken and the selectron has a much greater mass than the electron, thus mass becomes also a factor for the presumed existence of gravity in selectrons! As we can perceive one goes around and around because these are simply hard issues to tackle coherently once we decide venturing beyond mainstream physics! The conjecture of no-gravity for the SM fermions means obviously the neutrinos (antineutrinos) are capable of interacting with the world through only the weak interaction. Such a claim has clearly a great impact on the role of the neutrinos in cosmology, given their expected abundance in this universe, and neutrinos cannot couple gravitationally to say galactic nuclei if we are right. If electrons don't gravitate then electrons or positrons falling into a blackhole cannot take place, at least due to the gravity pull. Likewise, processes such as $e^-+e^+\rightarrow\gamma+g$, with g the graviton, should be scrutinized in much more details. Laboratory experiments based on $e^-e^+\rightarrow X\rightarrow\mu^-\mu^+$ where X is a tensor particle resonance, like the massive KK graviton excitations predicted in the RS brane-world, must not take place in the lab if our fermionic no-gravity conjecture is right! Yet, there are too many theoretical subtleties to cope with due to Feynman vertex physics and extra dimensions before making any concrete conclusion. Thus far, and keeping it short, the D0 collaboration at the Fermi Lab has not seen any spin 2 KK graviton resonance decaying into a P-wave dielectron or a dimuon (or even an S-wave diphoton) final state!)

We are not yet done with our tittle-tattle, because now comes the real $e^+e^-$ annihilation at any energy to photon fields, and if item (1) at the end of this section's summary is viable there should not be any curvature before e+e-collision (ignoring the E&M coupling) and there comes a curvature after the pair annihilation because all energy is now bosonic in quality, and seemingly there is no superspace structure, to guide us, and no curvature conservation law at hand before and after the collision to steer us (although I have not worked out the details, I have the sentiment that perhaps the superspace formulation of our principal conjecture may be worth studying)! The situation of course is more amicable if e+e- goes to quark-antiquark production (which is allegedly deprived of gravitational masses) until hadronization takes place, if the total fermionic energy is right, and then the familiar gravity emerges outside the hadrons! According to GR the electromagnetic field around an electron curves the spacetime, but according to (1) of the summary, and in combination with QED's leptonic virtual pair depiction, this should not happen for any energy value, and even so at very high probing energies, where according to orthodox gravity the virtual graviton dressing of the bare mass is substantial. Is all this a vague hint to us that we should look at a milder explanation for wanting $m_g$=0? For example, search for a fermionic metric that is on the average Minkowskian, or flat, within the cloud spacetime endowed with some linear extension, so that test particles probing these fermions do not feel gravity on the average, regardless of their energy and nature? Naturally easier said than done!

One can also argue, in contrast to the above spat, that electromagnetic energy in the cloud is electromagnetic energy and it should not matter whether the bare charge source resides on a bosonic or a fermionic SM particle bare mass. Thus there is nothing weird for this field, say, surrounding the bare electron, to pair produce virtual bosons, which in turn curve the spacetime around while the bare electron source ($m_g$=0) does not feel the curvature. (We keep in mind that the orthodox doctrine says that a free particle following a curved geodesic still experiences gravity self-coupling in the absence of external forces, and that the E&M waves by themselves interact with the spacetime curvature; e.g., we have: $\Box A^\mu - R^\alpha_{\ \beta} A^\beta = -4\pi J^\alpha$.) This interaction is not felt by $m_g$=0, but what about the inertial mass $m_i$ which is not zero, and may be in great part due to electromagnetism! A gross violation of the WEP, as we claim it, also creates problems for using the powerful tools of the covariant wave equations, see **[18].**

At the quantum level then any coupling of graviton to the SM fermions must be absent and that means, e.g., in calculating Feynman diagrams things like the graviton-SM fermionic vertex must not exist at the tree level. There are conjectures here and there in the literature that spin ½ particle motions in an external gravity field is curvature dependent, contrary to the orthodox belief that all particles move in the same way, and cannot be excluded by any choice of coordinates, a result that is obviously in defiance with the equivalence principle! Likewise there are published conjectures that there are no consistent cross-interactions possible among different gravitons in the presence of Dirac fields! The bremsstrahlung from electron-electron collisions in stellar interior is commonly believed to be one good source of gravity production. Well, it won't be so "easy" for the photoproduction of gravitons to take place if our no-gravity conjecture holds for the electrons! Our emphasis on the no-gravity tail was mostly on free (or quasi free) leptons and quarks, so the possibility of spontaneous emission of graviton by the decay of say a hydrogen atom from the 3d to 1s state may still take place. But obviously what one does not want to have around, if our conjecture is right, are things like the classic mini blackhole electrons with huge surface gravity that by virtue of the Hawking radiation (an ignoring extra dimensions) should evaporate by radiating in about $10^{-106}$ s! The $m_g$=0 setting presumably takes care of that! There are some suggestions in the literature that gravity may provide the stabilization that is needed for a classic electron that otherwise tends to blow up due to its repelling negative charge elements. How should one deal with this issue, or the more complex orthodox issue of the off-shell electron vacuum loop coupling to external gravitons (that some suggest can give rise to an effective cosmological term), while the on-shell case does not gravitate?

It was suggested long ago that the zero point energy (ZPE) of the fields may be connected to gravitation and inertia, and there are articles on the web on this alleged connection that readers can find easily. If so, then how do we get from "there" to the $m_g$=0 conjecture of here for the SM fermions? It is certainly of outmost importance to ultimately resolve the issue of (that most probably will come from experiments) whether gravity is a fundamental d.o.f in nature, or perhaps a manifestation of the other nongravitational forces in some way (in analogy to e.g., the nuclear force being affiliated to QCD, or the van der Waals force among the neutral molecules related to the E&M force), forcibly downgrading the geometrical approach of Einstein to gravity! On the other hand if the geometry of spacetime is so pivotal to fundamental physics then it is only natural to raise the possibility, as some already did, that the observed fundamental scale hierarchies in physics be also related directly to geometry, obviously an easy question to ask but too early to answer!

Obviously, there is a world of mystery to tackle within this contentious scenario of ours, and the messages are sometimes tempting (especially when considering cosmological ramifications of this scenario) and sometime frustrating, and even more so when we have not found any satisfactory orthodox type rational behind our $m_g=0$ conjecture. The full triumph of the geometrization of gravity by Einstein relies on the strict assumption $m_i=m_g$. Otherwise getting rid of the orthodox gravity as a force acting on the mass is no longer justified and the geodesics provided by GR as being independent of any mass parameter is also worthless! The point we are trying to make is the $m_g=0$ conjectures for the quantum world of the SM fermions is a blow to the entire wisdom behind GR and its Newtonian limit! Yet, we keep in mind that all these wild thoughts can become serious thoughts if one day somebody in the Lab shows that indeed electrons do not fall, or fall too "unusually" under gravity!

Let us begin fresh all over again by considering an isolated electron and once more ask the palpable question: does the electron curve the spacetime around it, even assuming it does not fall under externally prescribed gravity? The answer is affirmative if we adhere to GR, which says both mass energy and the surrounding E&M field energy of the electron with the observable charge -e contribute to curving spacetime locally. A prototype metric here, following GR, is, e.g., the Reissner-Nordstrom (RN) metric, where the line element is now given by $ds^2=-A(r)dt^2+A^{-1}dr^2+r^2d\Omega^2$, with $A(r)=(1-2m_e/r+e^2/r^2)$ in units $G=c=1$. Although this is a metric designed for non-rotating charged blackholes deprived of angular momentum, we can use it for our electron study purposes in the limit of zero gravitational mass, because in this case no real valued physical event horizon ever appears to make it look like a blackhole. The nonvanishing Ricci tensor components are in this case $R_{00}=-Ae^2/r^4$, $R_{11}=-e^2/Ar^4$, $R_{22}=e^2/r^2$ and $R_{33}=e^2\sin^2\theta/r^2$, and the scalar curvature $R\sim e^2$. Let us now try to get a bit closer to the orthodox doctrine by saying although the electron bare mass doesn't gravitate ($m_g=0$) but its electromagnetic field nonetheless contributes to the local curvature. (As mentioned before, the witty observation here is that the E&M contribution when M is not zero has the tendency to reduce, and not enhance, in contrast to the GR dogma, the curvature R. In our case, though, with zero $M_g$ the only contribution to R is the charged field!). Then the function A(r) in the RN metric is simply $(1+e^2/r^2)$. Since mass is the protagonist moving along the geodesic and we want it to be a free worldline in the flat spacetime, and because its E&M field gravitates, as seen in the above Ricci tensor components with $A=1+e^2/r^2$, then this is the case where the Einstein field equations can be kept with $T_{\mu\nu}$ pertaining to only the electron self E&M field and the metric is the reduced RN metric, while the contribution from the affine-connection that goes in the making of the geodesic equations is to "somehow" fade away! Obviously this kind of idea creates some interpretational difficulties over here! One may argue that this paragraph is simply contradictory from the start in the following sense. Since we killed the WEP to begin with then what is the point of looking for the RN metric in the first place! This argument is at best weak. E.g., nobody can say, i.e., experimentally speaking, that massless photons cannot have an infinitesimal charge, we have only upper limits for the photonic charge that is super minuscule, and can still use the above massless RN metric! The reality however goes well beyond all the classical arguments discussed over here because they do not include any quantum effects and back reaction on the metric! (Those interested to continue exploring the RN analyses may want to work on the derivation of the affine-connections with the above metric and then compute the geodesics.)

Let us reconsider the naïve GR-RN metric with G and c shown explicitly, and set up few games in the context of variable G, but unlike the previous paragraph we shall no longer set $m_g=0$, meaning we shall honor the WEP. The line element in this case is given by: $ds^2=(1+F)c^2dt^2+(1+F)^{-1}dr^2+r^2d\Omega^2$, where $F=(-2M/c^2r+Q^2/c^4r^2)G$ and no cosmological constant is considered. Here $Q^2$ can be viewed as the sum of the square of an electric and a magnetic charge. In the *first* game we shall assume F is a pure constant $h_{oo}$, obviously we can do that only if G is a variable (and clearly we are now confronting a different ball game)! Then $G(r)=h_{oo}c^4r^2/(Q^2-2Mrc^2)$ which is interesting in that G is also related to the E&M charges! As seen, the metric is deprived of an event horizon and is also singularity free provided $h_{oo} \neq -1$: $ds^2 = (1+h_{oo})c^2dt^2 + (1+h_{oo})^{-1}dr^2 + r^2d\Omega^2$. In addition, we can redefine a constant light speed in this space as $c_o = c(1+h_{oo})^{1/2}$ so to end up with $ds^2 = c_o^2dt^2 + (c/c_o)^2dr^2 + r^2d\Omega^2$. Another way of interpreting this metric is a flat space scenario along with an effective "light" speed (we can dub this speed the gravity speed in the RN metric!), and moreover we have also defined variable G in terms of $Q^2$ and M (i.e., by using the RN metric). The variable G is well behaved at r=0 and is zero, but at $r=Q^2/2Mc^2$ (about the classical radius of M) it is singular (whose sign depends on the sign of $h_{oo}$). The singularity though pertains to G and not the spacetime (event horizon), where the geodesic is simply $d^2x^u/d\tau^2=0$ in the absence of spin terms! At distances $r>>Q^2/2Mc^2$, $|G|$ is proportional to r (which is always so for the chargeless case). An alternative to the above is to demand G to be also zero at large separations. This can be accomplished by, e.g., setting, $h_{oo} \sim e^{(-r/ro)}$ and then take it from there. Another variant is to choose a form for G that vanishes at both zero and infinite separations so that the spacetime is flat at both zero and large length scales, and this is what we shall find towards the end of this section. One plausible choice for G may be the following type: $G(r)=(b/r)e^{-ro/r}$ where $r_o$ and b are constants, then $h_{oo}(r)=(b/c^2r^2)(-2M+Q^2/c^2r)e^{-ro/r}$. Clearly $h_{oo}$ vanishes at r=0 and at infinity. At the weak field limit the acceleration is given by $\mathbf{a}=d^2\mathbf{x}/dt^2=-(\frac{1}{2}c^2\partial h_{oo}/\partial r)\check{\mathbf{r}}$, where $\check{\mathbf{r}}$ is the radial unit vector. One can now easily compute a single r-value for which the acceleration vanishes, and likewise one can entertain the possibility of a maximal acceleration and choose it as $2Mc^3/\hbar$, as was speculated by few in the 1980s, to fix $r_o$ (details omitted). Finally, we may want to ask whether there is an event horizon here. That will clearly depend on whether or not the expression $(b/c^2r^2)(2M-Q^2/c^2r)e^{-ro/r}=1$ admits an acceptable solution(s) for r!

The above discussion pertained to game one, you may recall. In game two we go a step further and include in the RN metric also the first derivative of G(r). Meaning, this time the line element includes $h_{oo}$ that itself is given by $(-2M/c^2r+Q^2/c^4r^2)G+\gamma(M/c^2)\partial G/\partial r$ and if we now impose, as we did in the above, an expression for G(r) as $G(r)=(b/r)e^{-ro/r}$ we get $h_{oo}=(a/c^2r^2)[-M(2+\gamma)/c^2+(1/c^2r)(Q^2/c^2+\gamma Mr_o)]e^{-ro/r}$. The case of $\gamma=-2$ is of special interest, but other options are also possible. There is plenty more to say on how variable G can possibly alter, and all this classically we note, the orthodox RN blackhole topics which may range from information paradox to entropy, evaporation temperature$\sim$1/G, evaporation time$\sim$G$^2$, coupling to matter, RN+CC term, quantum correction to RN metric, and so forth.

Continuing with the earlier discussion for the case $m_g=0$ with constant G, one may argue that it is perhaps best to use a more relevant metric for an object endowed with spin than the RN metric deprived of any angular momentum. There is such a metric around and it is the textbook Kerr-Newman (KN) metric, admitting a ring singularity, that represents the most general metric for spinning charged blackholes with parameters M, e, and a=J/m, where J is the angular momentum. The complexity of using this metric for us stems from the requirement $m_g=0$ (where the event horizon becomes imaginary if G is constant). In this limit we have: $g_{11}=\cos^2\theta$, $g_{22}=g_{33}=$infinity, $g_{34}=g_{43}=0$, $g_{00}=-1$, but neither the geometrical radius nor the null-surface radius are real valued and finite-though one can make these radii zero provided one can live with an infinite and negative Newton's constant around the electron which is given by the relation $G\sim-s^2(\hbar/mc)^2(c^4/e^2)\to \infty$ as m$\to$0, where s is the spin (this $\infty$ however may not be such a tragedy for the electron, given $m_g=0$!). The overall inverse KN metric though is not well defined in the massless case. We take this as an indication that the singular metric and its inverse for the $m_g=0$ charged electron (endowed with an electromagnetic self field) may represent a world that is much more involved than the flat spacetime world we promoted earlier. If the SM fermionic matter fields are to be more primary than the bosonic particles (which is not unreasonable) then having a flat background spacetime, as we like to have, for the fermions is clearly not the result of the orthodox GR as, e.g., prescribed by the KN metric, which is surely incomplete. After all, most well studied metrics, like KN, assume symmetric $T_{\mu\nu}$, no torque term corrections, that may be necessary, and do not include torsion (which is expected for particles with spin)! The whole Kaluza-Klein approach to gravity in the higher spacetime dimensions, yielding the Einstein gravity, and say the E&M, in 4D is therefore unrealizable for the SM fermion if $m_g=0$! Incorporating variable G in the KN metric is certainly interesting but is beyond our present scope.

Despite all the arguments we've been having, and possibly can add more constraints to the previous ones, and despite the possibility of resorting to different geometries like the Riemann-Cartan spacetime as the fermionic spacetime background-which incidentally also implies a contact coupling between spinning particles, admitting attractive coupling when spinning in opposite directions and repulsive coupling when spins are parallel-we cannot say how all these arguments can clarify why the SM fermions do not fall under gravity. None of these models are really designed for such a purpose! Relativistic QM teaches us that the spin assigned to a point elementary particle (as oppose to spin concept in GR) is purely a quantum phenomenon with no classical analogue. So, any attempt to find the effect of spin-curvature coupling at the classical GR level must sound absurd, and an effective theory at that classical level is also of little utility and probably not unique. Only full quantum gravity can shed light on spin curvature coupling but regrettably we still have a long way to go! GR says free falling particles must follow the geodesics lines, and for fermions with spin this scenario is at best vague. Because of the spin-curvature coupling, which may tend to pull off the fermions from the geodesic worldlines, the linear momentum 4-vector and the four-velocity may not be collinear. So, the bottom line is this: we have no complete understanding of the spinning particles and their motion in GR, but there are people who are still working on this. (The orthodox tool to cope with fermions is of course the vierbein, in place of the metric, to connect global $M^4$ to local frames in say $R^4$, in our case though we say this construct is extraneous because the SM fermions either do not gravitate or defy greatly the conventional scheme!) There have been past attempts, for instance, to model, and in the classical language of GR, specialized objects (called geons) with the transformation properties of the spinors. Such attempts seemed meaningful provided a nontrivial topology of spacetime manifold is invoked within a fermion which is not time orientable. The latter constraint may then be taken as the onset of quantum phenomena and one can then take it from there; for more insight see Hadley's article [19] and the reference therein. An electron is also a quantum particle and accordingly one may want to check how gravity can affect it by a purely quantum phenomenon, as opposed to a direct fall. An electron interferometer of a kind may be used to check whether the gravity potential at different (yet close) heights can induce a phase shift which is proportional to the product of its inertial by its gravity mass $m_i m_g$. Such a technique has been successfully used in the case of neutrons and it is confirmed that indeed for the neutrons $m_i = m_g$. The electron being much lighter though will lead a phase shift of about 3.37 million times smaller than the neutron under identical conditions and that is desperately small for a detection, a muon is certainly a much better candidate for the job, if technically possible, to test our conjecture.

Whatever we may be able to devise in terms of semi-classical GR or come up with other scenarios, the ultimate scenario for spin-curvature coupling is not going to be trivial. So, we may ask if in the absence of full quantum gravity there is anything we can do to at least get some understanding of how fermions behave under gravity within the context of GR. The answer is yes, we can do plenty! An orthodox answer, e.g., among many, for a spin ½ particle $\Psi$ is the Dirac equation in curved spacetime background (discarding nonrenormalizable supergravity and the finite string theory which are also "things" we can do). The Dirac equation in curved spacetime deserves certainly some credit, even though it is no substitute for full quantum gravity. (Consistent quantum gravity must ultimately incorporate also the SM particles and explain its more than dozen parameters, and that means it must forcibly contain also the fermions that our very conjecture negates!) Assuming a massless spinor for simplicity, the Dirac equation, after using the covariant derivative $D_\mu = \partial_\mu - \frac{1}{2} i \omega_\mu{}^{ab} \Sigma_{ab}$, which brings in the torsion and the spin connection $\omega$, is $\gamma^\mu D_\mu \Psi = 0$. By squaring the Dirac equation the effect of the curvature R becomes perceptible: $g^{\mu\nu} D_\mu D_\nu \Psi - \frac{1}{4} R \Psi = 0$. As you can immediately see reducing this to a KG type equation in flat spacetime, following our $m_g = 0$ conjecture for spin ½ particles, is quite nebulous! The classical limit of the latter equation is probably good enough to shed some light on the classical dynamics of a spinor in curved space, but even that is neither a trivial limit nor it is unambiguous. In simplistic terms, the appearance of Torsion is also a manifestation of the back-reaction of the spinor fields, which in turn affect the geometry of the spacetime. One can imagine of a more extended geodesics equations for the spinning particle, as explained briefly in Weinberg's old textbook on gravity, of the form: $d^2 x^\lambda / d\tau^2 + \Gamma^\lambda{}_{\mu\nu} dx^\mu / d\tau dx^\nu / d\tau + f(x) S^\rho R^\lambda{}_{\mu\nu\rho} dx^\mu / d\tau dx^\nu / d\tau = 0$, where $S^\rho$ is the particle spin and f(x) is an unknown scalar function. The last term, normally neglected, is of importance only when the particle is "larger" than the characteristic spacetime dimensions of the gravitational field around. But in the absence of curvature the "free" geodesic, at least classically, is still $d^2 x^\lambda / d\tau^2 = 0$, even with spin. But a test fermion moving near any macro matter is moving in a curved spacetime, so how are we to get flat spacetime geodesics (worldlines) for such a test particle?

SM fermions moving in the curved Riemannian background $R^4$, generated by any macro matter, yet moving always on straight worldlines requires erecting "global" inertial frames in $R^4$ and that is no trivial matter, or perhaps is even nonsensical-momentarily co-moving inertial frames wouldn't do! It may be that the global Lorentz covariance in flat spacetime, or local LT in curved spacetime, is after all a broken symmetry! If so, how does this help validate our fermionic lack of gravity conjecture (at the end of this section we shall present some ideas on this issue)? There is obviously no single answer to give, or even the right question to ask, one can only speculate. If global LT is violated it can spell trouble for, e.g., global supersymmetry because the latter is based on enlarging the Poincare' algebra by also including spinorial generators, in addition to the bosonic ones. A more mundane side to this story is that the Lorentz group admits a spin ½ (spinorial) representation. So if the Lorentz invariance is to be broken then spinors can no longer be candidate objects for the representation of the Lorentz group. Yet, in a sense this may free us and prompt us to do some rethinking on what to do with the spin ½ particles in the context of GR, which is locally Minkowskian, or any modified version of it. Again all these sketch out issues are encouraging signs from here and there that there may be something factual in our principal conjecture after all, but it is still not clear how all this can yield a better understanding of this conjecture.

Must we seek, for a change, an understanding of our conjecture via an exploration into the very basics of the continuum spacetime, that is the metric, and try our luck once more? Perhaps! Here is a phenomenological example of this idea which amounts to writing down the total spacetime metric as the sum of two "sub" metrics: $g_{\mu\nu}(tot)=g_{\mu\nu}(B)-4l^2(x)s|1-s|\partial_\mu\Phi\partial_\nu\Phi$. Here s is the spin, and we are considering only a simple system, l(x) is some variable length scale, and as before $\Phi=G(x)/Go$. As seen, the second metric term survives only when the (SM) fermions are present, and moreover variable l(x) may be tailored in such a way to cancel any (would be) genuine metric singularity (e.g., balckhole horizon) stemming from $g_{\mu\nu}(B)$. The letter B in $g_{\mu\nu}(B)$ symbolizes the bosonic SM fields, hadrons, nuclei, atoms and macroscopic matter in general all curving spacetime. Of course there are many technicalities that we must sidestep so to move on. But it is clear that if all we had around were of "B-field" types then the above decomposition of the total metric achieves nothing, but things are different if fermions (say spin ½) are included. In this case we may face two plausible options. One option is to impose a flat metric (e.g., the Minkowski or the de Sitter metric) for the total metric $g_{\mu\nu}(tot)=\eta_{\mu\nu}$-in conformity with the observed flatness in cosmology-so to get $g_{\mu\nu}(B)=\eta_{\mu\nu}+l^2(x)\partial_\mu\Phi\partial_\nu\Phi$. Because we expect gravity to be always finite and weak, the latter metric form reminds us of the usual metric decomposition that is adopted for linearizing any GR metric in the weak field limit (written as $g_{\mu\nu}=\eta_{\mu\nu}+h_{\mu\nu}$) for quantization and the gravity wave propagation purposes! (In the absence of the B-fields the "bare" fermionic metric is the flat metric given by $\eta_{\mu\nu}=-l^2(x)\partial_\mu\Phi\partial_\nu\Phi$, and the inverse metric is $\eta^{\mu\nu}=-4l^2(x)\partial^\mu\Phi\partial^\nu\Phi$, and after some simple algebra we find $\partial_\mu\Phi\partial^\mu\Phi+l^2(x)=0$ and $\partial_\mu l\partial^\mu\Phi=0$. The case l(x) being a constant, $l_{o,}$ is a viable solution. Among the three possible choices for $l_o$ in particle physics: $\hbar/m_lc$, $e^2/m_lc^2$ and $m_gG_N/c^2$, the first one may be more consistent than the others, especially in the sense of connecting to quantum mechanics (of course in that case massless fermions must be ruled out, otherwise we get $\eta_{\mu\nu}=0$!) On cosmological grounds $l_o$ can be chosen as $\Lambda^{-1/2}$ where $\Lambda$ is the cosmological constant, and now related to variable G in the flat space of the naked fermions. Anyway, one can now choose a suitable "bosonic" metric (say the FRW metric) and then find the connections and the curvature tensors, the scalar curvature, and what have will, and then get involved with sorting out the many different technical issues; and that much for the first option.

There is also a second possible option for tackling our principal conjecture, and still in the context of modified GR that respects both covariance and the EP. That option goes under the name of *bimetric* gravity. This one unlike GR begins with two metrics and the gravity speed is not identical to the light speed. Much has been said on this issue in the literature for decades, and I will not repeat them here. But the general idea (or certainly a modified version of it) may be of some value for tackling our principal conjecture of the "gravityless" fermions moving in a curved spacetime that we questioned before. Basically, at each point of the spacetime one has a (physical) curved spacetime metric for gravity, and other inertial forces, and a flat-space (background) metric for describing the inertial fields, and in the absence of gravity both metrics become identical. There can be a preferred cosmic frame represented by a unit 4-vector pointing, e.g., in the direction of the cosmic time flow to aid the overall formulation of gravity as a field theory, and in par with the other three quantized fundamental field theories (this being the major motivation for such theories). Undoubtedly, a constraint of a kind is warranted so to fuse our principal conjecture with the bimetric theory that, e.g., says if one wants to use a distorted version of these theories, the flat-space background must only contain the SM matter fields (recall this is the extreme case of our principle conjecture whereby fermions are completely deprived of gravity, but a milder case is also possible in the context of variable G, where fermions still feel the "variable", albeit weak, gravity) and the Riemannian manifold is for the rest of the bosonic, hadronic and/or the macromatter in general! The bimetric theory is an attractive option to be pursued by us.

There is no denial that complexities are abundant in matters of relativistic gravity and thus one may envy all the oversimplification effects that our conjecture entails in stripping altogether gravity as a universal force for all the SM fermionic particles and be done with it (I am of course referring to "point" fermions here and that may include also the SUSY fermionic sparticles, the leptoquarks, and especially the neutrino species that I've not discussed before, having many applications in cosmology and supernovae explosions and the issue of their arrival time on earth compared to photons and the gravitons after a supernova explosion). The conjecture, if at all true, simplifies nature's complexity a lot, specifically at the SM level. It makes it rather senseless to worry about hard to tackle issues like the spin-curvature coupling and the geometric nature of the spinor fields, and so many other field theoretic subtleties, the gravitons pair producing leptons, etc. If what we posit has some merit then no direct graviton exchange should take place between the SM fermions and the rest of the particles. Both gravitational Compton effects and photoproduction of gravitons by, e.g., electrons should not take place, etc. Although the general idea put forth here is tempting the devil is in the details and we have no exact clue, apart from an experimental confirmation, to justify our conjecture! A conjecture of this kind has far reaching consequences. For example, the idea of placing all the SM particles on branes (the brane-world scenarios) and then requiring only gravity to travel in the bulk and serve to communicate with the brane fermions may have to be revised entirely if these fermions are to be deprived of gravity!

Thus far we've speculated a lot on the possible lack of gravity for the fermionic fundamental fields but have not been able to substantiate our conjecture on any firm ground, perhaps a suitable bimetric theory can be ultimately formulated but for now we are not there! Mitigating totally, or in part, gravity for the latter SM particles may be related to many factors. Here is an incomplete list: (1) vanishing $m_g$, (2) variable G, where gravity exists at large separations while vanishing at small separations, (3) exotic spin effects reducing or canceling curvature (!), (4) exotic topologies, perhaps in the context of GR or its modifications, (5) Newtonian gravity modification using variable G, if gravity is always finite and weak, (6) the gravity constant G for fermionic SM particles is constant but much smaller than $G_N$ (see next paragraph), (7) the gravity constant G may be an oscillatory function of time in the rest frame of a fermion, averaging to zero during each quantum cycle, (8) other fermionic effects related to fundamental constant spacetime variations capable of canceling gravity at any separation around the alleged fermions, (9) a gravity shielding mechanism of a kind not yet formulated, and finally (10) the possibility of variable c having something to do with the cancellation of gravity for the SM fermions (we shall deal with this one shortly).

(Extra Note: Case (6) of the above is somewhat amusing and it is discussed in below for the fun of it because it may be of interest to early college students. We'll discuss it here not because it is right nor has any chance of competing against today's hard core classical gravity and it quantized version, but rather to show how one can do so much with only few number of elementary assumptions! Let us enlarge and go beyond (6) by envisaging a primitive type scenario where the total electric charge (shown by $q_\pm$) of any SM massive particle of spin s is comprised of two components: $q_\pm = \pm e\text{-}4s \mid 1\text{-}s \mid m\sqrt{G_f}$, where the charge e can be fractional in case of quarks and $G_f$ is the fermionic gravitational constant (that in principle can be different for different fermions and also different from $G_N$!). This decomposition implies an inherent asymmetry between the + and the − charged identical particles thanks to "gravity" and may have a role to play in tackling the matter-antimatter asymmetry in the universe (at least that's the primitive idea!). Obviously a fermion and its antifermion annihilating into say a photon, will imply the photon must have an electric charge equal to $-2m\sqrt{G_f}$ if and only if the charge conservation is assumed (?), but in this case the photon's tiny charge gives also a hint on its origin of creation via m. Experimentally, the photon charge must be

$<5.10^{-30}e$, so for e+e- annihilation this implies $G_f<1.74.10^{-24}$ dyn.cm$^2$g$^{-2}$ which is many orders less than $G_N\sim6.7\times10^{-8}$ dyn.cm$^2$g$^{-2}$! But what is the advantage of this? The amusing part is the elimination of gravity in the 1/r$^2$ theories of coulomb and the Newtonian interactions. E.g., for two identical masses the total coulomb coupling+the normal gravitational coupling (which is $\sim-Gm^2/r^2$) shown by $U_{tot}$ amounts to: (a) $U_{tot}=-e^2/r$ for q+q- coupling, thus no net gravity and (b) $U_{tot}=(e^2\pm2em\sqrt{G_f})/r$ for q-q- or q+q+ coupling respectively, with interesting results that are self evident, and as seen the usual Newtonian gravity force -$G_fm^2/r^2$ is eliminated in case (a). The cross term $2em\sqrt{G_f}$ r in (b) is somewhat inspiring in the sense of parity violating weak interaction, but its magnitude may be too small to compete against the weak coupling (besides, it is absent in the case of q+q-, and say the neutrino coupling!) unless $G_f$ is a variable quantity of difficult-to-predict nature! This primitive scenario does not contradict any observation. For example, experimentally $|q_{e+}+q_{e-}|/e<4.10^{-8}$ which is many orders of magnitude larger than the predicted value $2m\sqrt{G_f}/e<5.10^{-30}$! This scenario has also an impact on the Dirac type magnetic monopoles, which we'll omit discussing here, and the gravitational radiation, under the banner of an "electric" charge, when a particle is accelerated (the tricky thing though is whether the radiation from the gravity charge term is electromagnetic or is a gravity wave?). That may be a matter of interpretation, or is it? The reasonable thing to assert is the extra piece $m\sqrt{G_f}$ in the overall charge is indeed an electric charge whose magnitude happens to be given by $m\sqrt{G_f}$, thus all charge radiation is electromagnetic. It is clear that all the above arguments remain unchanged if we used $G_N$ instead of $G_f$. The only inconsistency that we can point to by using $G_N$ is the photon charge -$2m\sqrt{G_N}\sim10^{-21}$e. And if the earlier limit on the photon charge is taken seriously, then one conclusion may be the extra electric charge -$m\sqrt{G_f}$ is not a conserved quantity (and then classifying it as an electric charge is dubious). But what is the advantage of using $G_N$? One nontrivial answer is the possibility of finding a massive positive fermion that has a zero electromagnetic coupling to its own kind (of course all arguments made here besides being primitive are also classical). Such a fermion will have a charge ½e and a mass m=½M$_P\sqrt{\alpha}$, where M$_P$ is the Planck mass and $\alpha^{-1}\sim137$ is the fine structure constant. A massive neutral fermion can also exist with upper mass M$_P\sqrt{\alpha}$, however this particle has gravitational coupling. The above scenario also says any particle with mass larger than M$_P\sqrt{\alpha}$ must necessarily be of negative charge! According to this scenario the gravity charge piece also contributes to the electron magnetic moment, albeit infinitesimally. Finally, let me also add that one can modify the earlier scenario by proposing a complex electric charge of the form $q_\pm(r)=\pm e-2i|1-s|m\sqrt{G}(r)$ so to unify electromagnetic and gravity in one shot (the overall coulomb coupling in this case includes both the usual electrostatic and the gravity interaction) in the context of the 1/r$^2$ theory, with the understanding that the only observable effects are the real parts. Have fun!)

**-Can variable light speed cancel scalar gravity?** Let us make another attempt, albeit somewhat informal, on how to possibly cancel "scalar" type gravity from the SM fermionic rank by invoking variable speed of light (VSL). As we'll show any VSL will ultimately depend not only on the momentum and the coordinates, but also on the mass parameter in a manner that for massless particles one recovers the universal c. As for the general topic of VSL, it has a long history and I've already said a bit on that earlier. But until relatively recent times not much significance was attached to it, in part because of the lack of any experimental discrepancies with the SR postulate of the constancy of the light speed, and in part due to the lack of any desire for confronting the foundations of special relativity.

(The author while toying around with the idea of variable ħ in the mid to upper 1970s, and still a graduate student, also proposed, and then derived, the concept of maximal speed for elementary particles different from the light speed c. He was able to show, by using certain modified chiral Lorentz transformations, that each elementary particle of rest mass m has a maximal speed that is unique in all inertial frames given by $c(1-2m/M_P)^{½}$ thus yielding a unique upper unification of mass for all particles at their maximal speeds given by the Planck energy [20]. Particles with masses about ½M$_P$ were to be always motionless and had zero maximal speed, which I called etherons to represent the quanta of the cosmic vacuum. Regrettably, the idea was considered too speculative at that time, despite private encouragements from few referees, for it to be published in prominent physics journals! A later version of it however was recorded in the 80s at the SLAC preprint center. But times have changed and nowadays we are tolerating possible Lorentz invariance violations and fundamental constant variations as functions of energy, or spacetime, for usage in micro or macro world studies, and variable c is just one of them. Plenty of modern materials exist on this subject and the interested readers may want to consult **[21]**.)

In an expanding universe one can think of a preferred frame pointing in the direction of the cosmic time flow and by using the Robinson-Walker type metric it was shown by several peoples that the postulate of the universality of the light speed is contradicted. For our "gravity" purpose though we expect the variable light speed, denoted as $c_x$, to differ only infinitesimally from the universal light speed c. And the reason for that is the gravitational potential in our treatment is going to be always finite at all length scales, unlike the Newtonian one, and numerically close to the difference $c_x^2-c^2$. Most already proposed models of variable c have no difficulties in devising the modifications of the Lorentz transformations (LT) at the infinitesimal local level, but understandably construction of the global LT is not likely. Meaning, for a generic variable $c(\mathbf{x},t)$ theory global coordinates cannot be set up straightforwardly in $M^4$ (e.g., the time axis having a form $c_x t$ is simply incongruous, though something like $tc(t)$ may not be illicit!) yet local Lorentz transformations may be constructed [22].

*-Our first LT scenario involving VSL and its immediate downfall:* Since we are addicted to Special relativity, and cherish its commanding applications in all nongravitational forces, in below I shall use a trick within the context of the modified LT, but shortly after I will abandon it altogether in favor of something better to follow. The starting idea is how one can stay close to the SR covariance of nature's laws, naturally in inertial frames, and yet invoke accelerating frames and the variable light speed? This contradictory menu may be implemented provided we make use of the simple aforesaid "ploy" that has an affinity to the Einstein's elevator gedenken experiment! Throughout our presentation we shall remain within the context of the Minkowski space, even though there are other flat spaces around (Rindler, Milne, de Sitter, etc). Begin with two inertial frames K and K' in uniform relative motion and consider a generic Lorentz boost expression in direction of a unit vector $\mathbf{n}$ to get the familiar LT: $t'=\gamma(t-\mathbf{v}.\mathbf{x}/c^2)$ and $\mathbf{x'}=\mathbf{x}+(\gamma-1)(\mathbf{n}.\mathbf{x})\mathbf{n}-\gamma\mathbf{v}t$, obviously there is no preferred frame over here. Now comes the simple extension or ploy: replace everywhere in the above global LT the constant ratio $\mathbf{v}/c$ by $\mathbf{v_x}/c_x$ (meaning set $v/c=v_x/c_x$) where $\mathbf{v_x}$ is now the variable boost velocity, due to some inertial force, and $c_x$ is the VSL! Clearly, $\gamma$ is not affected, t' is not affected, because $\mathbf{v}.\mathbf{x}/c^2=\mathbf{v_x}.\mathbf{x}/cc_x$, the only affected term is $\mathbf{v}t$ in $\mathbf{x'}$ becoming $\mathbf{v_x}t$. We write the latter as $\mathbf{v_x}t=\mathbf{v}t+\mathbf{v}(c_x/c-1)t$ and then impose $\mathbf{v}(c_x/c-1)t=\boldsymbol{\alpha}$, where $\boldsymbol{\alpha}$ is a constant length vector in the direction of the boost. This gives $c_x=c(1+\alpha/vt)$ (and $\mathbf{v_x}=\mathbf{v}(1+\alpha/vt)$) and a suitable candidate for $\alpha$, that has relevancy to only spacetime and not particle masses, is the Planck length $l_P$. Thus, we can take $\boldsymbol{\alpha}=l_P\mathbf{n}$ and $c_x=c(1+l_P/vt)$ to find: $\mathbf{x'}=\mathbf{x}+(\gamma-1)(\mathbf{n}.\mathbf{x})\mathbf{n}-\gamma\mathbf{v}t+\gamma l_P\mathbf{n}$. Although this may look fine, the formulation is deficient for several reasons: the fundamental length is not LT invariant, the formulation is too superficial when it comes to clock synchronization at $\mathbf{x}=0$, $t=0$ and $t'=0$ in that both origins can not coincide and the synchronization in the K' frame must be done at spatial $\mathbf{x'}=l_P\mathbf{n}$, besides the simple light cone in say K' looks awkward in the (now absolute) K frame, and $c_x$ also diverges at $t=0$, etc. So these are enough reasons for us not to try to fix (e.g., by invoking a local time $t(\mathbf{x})$, so that $\mathbf{v}t(\mathbf{x})=\mathbf{v_x}t$), or make interpretation of these results in favor of abandoning the approach altogether, while retaining certain features of it for usage in an improved special relativity proposal as outlined in below!

**-Double upper speed relativity (DUSR).** We shall now introduce in an informal way a class of modified LT that do not suffer from the above shortcomings involving two upper speed limits. One is the standard and the universal light speed c, to set up: the Minkowski spacetime background with the orthodox metric (for us (-1,1,1,1)) and coordinates $x^\mu=(ct,\mathbf{x})$ and the spacetime scales. And the other one, shown by $v_m\leq c$, is standing for an upper constant speed assigned to a given mass $m_o$ of an elementary particle belonging to the SM. We anticipate the $v_m$s to depend also on the inertial rest masses $m_o$! Inertial observers in relative uniform motion in two inertial frames K and K', measuring given particle properties will be restricted in their uniform speed v in $M^4$ by $v_m$ as the maximal speed, while extracting impersonal information from the particles (in the collider physics jargon, e.g., this may mean a maximum rapidity!). The speed interval $v_m\leq v\leq c$ is not accessible for any meaningful physical communication of information between the K and the K' observers-because now the factor $\gamma=(1-v^2/v_m^2)^{-1/2}$ is imaginary (Tachyonic)-unless $m_o=0$ and in that case $v_m=c$! The mass dependency of $v_m$ obviously requires invoking another mass scale for dimensional reasons. We take that mass scale, shown by $M_P$, to be a universal mass applicable to all SM elementary particles, and one reasonable, but not necessarily unique, choice is the Planck mass. So unlike SR our DUSR also knows something about the gravity coupling G and h, albeit in the passive (eventually kinetic) sense (in contrast to GR where G enters the game to promote dynamics!). We expect in the limit $m_o=0$ to find $v_m=c$, and in the limit $M_P=\infty$ to recover special relativity, with again $v_m=c$. It is also obvious at this start that since GR formulated in $R^4$ is locally SR in $M^4$, a modification of GR may become imperative if the modified GR is to be locally DUSR with G already in its making! But such extension contradicts many deep rooted and well-known doctrines that are imbedded in the GR formulation, so for the time being we sidestep these to move on with the primary DUSR formulation.

Any elementary particle in DUSR has its own maximal speed in $M^4$, and the $v_m$ for composite particles (such as hadrons made of quarks) are more unwieldy to sort out. But it is reasonable to assign an inertial mass to the overall composite entity close to the sum of the inertial masses. However, the inertial mass is normally smaller than the sum of the individual masses due to the binding energy, and in that sense it is not a straightforward matter to assign an inertial mass representing the composite entity, so in below I will only concentrate on single (SM) particles. The modified LT that we are proposing for DUSR are as follows: (1) $\mathbf{x'}=\mathbf{x}+(\gamma-1)(\mathbf{n.x})\mathbf{n}-\gamma(c/v_m)\mathbf{v}t$ and (2) $t'=\gamma(t-\mathbf{v.x}/cv_m)$. The interval invariance $S^2=S'^2$, taken from the origins, fixes gamma as $\gamma=(1-v^2/v_m^2)^{-1/2}$, thus we find $S^2=c^2t^2-x_\parallel^2-\mathbf{x}\perp^2$, and this is expected, given the particles are moving in the Minkowski spacetime (and as usual the light wave front moves with the light speed c). Consequently, the proper time $d\tau$ along the trajectory is the same as in traditional special relativity: $dt(1-v^2/c^2)^{1/2}$. The matrix elements $\Lambda^\mu{}_\nu$ for the homogeneous Lorentz group are also the same as in SR, except c is replaced by $v_m$. Namely: $\Lambda^0{}_0=\gamma$, $\Lambda^0{}_j=\Lambda^j{}_0=\gamma v_j/v_m$, $\Lambda^i{}_j=\delta^i_j+v_iv_j(\gamma-1)/v^2$, and $\det\Lambda=+1$. (The above modified LT become the identity SR LT upon imposing the identity $\mathbf{v}/v_m=\mathbf{v}_{SR}/c$ and different particle masses with the same "$v_m$-observer" moving with constant velocity $\mathbf{v}$ are seen by the "c-observer" as having infinitesimally different $v_{SR}$!)

Now, the velocity addition rules are: $u'_\parallel=(u_\parallel-vc/v_m)/(1-vu_\parallel/cv_m)$, and $\mathbf{u}'\perp=\gamma^{-1}\mathbf{u}\perp(1-vu_\parallel/cv_m)^{-1}$. Or in vector form $\mathbf{u}'^2=[(\mathbf{u}-c\mathbf{v}/v_m)^2-(\mathbf{u}x\mathbf{v}/v_m)^2]/(1-\mathbf{u.v}/cv_m)^2$ with the boost $\mathbf{v}=\mathbf{n}v$. Several surprising results emerge when v or u are near $v_m$ (and when $\mathbf{u}=c\mathbf{v}/v_m$) that may differ drastically from special relativity with some possible deep consequences in particle physics and toy models for studying flat space cosmology, but given the simplicity of the math I'll omit discussing further these consequences.

The momentum $P^\mu$ and $P'^\mu$ are related by our modified LT ($P'^\mu=\Lambda^\mu{}_\nu P^\nu$) analogous to the spacetime transformations: $\mathbf{P'}=\mathbf{P}+(\gamma-1)(\mathbf{n.P})\mathbf{n}-\gamma(P^0/v_m)\mathbf{v}$, $P'^0=\gamma(P^0-(\mathbf{n.P})\mathbf{v}/v_m)$, and consequently we find $P'^{02}-\mathbf{P'}^2=P^{02}-\mathbf{P}^2=Q$. The letter Q here is an indication that different choices are possible. For example, the three simplest forms for Q are: $m_o^2c^2$, $m_o^2v_m^2$, and $m_o^2cv_m$, we'll get to these in below.

To get to the kinematics we define a 4-vector velocity $s^\mu$ with the principal constraint $g_{\mu\nu}s^\mu s^\nu=v_m^2$, and use the definition $s^\mu=(dt/d\tau)(v_m,v_m\mathbf{u}/c)$, thus the 4-vector momentum (we shall stage the setting in flat spacetime and assume $P^\mu$ and $s^\mu$ are parallel) is then $P^\mu=m_os^\mu$. It is clear that $s^\mu$ is not identical to the orthodox 4-vector velocity in SR $u^\mu=(dt/d\tau)(c,\mathbf{u})$, implying $u^\mu u_\mu=c^2$, in fact $s^\mu=(v_m/c)u^\mu$. In terms of components we have $P^0=\gamma^c m_ov_m$ and $\mathbf{P}=\gamma^c m_o(v_m/c)\mathbf{u}$, where $\gamma^c=(1-v^2/c^2)^{-1/2}$ is the usual gamma factor of SR (not to be confused with the earlier $\gamma$ we've used for our spacetime LT), giving $P^{02}-\mathbf{P}^2=m_o^2v_m^2$. The latter imply that there are *maximal* $P^0$ and $\mathbf{P}$, which we denote by $P^0_m$ and $\mathbf{P}_m$ respectively, given by: $P^0_m=\gamma_m m_ov_m$ and $|\mathbf{P}_m|=\gamma_m(v_m^2/c)$, where $\gamma_m=(1-v_m^2/c^2)^{-1/2}$. The acceleration 4-vector is $a^\mu=dP^\mu/d\tau$, therefore it should be evident that there exist also an *upper acceleration* at $v=v_m$ that can be readily calculated (details omitted)! At this stage we must look at several options we have in order to define the moving energy E! The "simplest" of the options for defining the energy are as follows: (1) $E=P^0c$, (2) $E=P^0v_m$, and (3) $E=P^0(cv_m)^{1/2}$ (you may want to try also a generic form $E=P^0c^\alpha v_m^{(1-\alpha)}$). On the other hand, we also have the problem of deciding how we want to define the upper maximal energy for a given particle. Again, the three most reasonable options are: (1´) $E_m=M_Pc^2$, (2´) $E_m=M_Pv_m^2$, and (3´) $E_m=M_Pv_mc$. Clearly we have many possibilities over here, nine to be exact, and must choose one. Each combination produces a distinct value for $v_m/c$, but few are identical.

The results of combining the above 1-3 with 1'-3' are as follows: 1+1' see (a), 1+2' see (b), 1+3' see (c), 2+1' see (d), 2+2' see (e), 2+3' see (f), 3+1' see (g), 3+2' see (h), and, finally 3+3' see (i):

(a)      $v_m=c(1+(m_o/M_P)^2)^{-1/2}$
(b)      $(v_m/c)^2=\frac{1}{2}[1+(1-(2m_o/M_P)^2)^{1/2}]$
(c)      $(v_m/c)=(1-(m_o/M_P)^2)^{1/2}$
(d)      $(v_m/c)^2=(M_P/m_o)^2[-1+(1+(2m_o/M_P)^2)^{1/2}]$
(e)      $(v_m/c)=(1-(m_o/M_P)^2)^{1/2}$
(f)      $v_m=c(1+(m_o/M_P)^2)^{-1/2}$

(g)     $(v_m/c)^3+(M_P/m_o)^2(v_m/c)^2-(M_P/m_o)^2=0$, for $m_o \ll M_P$ we find $v_m/c \sim 2(1+(M_P/m_o)^2)/(3+2(M_P/m_o)^2)$

(h)     $(v_m/c)^3-v_m/c+(m_o/M_P)^2=0$, for $m_o \ll M_P$ we find $v_m/c \sim 1-\frac{1}{2}(m_o/M_P)^2$

(i)     $(v_m/c)=\frac{1}{2}(m_o/M_P)^2[-1+(1+(2M_P/m_o)^2)^{1/2}]$.

As noted, the common feature in all these nine cases is that for either $m_o=0$ or $M_P=\infty$ orthodox SR is recovered. My two choices are (a) and (e). Although (a) makes not only the upper inertial masses universal but also the energy, I think the latter universality, though tempting, especially for quantum gravity purposes, is simply too much to ask. Thus, my final choice is (e), where the energy is defined as $E=P^0 v_m$ with $v_m=c(1-(m_o/M_P)^2)^{1/2}$, which also implies no particle can have a rest mass in excess of $M_P$. So we have now both the concept of upper speeds and a common upper rest mass in our formulation! The rest energy is $m_o v_m^2=m_o c^2(1-(m_o/M_P)^2)$ and as seen it is also symmetric under the sign exchange of one or the two masses. Lastly, an important relation that is of common usage is $E^2=|\mathbf{P}|^2 v_m^2+m_o^2 v_m^4$.

By choosing $M_P$ as the Planck mass we find $v_m^2-c^2=-Gm_o/(\hbar/m_o c)$, which is recognized as the Newtonian gravity potential of a particle about its Compton wavelength! It is obvious that we are finding fine traits over here. Needless to say invoking maximal speeds (a kind of cutoffs in E and P) has also a direct impact in evaluating: Feynman diagrams (e.g., the loop diagrams), formulating the Dirac equation, GR (e.g., the perfect fluid model), QFT in general, cosmology, if regarding the SM particles!

Although in formulating DUSR we relied on the concept of an upper speed for any massive particle in $M^4$, we did not discuss the dynamic reason (if any) for such speed limitation. From the earlier findings we gather that $v_m \neq c$ has something to do with gravity, or self-gravity of a kind, in the Minkowski spacetime. But how can we claim self-gravity limits the upper speed in $M^4$? This now becomes the domain of guesswork; e.g., there may be an eternally static and unobservable ether permeating the $M^4$ spacetime and interacting gravitationally in the direction of motion with any particle mass to reduce its upper speed (being c in SR) to a value below c, a kind of viscous fluid or something! Such a speculation is simply too sketchy and very much along the traditional expectations for mechanical ether. But if there is ether in $M^4$ what is its composition? By looking at the relation between $v_m$ and c we realize that any massive particle of mass $m_o=M_P$ will have a zero upper speed $v_m$, and thus zero energy and momentum no matter what (i.e., zero energy momentum tensor altogether, which in GR is reminiscent of vacuum!) in all inertial frames! Long ago I called these illusive particles *etherons* [20]. Point etherons are deprived of motion and are hidden, so to say; to all observers (the $\gamma$ factor is also no longer meaningful). Yet, etherons have rest mass and according to the Newtonian view of gravity (e.g., the local gravitational Poisson equation) etherons can couple gravitationally to particles, and we must now add without ever moving. Because etherons are deprived of energy and momentum and are classically frozen, an arbitrary "free" particle motion of a kind in this static ether is unaltered by possible collisions with the etherons because the conservation of its $P^\mu$ is guaranteed, given $P^\mu(etherons)=0$ in any frame. In other words our ether, populated by the etherons of mass $M_P$, is "mechanically" transparent to any real particle traveling in it. Extensions of these sterile etherons to more dicey particles are also possible (e.g., treating them as Planckian blackholes, endowed with charge, spin, magnetic charges, etc!), but we shall not tackle these issues here.

(A dandy inquiry may be this: do the etheron mutual distances increase as the universe expands, or what? If the former was true then at the big bang era the universe must have been dominated by the total ether mass. At the opposite end, if we insist that etheron mutual distances must remain constant at all times (assuming stable) then in an expanding universe we may have to entertain the idea that as the universe expands its active mass content gradually converts into more and more sterile etherons, until the universe depletes its active "physical" mass content sometime in the future (this is however much more of an involved issue which I'll omit here)! Is this a kind of dark energy (or rather dark matter, since etherons don't have rest energy) scenario? Currently we know of only four fundamental interactions, but what if there are extra forces so to make the etherons more active, interaction-wise speaking, transcending gravity? Well, in that case obviously things change!)

Any etheron having apparently coordinates $x_\parallel$ and t in the K frame, will have imaginary coordinates $x'_\parallel$=ict and $t'=ix_\parallel/c$ in the K' frame, thus nonsensical in the real representation of the Lorentz group (also the velocity addition law yields absurdity for the etherons)! Although classically the very massive etherons, which are the quanta of the classical vacuum, are motionless, quantum mechanically there can be position and momentum fluctuations of the etherons, due to the uncertainty principle. However, there is also a lack of localizing such quanta classically before even invoking QM because it costs no energy to produce them in small or large numbers (note: had we adopted the maximum energy for all particles as $M_P c^2$ instead of $M_P v_m^2$ the story would have been very different and etherons would have had rest energy as well, and the GR field equations could have been used for the tackling of the ether!). The message is we have no unambiguous way of formulating the classically motionless etherons quantum mechanically (e.g., what are the etheron observables to measure?). It is amusing however to note that one can derive a momentum-coordinate uncertainty look-a-like relation for the etheron by doing the following: define a tiny virtual momentum for the etheron as $\Delta P = M_P(c-v_m)$, and a tiny coordinate uncertainty $\Delta L = \hbar/M_P c$. Multiplying both, and after specializing to cases where $m_o \ll M_P$, we find $(\Delta P.\Delta L)/\frac{1}{2}\hbar \sim (m_o/M_P)^2$, where $(m_o/M_P)^2$ is indeed the term distinguishing $v_m$ from c and so this may be a middling explanation of why $v_m$ is not c in the context of an unobservable ether fluctuation. A fluctuation obviously prompted by the motion of a particle of mass $m_o$! The story of ether in physics (Maxwell, Lorentz-Poincare, etc) is of course centuries old, but is neither of interest here nor has much of relevancy to our scenario. Suffices to say that our ether does not seem to be of the type already discussed in the literature, and unlike most discussed etheron medium, it does not rely on rather artificial constructs (such as the continuity equation for the etheron density, the Navier-Stokes equations, and so forth) to model the universal ether. In the old days the ether picture was mostly mechanical; in more recent times some people are invoking the perfect fluid model where the ether pressure has a thrusting effect on matter. Interested readers may want to consult [23] for many details.

-**Accelerated frames and DUSR?** The raw objective in here is to devise global relativistic coordinate transformations capable of accommodating both inertial and accelerated frames of a peculiar kind! Sounds absurd? Perhaps not, we already have such transformations! Simply take a look at the modified DUSR Lorentz spacetime transformations, which depict the underline physics of inertial observers in uniform relative motion up to a maximal speed $v_m$ per particle, and then make the following intriguing observation. The modified LT are unaffected under the replacement of the constant ratio $v/v_m$ by $v_x/v_{mx}$, where $v_x$ and $v_{mx}$ are both variable speeds, provided, as earlier, we set $v_x = n v v_{mx}/v_m$, where the sub index x refers to the spacetime coordinates of the origin of the K′ frame as measured by the K frame observer. This observation is an interesting one because $v_x$ being a variable frame velocity signals that we are now dealing with linear accelerating frames, albeit in some well prescribed way, in the direction of the boost. Yet one can still invoke inertial (non-accelerating) observers of the DUSR, and the spacetime being flat allows one to speak of a global time variable as in ordinary LT; the key ingredient of course is the variable upper speed $v_{mx}$! A peculiar accelerating linear motion of this type may be of help in relating gravity to variable upper speed variations, and perhaps of help on nongravitational issues, like resolving the twin paradox, where one twin has to certainly experience acceleration during his round trip voyage from the inertial Earth! (Comment: In the Einsteinian view the link between acceleration and GR is the EP, so one considers the diffeomorphism on the Minkowski space (Diff(M)) which we know includes the Poincare' subgroup, and that is what most people consider.)

The interpretation of "accelerated" versus "inertial" observers of DUSR is not a trivial one for, e.g., observer K because of $c_x$ not being constant**.** Normally, in ordinary SR with constant c mutually accelerating observers cannot find well defined global coordinate transformations to communicate information and share their spacetime coordinates, even possibly in a single frame. What we gather from DUSR, on the other hand, and that is perhaps nontrivial, is the possibility of accelerated frames communicating information at the global spacetime level-even though they use, e.g., different local time coordinates-without the formation of an apparent horizon, so it appears, provided one invokes variable $v_{mx}$ and specializes only to noninertial linear frame motions subject to $v_x = n v v_{mx}/v_m$, this relation however acts as a hidden symmetry within the context of our modified LT! What we observe here is not necessarily a unique noninertial frame motion for a given $v/v_m$ ratio, but rather a family of noninertial frames of distinct variable upper velocities, or light speeds, as seen shortly, and frame velocities, even if $v/v_m$ is fixed, and the transformation between two such frames may be related by some (unitary?) group of a kind.

Now we want to ask: who is the controlling agent for adjusting the variable light speed values from one "x" to another while preserving the above equality? The K' observer, or is it the K observer? It is normal to expect that the measurement of the VSL is done by the noninertial observer K' with his clock and rod, but the latter is not the agent for such a change in the first place! Hence, this is what we shall do: we shall assert that the change of $c_x$, if it is to occur, as we presume, can only happen due to a unique local physical agent of a sort operating during the acceleration of the frames and that is our friend gravity! Thus, nongravitational accelerations, such as an "electron-rocket", caused by nongravitational forces should not *cause* variation of c and in this case we are simply back to the standard understanding of the accelerated frames in ordinary SR (see the note added further in below)! In the presence of gravity and other forces then again the VSL will occur due to gravity alone!

(We remark that our line of thought here is close to Einstein's "once upon a time" view that VSL was related to gravity. During 1907-1911, only few years before the advent of GR in 1915, he conjectured that $c_x$ is related to the gravity potential $\Phi_x$-as evaluated with respect to a point (for us the K frame's origin) where the light speed is $c_o$-by the relation $c_x = c_o(1+\Phi_x/c^2)$. Later on he traded in (loosely speaking) the VSL concept with gravitational time dilation, and the rest is history in that (tensorial) gravity became the spacetime "norm" in form of the non Euclidean metric, which in turn gave rise to the famous relationship $g_{00} \sim -(1+2\Phi_x/c^2)$ in the weak field limit. Ultimately, in modern times we treat c as a constant in GR and replace gravity with curved spacetime metric. It is then understood that under no coordinate transformation the curved spacetime metric can ever be converted to a Minkowski metric, and in terms of physics this means "gravity" or the EP cannot be eliminated by any coordinate transformation from say a noninertial to an inertial frame.)

There is an issue left at this point that needs to be addressed. The relation $|v_x/c_x| = |v/c|$ at first appears one sided in that the K frame is treated as special, and also we have not said anything about $|v_{x'}/c_{x'}|$! Because gravity is viewed as a "field", and as the agent for causing variable c, it operates both ways, then to restore the symmetry for variable c in both K and K' frames we may set $|v_x/c_x| = |v_{x'}/c_{x'}| = |v/c|$, a relation that is also expected independently from the DUSR Lorentz transformations from either K to K' or the reciprocal transformations from K' to K. But this can cause a severe restriction on the functional form of $c_x$ or $v_x$ to the point of triviality! For example, in the simplest case, a trivial expression for these two speeds can be envisaged by making use of the *invariant* proper time τ, as $v_x \sim vf(\tau)$ and $c_x \sim cf(\tau)$, and obviously the same can be chosen for $v_{x'}$ and $c_{x'}$ with $f(0)=1$. Now, is this good or not so good I cannot really tell, and likewise I cannot tell whether the idea of a special frame like K is good or not so good, because after all the ratio $|v_x/c_x|$ is a constant equal to $|v/c|$ that conceals the variation of the individual $v_x$ and $c_x$ (or independently their prime analogue) at the LT level. (Note: since we are not violating any Poincare' invariance, so it seems, then any special frame motion must remain unobservable.) Therefore, one may wander whether there is any need to repeat, or impose, the same story for the $v_{x'}$ and $c_{x'}$ and by doing so severely restrict the functional forms of these variable quantities? In other words, there is no a priori obligation anywhere to impose a direct spacetime correlation between say $|v_x|$ and $|v_{x'}|$, the only requirement is $|v_x/c_x| = |v_{x'}/c_{x'}| = |v/c|$, a relation that can mean many things. The general nature of our relation $|v_x/c_x| = |v/c|$ is made more evident by noting that even complex values of $v_x$ and $c_x$ are admissible, at least in principle! For example, write $v_x = v + if(x)$ and $c_x = c + ig(x)$, and require their ratio to be real, what you get is an automatic identification $v_x/c_x = v/c$, provided $f(x)/g(x) = v/c$, so in this case the noninertial observer has an imaginary acceleration in $M^4$! Again whether this has any physics behind it or not I cannot tell. (Note: though not relevant to our case, it is worth reminding that there have been past, and still ongoing, attempts to complexify the Minkowski spacetime for the purpose of the Hamiltonian formulation of gravity in phase space by employing certain complex variables known as the Ashtekar variables and the focus of this formulation is on the evolution of "spatial" geometry (or more precisely geometrodynamics which is an interesting subject on its own).) Verbally, a relation like $|v_x/c_x| = |v/c|$, which we conjecture contains the seed of "gravity" in it, means whatever a would-be accelerating observer gains locally in velocity (momentum, energy) is "lost" or compensated by the local rescaling of the light speed, and this kind of peculiar frame acceleration is not to be confused with ordinary acceleration of massive particles in spacetime due to some other force!

(**Extra note:** The treatment of a NIF (noninertial frame) in orthodox SR with VSL=constant relates the proper time T of an accelerated observer (in our case all observers are pointwise) to the time t as measured in an inertial frame (more precisely moving in a momentarily co-moving IF, MCIF). E.g., in case of constant proper acceleration $a$ one finds t=(c/a)sinh(aT/c). The topic dealing with the relationship between a generic accelerated clock and the one attached to the corresponding MCIF, and more like issues relating a NIF to its corresponding MCIF, is known as the "clock postulate", which is somewhat more involved than its title conveys. In simple terms, the clock postulate states that an accelerated clock's rate is equal to the clock rate in the MCIF-thus is independent of its acceleration-which for a brief moment of time has a zero relative velocity with respect to the NIF, and infinitesimally later you have another MCIF with another inertial clock that again matches the NIF clock at that moment, and so forth. The clock postulates is not just a statement about time rate, it also says the length contraction of a moving object and its relativistic mass are independent of the acceleration. (But note: The timing rate of a moving clock is certainly sensitive to its speed value.) The accelerating moving clock counts its time in one way and an inertial clock counts is time in another way, yet the ratio of the rate of the inertial to the rate of the accelerated clock (which is equal to the gamma factor $\gamma=(1-v^2/c^2)^{-1/2}$) is still the same gamma factor ratio than when the moving clock is moving uniformly without ever accelerating. Finally, we draw our attention on the absolute fact that in all SR matters (including the issue of inertial versus noninertial observer matters) the light speed c is always a constant, in sharp contrast to our case of variable c.)

Undoubtedly, there is a genuine interest in understanding the nontrivial relationship between quantum mechanical predictions and accelerated observers making such measurements. The standard Lorentz invariant QM measurement theories, generally respecting the linear superposition principle of the state vector, involve only inertial observers, which is only an idealization of the real world that must in all likelihood involve noninertial observers of a kind. Our proposed DUSR was designed with an eye on facilitating information communications between a special class of accelerating observers and their MCIF inertial counterpart, but this makes sense only if the light speed is allowed to be a variable, albeit very close to c for all *gravity* issues. Thus, at some stage in the future we must tackle the issue of how the specific noninertial observers of DUSR deal with the quantum measurement theory! DUSR is a theory that not only admits orthodox inertial observers but apparently also, and independently, noninertial observers of a kind (maybe of an abstract kind, otherwise physical observers cannot change the value of c locally in the vacuum just like that!) and the subtle point is how are these physical and unphysical observers related and I see here some analogy with the original development of GR confronting the Newtonian gravity version! The assertion that $c_x$ is to be always close to c says the ratio $v_x/c_x=v/c$ (when v/c is purposely fixed for a given quantum particle) implies also that the temporal changes of $v_x$ (acceleration) must be always mute under gravity. Or in other words the time scale for any velocity change must be long compared to the characteristic (quantum) time of the particle's internal oscillation $\sim\omega^{-1}$. Such a situation cannot occur for say the electromagnetic couplings, involving waves and not having a weak coupling constant, but it is indubitably compatible with gravity coupling, which is to be always week in our models. The generic case of acceleration where v/c is not fixed but can vary continuously in infinitesimal discrete steps is expressed by the local equivalence between pointwise momentarily commoving inertial observers in the Minkowski spacetime and identical accelerated observers still in $M^4$. The infinite numbers of discrete changes of v/c (referring to MCIF observers) allow the treatment of a generic acceleration while keeping $c_x$ always close to c, or $c_x$=c when gravity is neglected.

Accelerated observers in SR, where the proper acceleration $a$ of an object (felt by the object itself) is defined as -$(ma)^2=(1/cdE/dt)^2-(d\mathbf{P}/dt)^2$, are quite distinct from their inertial counterparts in terms of what they can see, and even more so when quantized fields are involved in the Minkowski space background. The famous Unruh effect is one example of the shocking difference between what inertial observers see (or rather fail to see!) as the T=0 K vacuum of an existing background QFT, and what an accelerating observer (e.g., a uniformly accelerating spin ½ detector), undergoing an acceleration a, sees as a vacuum flux full of bosonic and fermionic degrees of freedom at temperature KT=ħa/2πc, consequently uniformly accelerating observers of distinct constant accelerations see distinct vacua! Another example showing the profound difference between an IF and a NIF is the decay of the particles. E.g., a particle like a proton that is stable in an IF appears decaying in an accelerated frame!

Since we have already chosen $v_m=c[1-(m_o/M_P)^2]^{1/2}$, we have to now ask which of the three quantities $(c, m_o, M_P)$ is the one to vary in the $v_{mx}=v_m(\mathbf{x},t)$ version? I'll be choosing $c$ for the task to stand for our variable light speed $c_x=c_x(\mathbf{x},t)$ because this is not unexpected in noninertial frames anyhow, and ignore the other possible options where one, or more than one, mass can vary. Yet, $M_P$ variation is interesting because it can be tide to variable G (assuming $\hbar$ and c constants), and finally variable $m_o$ has the intricacy of its own and is of interest to me (I've said few words on this much earlier). (Another hybrid option not elaborated over here is, e.g., for the light speed and $m_o$ to be both variables but such that $v_m$ (and thus $v/v_m$) remains a constant! If you do a bit of work in the static limit you may connect $c^2(\mathbf{x})-c^2(0)$ to self-gravity with variable G and in this way deduce $m_o(\mathbf{x})-m_o$!)

Anyway, the choice is made and is: $v_{mx}=c_x(1-(m_o/M_P)^2)^{1/2}$ and we can now use freely $\mathbf{v}/c=v_x/c_x$. We recall the orthodox LT can be written in general as: $x'_{\|}=x_{\|}\cosh\theta-ct\sinh\theta$, $t'=t\cosh\theta-(x_{\|}/c)\sinh\theta$, where constant $\theta$ is related to v as $v/c=n\tanh\theta$. In case of DUSR this becomes $v/v_m=n\tanh\theta$. Obviously, the Minkowski metric is the outcome at local levels as long as $\theta$ remains a constant, and that is also true for our peculiar setting $v/v_m=v_x/v_{mx}$, but it is no longer true if one looks at extrapolating the above LT to the case where $\theta$ is a generic spacetime (and generally nonlinear) variable. Imposing nevertheless $dS^2=dS'^2$ can yield a particular solution for $\theta$ given by $\theta=\text{Ln}[(ct+x_{\|})/(ct-x_{\|})]$, which in turn gives $v_x/c_x=2ct.x_{\|}/(c^2t^2+x_{\|}^2)$. This ratio however is not Lorentz invariant (except when $ct-x_{\|}=0$ yielding v=c) and in this case the concept of an active preferred frame may be invoked. But what usage the latter finding may have for us is not clear at this stage. One special example of spacetime transformations that yield a globally invariant Minkowski interval with respect to the origins of the K and the K' frames $(S'^2=S^2)$, without explicitly showing a frame speed $\mathbf{v}$ in the transformation formulas from K to K', can be written as: $x'_{\|}=(S_{\|}/ro)(x_{\|}-ct\sqrt{1-ro^2/S^2_{\|}})$ and $ct'=(S_{\|}/ro)(ct-x_{\|}\sqrt{1-ro^2/S^2_{\|}})$, where ro is some constant length and $S_{\|}^2=c^2t^2-x_{\|}^2$. One can now invoke the instantaneous frame speed $v_x$ and the variable light speed $c_x$ related by $v_x/c_x=(1-ro^2/S^2_{\|})^{1/2}$. Of special interest is the case where $v_x/c_x$ is a constant equal to $\mathbf{v}/c$ which, in turn, yields $S_{\|}=ro/(1-v^2/c^2)^{\frac{1}{2}}$ and equating this to $c^2/a$ (or alternatively to $v_m^2/a$, with $a$ a constant proper acceleration) gives $ro=(c^2/a)(1-v^2/c^2)^{\frac{1}{2}}$. And this, in turn, yields a minimal length at $v=v_m$ as $r_{min}=(c^2/a)(m_o/M_P)$. Now, if the smallest possible length is taken as $\hbar/M_Pc$ then the maximal acceleration for a particle of mass $m_o$ is $a=m_oc^3/\hbar$ (or rather $m_ov_m^3/\hbar$ so to vanish for the etherons). For an electron this is about $3\times10^{31}$ cm/s$^2$! (Incidentally, in the orthodox approach to constant acceleration people usually use the 2D Rindler coordinates to study uniform acceleration in $M^4$ and that can be also extended to DUSR but I shall not pursue this.)

Consider our general ansatz $\mathbf{v}/v_m=v_x/v_{mx}$ or simply $v_x=c_xv/c$. Then unless I am missing something the latter equation does not imply that $c_x$ must be always smaller than c. If $c_x$ is superior than c at some locations then all we gather is $v_x$ must be also superior than v at those locations, and vice versa. The previous putative limit on $c_x$ to be always close to c is only viable for all gravitational considerations, but may not be viable for all other nongravitational forces in the context of more conventional VSL theories.

Let us now mull over to the Newtonian limit (the relativistic format will be discussed shortly in the Hamiltonian context) and discard all relativistic corrections from the gamma terms and also consider the static limit, then we have $\mathbf{a}\sim d v_x/dt=v/c.dc_x/dt=(v/c)(\nabla c_x.v_x)=\frac{1}{2}(v^2/c^2)\mathbf{n}(\mathbf{n}.\nabla c_x)$. (In cosmology something like $dc_x/dt\sim-c_xHo$ may be tempting where Ho is the Hubble constant.) Next, assume the noninertial frame K', flying away from frame K, has a "test-particle" mass m at its origin, and likewise the frame K origin is endowed with a mass M>>m, and denote the variable distance between the two masses by r and, finally, posit that the noninertiality of the motion is due to a real force, the gravity force. Then from the above relation for $\mathbf{a}$ (not necessarily a constant) and in combination with the Newtonian gravity force between m and M we find $-G_N Mm/r^2=\frac{1}{2}mv^2\partial(c^2_x/c^2)/\partial r$, which is not too far off from the Einstein's proposed formula (1907-1911) presented earlier, but the departing points are totally different in view of our DUSR and its pivotal identity $v_x=c_xv/c$. (Note: according to GR though the moment you introduce the two masses spacetime is no longer flat! If the two masses where fermionic with vanishing gravitational masses, following our principal conjecture, then the metric is still Minkowskian and $\partial(c^2_x/c^2)/\partial r=0$, implying for such fermions $c_x$ is constant! If we don't use the $m_g=0$ conjecture then to keep $M^4$ you need to get rid of gravity by using variable c, and ultimately variable G.) A simple integration then yields a relationship between variable $c_x^2$ and the gravity potential, but care must be exercised. First of all we do not want $c_x$ to be too different from c, and that means we do not want it to diverge at say r=0! This is taken care of by invoking variable G(r), in place of $G_N$, which is finite at r=0. (Note: I will generally write the gravity potential not as an expression proportional to $-G_x/r$ but rather as my favorite expression $\sim G_x/r_o=G_N e^{-ro/r}/r_o$, which is *always* positive (!), and gives rise to the -1/r behavior for r>>ro, and vanishes at r=0 while tending to $G_N/r_o$ at r=$\infty$.) Secondly, Newton's gravity force is always attractive, so normally the gravity induced acceleration $a$ must be negative and eventually bringing to a halt the increasing separation between K and K' and ultimately resulting in the collapse of the two frames on each other (analogous to throwing up a stone from the ground and its ultimate return to the ground) which is not seen by the inertial observers moving with the uniform relative velocity **v**!

Such issues, and there are more, especially in the relativistic domains that I've not yet discussed, present serious contradictions and are contradictory to the existence of constant v by virtue of $\mathbf{v}_x=c_x v/c$! A more direct way of seeing better the situation is to make use of the nonrelativistic energy conservation law now taken as: $E_{tot}=\frac{1}{2}mv^2=\frac{1}{2}mv_x^2+U_{pot}(r)=\frac{1}{2}mv^2(c_x^2/c^2)+U_{pot}(r)$ so that the potential energy is $U_{pot}(r)=\frac{1}{2}mv^2(1-c_x^2/c^2)$. Thus, the gravity potential is $\Phi=U_{pot}(r)/m=\frac{1}{2}v^2(1-c_x^2/c^2)$ and is positive because $c_x<c$! A reasonable remedy to all these is to conjecture, and once for all, that the major role of variable $c_x$ is to *cancel* gravity for the SM particles (and we want these particles to have fermionic degrees of freedom) at all separations, including zero! This can be done only with the gravity potential that uses our favorite variable G expression given in the above. But I will not bother working the details over here because the fermionic nature of our conjecture cannot be seen from the above treatments, and we want to move on to the next topic that may be a more rigorous approach to VSL and gravity. Suffices to say that in sharp contrast to fermions, in case of bosons one may want to generate gravity, and not cancel it, by using variable light speed! The interest in DUSR, which is a global LT of a kind with upper particle speeds included, in connection to noninertial frame motion coupled with VSL, which is a hidden Lorentz coordinate transformation symmetry of a kind, suggests that perhaps gravity phenomena do not need the local Lorentz invariance prescript of Einstein after all; this may sound too simplistic but is also worthwhile exploring further. It is good to remind ourselves that we are working throughout this section with only a minimal set of tools and yet we are capable of speculating on how fermions (inherently quantum mechanical in nature) should behave under gravity without even resorting to the proper action formulation of this problem involving spinors, the spin connection, and all that. Despite this technical limitation, and this being an informal section, let us continue and see how far we can stretch our investigation of gravity for the fermion by using this time the Hamiltonian approach as outlined further in below.

Up to now I have outlined the general idea of variable $c_x$ in the context of: DUSR and specialized noninertial linear moving frames and the cancellation of gravity for the SM particles in $M^4$. Obviously many hot issues remain to be confronted in such formalism, especially in cosmology. We may ask, e.g., whether there is a need for spontaneous symmetry breaking of the Lorentz invariance due to say inflation in the early universe if we invoke our accelerated frame motion recipe along with variable c? Based on what we've learnt so far we shall in below look at variable $c_x$ at the local level from a Hamiltonian perspective pertaining to a fermion under a gravity potential energy and see what possibly takes to impose free motion on such a (fermionic!) particle. Let me also add that for particles of spin $s^\mu$ we can envisage an interaction term due to the spin-acceleration coupling of the form $s^\mu a_\mu \sim \mathbf{s} \cdot \mathbf{a} - \frac{1}{2}(v^2/c^2)(\mathbf{n} \cdot \nabla c_x^2)(\mathbf{s} \cdot \mathbf{n}) = \frac{1}{2}(v^2/c^2)(\mathbf{n} \cdot \nabla c_x^2)s_\parallel$ where $s_\parallel$ is the spin component along the direction of the acceleration. Although such a term may exist, it is not clear what exact role this interesting coupling can play in our approach to canceling or generating gravity via variable light speed (e.g., can we set $\mathbf{s} \cdot \nabla c_x^2 \sim$ gravitational potential energy, or something similar along this line?). If the light speed variation in flat spacetime is to have a close relationship to generating or canceling gravity for the SM bosons and fermions then is this a vague hint on the impossibility (or irrelevancy) of ever detecting the local light speed variations in experiments (a diffeomorphism invariant $c_x$?), especially in the static cases where the light speed varies only spatially? (Incidentally, another way of seeing a possible affinity between variable c and gravity is to interpret the infinitesimal energy difference relation $m(c_x^2(x_1)-c_x^2(x_1+dx))$ as the universal coupling "at-a-distance" of light to all matter, similar to the universality of gravity field coupling to matter and the Higgs particles (if extant) coupling to matter! E.g., the Higgs coupling to fermions (the Yukawa coupling) is proportional to the fermionic mass $m_f$ and is given by $g_{fH}=m_f/<v>$, where $<v>$ is the Higgs field VEV and is about 246 GeV. But in the following I shall stay away from the Higgs and the like "stuffs"!)

**-The Hamiltonian Approach to $c_x$ and Gravity "field" Cancellation:** In what follows we shall trade in our earlier crude ideas on: variable c, frame acceleration, and variable G in favor of a more established approach founded on local LT invariance, variable $c_x$, and the (locally invariant) relativistic radial Hamiltonian. We consider the Hamiltonian of any particle of mass $m_o$ under the influence of a spherically symmetric and static gravity potential energy U, which is due to an external gravity field (the static case may also be interpreted as a formulation on 3-manifolds, which over here are flat spatial hypersurfaces, at some fixed time t=to). The DUSR inspired Hamiltonian is chosen simply as $H(r,P)=(P^2c_x^2+m_oc_x^4)^{1/2}+U(r,P)$ and, according to what we've said before, it is expected the $c_x$ "field" values be always very close to c under gravity, no matter the circumstances. Moreover, any possible spin-acceleration coupling term is discarded in H at this stage. Now an important observation is in order: following DUSR we should have included $v_{mx}$ instead of $c_x$ in the above Hamiltonian but for the current presentation I will, for simplicity sake, stick to $c_x$ by assuming $m_o$ to be much smaller than $M_P$. Because what we really have in mind is to cancel gravity field via the variable $c_x$ field, and we desire this cancellation to apply to only the SM fermions, as promoted in this section, then what we shall do next is to linearize the Hamiltonian (in analogy to what Dirac did in 1928 for the quantum fermions in the context of relativistic quantum mechanics, and his ultimate usage of the 4x4 gamma-matrices, which we'll discard in our current treatment though it can be formulated but that's another story). Additionally, because we anticipate the cancellation of gravity the linearized Hamiltonian is to represent a "free" particle in $M^4$ under a constant potential, which at the classical level induces no force. With this said, the Hamiltonian formulation is: $H(r,P)=(P^2c_x^2+m_o^2c_x^4)^{1/2}+U(r,P)=Pc+m_oc^2+U(\infty,P)$, where U is the (gravity) potential energy (involving variable G), and P is the radial momentum of the particle, and as usual $d\mathbf{P}/dt=-\nabla H$, and $d\mathbf{r}/dt=\nabla_p H$ (to be precise here again the constant c is to be replaced by $v_m$).

By using the linearized Hamiltonian of the above we obviously find $d\mathbf{P}/dt=-\nabla H=0$, as desired for a *free* particle. On the other hand, the equality $(P^2c_x^2+m_o^2c_x^4)^{1/2}+U(r,P)=Pc+m_oc^2+U(\infty,P)$ implies that at P=0: $c_x^2(r,0)-c^2=(U(\infty,0)-U(r,0))/m_o$, which is a relation that can be used for studying particle orbital motion at fixed r (e.g., for the interesting case of Bohr's atom where r is quantized). The latter H-equality also implies $c_x^2(0,0)-c^2=U(\infty,0)/m_o$ (provided U(0)=0) and $c_x^2(\infty,0)=c^2$! The fermionic "master" solution for $c_x^2$ applicable to all r and P is: $c_x^2=(P^2/2m_o^2)\{-1+[1+(2m_o/P^2)^2(Pc+m_oc^2+U(\infty,P)-U(r,P))^2]^{1/2}\}$. (We note that in case of etherons $v_{mx}=v_m=0$, so P=0 for the individual etherons, in compliance with our earlier discussions, and their potential energy caused by any gravity source in the ether is always zero or a constant, depending on whether we consider bosonic or fermionic etherons forming a spin network!). An equation of this kind is a representation of $c_x^2$ in the (P, r) phase space, and of particular interest are then the surfaces of constant $c_x$ (in analogy to the equipotential surfaces in electrostatics) and their curvatures, all of which require knowing the potential function. For massless spinor particles (neutrinos) we find $c_x^2(r,P)=[c+(U(\infty,P)-U(r,P))/P]^2$ but if U is proportional to m=0 then we get $c_x=c$, as expected, however if a neutrino has a tiny rest mass then $c_x\neq c$ and its arrival time to earth from say a distant supernova explosion is different from the light time of arrival. In the ultra relativistic regimes where $P>>m_oc$ we find $c_x^2(r,P)\sim c^2+O(1/P^2)$ for all massive fermions. Also by taking the gradient of $c_x^2$ in the NR limit, where P and the potential energy (assuming always finite) are small compared to $m_oc^2$, we find the known Newtonian result: $\nabla c_x^2\sim -\nabla U/m_o\sim \mathbf{a}$, where $\mathbf{a}$ is the vector acceleration.

All the relations thus far obtained make obvious that $c_x\sim c$ at all values of r and P provided the potential energy is finite at $r\sim 0$ and that is exactly what we desire to have. So, our next task is to model a finite valued gravity potential for a point particle. We use the familiar integral form for the potential generated by a mass density $\rho$: $\Phi(\mathbf{x},\mathbf{P})=(1/r_o(\mathbf{P}))\int d^3x'\rho(\mathbf{x'})G(\mathbf{x},\mathbf{x'})$, where $G(\mathbf{x},\mathbf{x'})$ is the Green's function having unit of $G_N$, $r_o(\mathbf{P})$ is a constant length that can depend only on $\mathbf{P}$, and we shall also neglect time delay effects in this static limit, as we did throughout. Our choice for the static Green's function, to be finite at $\mathbf{x=x'}$, is an exponential form given as:

$G(\mathbf{x},\mathbf{x'})=G_N e^{[-r_0/|\mathbf{x}-\mathbf{x'}|-|\mathbf{x}-\mathbf{x'}|/l_0]}$. And for a point mass with $\rho(\mathbf{x'})=m\delta^3(\mathbf{x'})$ the resulting expression for the potential involving two length scales is $\Phi(\mathbf{x},P)=(G_N m/r_0)e^{-r_0/|\mathbf{x}|}e^{-|\mathbf{x}|/l_0}$ which is always *positive* (!), unlike Newton's gravity potential$\sim$-1/r, and vanishes at both $\mathbf{x}$=0 and $\infty$. Retaining or discarding the second (Yukawa type) exponential is the difference between wanting either a vanishing or a constant potential for r$>>$(r$_0$,l$_0$). One of the length scales (as an option) can be eliminated by setting r$_0$=l$_0$ (so that $\Phi$ is invariant under the exchange of r$_0$/|$\mathbf{x}$| to |$\mathbf{x}$|/r$_0$, a short-long scale "duality" feature, if you will, of great theoretical importance!). As for r$_0$ it can be chosen as the Planck length (I will not bother showing explicitly the P dependency of r$_0$ by assuming it is only a slowly varying function of P, but the exact form of the P dependency for $\Phi(\mathbf{x},P)$ can be computed for example for the bosons, see the next paragraph). I shall thereafter discard theYukawa term for simplicity. For separations r$>>$r$_0$ the potential is $\Phi(\mathbf{x})\sim G_N m_0/r_0$-$G_N m_0/|\mathbf{x}|$, which is consistent with the Newtonian gravity. Also our earlier finding for $c^2(\mathbf{x},P)$ at P$\sim$0 says: $c^2(\mathbf{x},0)$-$c^2=(G_N m/r_0)e^{-r_0/|\mathbf{x}|}$ which in turn yields $c^2(\infty,0)$-$c^2=(G_N m/r_0)$ and $c^2(\mathbf{0},0)=c^2$. This is certainly a consistent finding, but so is the case of retaining the Yukawa exponential, which implies $c^2(\infty,0)=c^2$ and $c^2(\mathbf{0},0)=c^2$, so I leave the choice open (see the comment in below). (Note: If one insists in including the spin s explicitly in the formulas then one modification to $c^2(\mathbf{x},0)$-$c^2=(G_N m/r_0)e^{-r_0/|\mathbf{x}|}$ may be $c^2(r)$-$c_0^2=4s(1-s)(c^2$-$c_0^2)e^{-r_0/|\mathbf{x}|}$ so that for s=½ we are back to the above, while for s=1 and s=0 bosons c(r) does not vary to impact gravity, but in light of the discussions in the second paragraph in below we discard such redundant spin factors!)

(**Remark:** there is an increasing awareness these days that gravity, even at the Newtonian level, must be modified (at least) at very large distances because of the so-called Dark-Energy (DE) problem. If the cosmological constant is to explain the DE-while being regarded as an overall vacuum energy of the familiar fermionic-bosonic quantum fields found in particle physics-then we find a huge discrepancy between what we predict quantum mechanically for the vacuum energy and what we observe in the universe. This huge discrepancy exists only if all the fundamental constants, including G, are treated as absolutely constants at all scales, and notably at immense distances. Our conjecture of "no gravity" for the SM fermions alleviates only partially the predicted value of the CC term from the QFT perspective. But the predicted value is still huge compared to the observation, even if we consider only a scalar field and that means predicting a large scalar curvature for today's universe, contrary to the miniscule curvature we observe! Thus, help must come from somewhere else. That somewhere for us has always been variable G, which we think not only affects what we see and measure today but may have had also a drastic impact on the very early universe and the inflationary models. The above favorite form $G\sim e^{-r_0/r}$ however is really more suited for the "IR" domains where r$<<$r$_0$. The modification of G (thus gravity) at very large scales r$>>$l$_0$ may require restoring the earlier $e^{-r/l_0}$ Yukawa piece (without necessarily assuming r$_0$=l$_0$?) so to ultimately yield a tiny curvature for the vacuum energy.)

For the SM bosons we may want to try something else, of course still within the Hamiltonian formulation, which can support our earlier conjecture that the gravity field for these bosonic particles, and in sharp contrast to the gravity cancellation for the SM fermions, comes about because of the variable c-field (or vice versa, if you will). And the following Hamiltonian formulation may do the job for describing such bosons: $H=(P^2 c_x^2+m_0^2 c_x^4)^{1/2}=(P^2 c^2+m_0^2 c^4)^{1/2}+U(r,P)$ (modulo the v$_m$-business discussed earlier!). As seen the complete c$_x$ can now be derived at once (details omitted). Visibly, in the nonrelativistic regime (P$<<$m$_0$c) we have c$_x^2$-$c^2\sim U(r,0)/m_0\sim\Phi(\mathbf{x},\mathbf{0})$, with $\Phi(\mathbf{x},\mathbf{0})$ as given earlier, while in the ultrarelativistic regime c$_x\sim$c+U(r,$\infty$)/P. (Clearly, an important step ahead is how to relate all these findings to the DUSR and its special class of accelerated frames, also needing c$_x$!) In the case of massless bosons the formula gives (c$_x$-c)P+U(r,P)=0, where P=h$\nu$/c, but this normally implies U=0 because c$_x$=c for m$_0$=0. So some modification of the Hamiltonian may be in order to explain the light bending under gravity, assigning a tiny rest mass to the "physical" photon may be tried as the starting point and matched to solar system data to see whether the tiny mass is acceptable.

Finally, there is another remaining issue we need to address: what is our result for d$\mathbf{r}$/dt=$\nabla_P$H? The answer from the linearized Hamiltonian is simply d$\mathbf{r}$/dt=c$\mathbf{n}$+$\nabla_P$U($\infty$,P), which in combination with the nonlinear Hamiltonian gives $\nabla_P c_x^2=[(P^2+m_0^2 c_x^2)^{1/2}/½(P^2/c_x^2+2m_0^2)][c\mathbf{n}+\nabla_P(U(\infty,P)$-$U(r,P))]$-P$\mathbf{n}$, so the last step now is to compute $\nabla_P c_x^2$ from the earlier master solution and equate it to the latter expression we've just computed to finally derive $\nabla_P(U(\infty,P)-U(r,P))$ for all r and P values (long details omitted).

**-Comment:** At this juncture I would like to make a general comment on the variable light speed issue. Once we proposed the relation $v_x/c_x=v/c$ in DUSR for accelerated frames then in all the discussions that followed until now we promoted the idea of $c_x$ as being the variable light speed. But a speck of pondering at the latter relation shows that there is no indisputable reason to make such a claim. E.g., is anything wrong calling $c_x$ the variable gravity speed? When we replaced c in the LT by the maximum particle speed $v_m$ to construct the DUSR spacetime transformations we were in a way breaking with the orthodox monopoly of the "light"+its "speed" value in shaping up these transformations. But by looking at the modified LT we still perceive the appearance of c (an example is $t'=\gamma(t-v.x/cv_m)$), although this time we may utter the appearance of the constant c is only for dimensional purposes (to, e.g., make the ratio $c/v_m$ dimensionless). Yet, it is to be recognized that for massless particles c coincides with the light speed, numerically speaking. We also note that throughout the above Hamiltonian formulations all that was really needed was $v_{mx}$, and this meant again that the constant c appearing in our modified LT need not be, *conceptually* speaking, identified with the "light" speed, even if numerically c is the light speed (we draw attention to the fact that the Maxwell's equations are invariant under the LT with c appearing in both system of equations)! Consequently, when we conjectured earlier that variable $c_x$ is to occur only for the gravity force and not necessarily for the three other forces this should not present a dilemma insofar as the masses M and m, assigned to the origins of the K and K' frames, remain chargeless. In the presence of charge we have extra complexities to confront. For one, the radiation of light by accelerated charges. Even if only one of the masses has a charge (recall most massive fermionic SM particles are after all charged) it will radiate (albeit weakly) due to its gravity pull, and in the nonrelativistic regimes the power $P$ radiated is proportional (according to Larmor's formula) to the square of the acceleration $a^2$. The delicate thing is now whether this acceleration is related or is derivable from our relation $\mathbf{v}_x=\mathbf{n}(v/c)c_x$ which now reads $\mathbf{a}=\mathbf{n}(v/c)dc_x/dt$? If the answer yes then $dc_x/dt=(c/ev)(1.5c^3P)^{1/2}$ with e the charge. Having $dc_x/dt$ scaling as $1/v$ is fairly suspicious, normally the higher v gets the more is the power radiated and not the other way around! Besides, since the relation $v_x/c_x=v/c$ is a hidden factor in our modified LT, the inertial observers are not observing any radiation (and this phenomenon is not the Unruh phenomenon!). If this radiation is to be compensated by say the drop in the particle's energy proportional to, e.g., $m(v_{mx}^2-v_m^2)$ then our previous claims that this kind of relation goes hand in hand with gravity is meaningless. Hence, we have to conclude that the formal acceleration associated with $v_x$ is not of type that should cause charged particles to radiate (besides, such a radiation can also perturb the Minkowski flat spacetime metric, according to GR, which we do not desire, even if the two charged masses have zero gravitational masses)! So in this sense we seem to be getting closer to Einstein's idea of free fall in local inertial frames that cause no radiation by charged particles (discarding the tidal force). In fact, if we follow the orthodox doctrine with constant c then ultimately our charged noninertial observers, as fermionic or bosonic mass points, must make a decisive choice in their NIFR on whether they admit moving under a universal gravity (which is not here for the SM fermions because of VSL canceling gravity), or are moving on free falling geodesics in curved spacetime that are locally Minkowskian (again not applying to the fermions)! The conventional dogma, however, applies only to the bosons, according to us, and to all the composite particles and macro matter in general.

In the early to about the middle of this section I tried to discuss in an informal manner, and by using classical language and very limited technical tools, various aspects of our principal fermionic conjecture regarding $m_g=0$, and then realizing that none by itself was sufficient to justify that conjecture, let alone explaining it. A completely new approach to gravity based on a multidisciplinary fusing of particle theory and spacetime "geometry", that we can collectively call "math", is needed for an in-depth tackling of that conjecture, all of course if we still insist in keeping it! The conjecture $m_g=0$ may be too bold to be true, but is good enough, and simple enough, to keep us wandering for the time being until the day an adequate experiment is performed to either confirm or dispel the whole idea! Seeing no electrons falling under gravity (or not seeing any SM fermionic gravity related effect in quantum experiments) will be a dramatic revolution in basic physics, and I think far superior than the more recent stringy revolutions leading to the M-theory that is still in the making. (We recall that the whole early motivation for going "stringy" was the inability of finding a concrete and "full" theory of quantum gravity in the context of point particles.) If the SM fermions are truly deprived of gravity then the question is how will this affect and justify the merit of our mighty string theory, especially by knowing that a fermion like the gravitino, the gauge particle of the graviton, does not gravitate?)

Shortly before submitting this paper while probing the recent literature I came across an article only few years old by Witterich [24] that captivated my attention because its emphasis was only on the spinors to generate gravity (loosely said, in that paper bosons and the vielbein existence are due to composite fermions) while not requiring any local but only global Lorentz invariance for its construction (thus an immediate possibility of using the DUSR instead of SR comes to mind). The theory is diffeomorphisms and globally Lorentz invariant but (unlike GR) not invariant, as just stressed, under the local LT. There is obviously no talk in that note of eliminating gravity for the fermions, but rather generating it by using only spinors, and there is no variable c either, nor the action contains the spin connection, and the vielbein in the action is only globally defined. Nevertheless, there is much more to this paper in form of technical bits and pieces that I feel can provide us with a distinct vision to better understand the weight of fermionic subtleties regarding gravity, and perhaps be also of help to decode the enigma of our principal conjecture (although I doubt it!). There is also an older paper by Friedman and Sorkin on generating spin ½ from gravity alone via nontrivial 3-manifolds [25]. The short abstract in that paper reads: "For a certain class of three-manifolds, the angular momentum of an asymptotically flat quantum gravitational field can have half-integral values". Therefore once again we note a wealth of possibilities and hints from here and there even in the existing literature encouraging a pursuit of any reasonably motivated idea in matters of gravity beyond GR. Perhaps, our no-gravity fermionic conjecture can be one of those reasonable ideas! Undoubtedly, our principle conjecture repeatedly confronts the mainstream approaches based on GR "plus". Consequently, all one can do is to wait for real progresses to be made in quantum gravity from the theoretical side in parallel with the experimental advances to ultimately shed light on the behavior of the fermionic SM particles (most likely the leptons) under gravity. Readers interested to know more on what is conventionally known about gravity coupling to intrinsic spin can consult the valuable article by B. Mashhoon [26]. Finally, let me add that it will be especially useful for our advance student readers to eventually collect and then list what all the experimental and theoretical consequences of setting $m_g$=0 for the SM fermions may be not only in various physics and cosmic ray related fields but also in early cosmology.

**-End of section partial summary:** We've put forth diverse ideas and options in this section for our gravity and variable G exposé pertaining to mostly the SM elementary particles. Undoubtedly it is hard to make an overall summary of so many assorted explorations, most of which did not led to any satisfactory conclusion. But perhaps the following three main options we've discussed targeting the SM matter fields stand taller: **(1)** kill the fermionic gravitational mass-or in a milder sense break the equality $m_i$=$m_g$, and perhaps even drastically in some cases. **(2)** Leave the weak EP alone and leave the fermionic-bosonic gravity disparity alone and concentrate in building an asymptotically free gravity theory using variable G-such that beyond some microscale separation the theory leads to orthodox GR, or some modification of it, with G "almost" constant or slowly decaying. **(3)** Discard the fermionic $m_g$=0 business and try to find a means to cancel gravity for the fermions, while generating it for the bosons; an example of a means used in this section was the VSL+DUSR in combination with variable G. And now I add **(4)** use both variable G and a violation of the EP to some degree and see what ensues. (It is reasonable to also ask if the next variable is to be the Planck constant, following VSL and variable G. My brief answer is yes and that's all I say!) Option (1) is the most exotic, clean and attractive one to my opinion, rendering quantum gravity irrelevant for the SM fermions, and is also simplistic, provided one can sort out its intricacies. Yet its faith is totally dependent on what future experiments say in form of "yes" or "no" answer, so at this stage it is only an idea and a reasonable speculation, albeit my preferred one. My final choice, at least from the operational stand, is item (2). Item (2) is less confrontational to mainstream gravity theories and is most likely easier to formulate than the other options, given what we now know from studying mainstream gravity and its action principle.

Option (2) is also operationally practical because: (a) it is a smooth approach to diluting gravity via variable G, and it too makes quantum gravity greatly irrelevant for the SM fermions. And its quantum corrections, according to us, are always weak because of vanishing G as r→0, even at very high-energies, and certainly in the asymptotically free regions. So a simple linearized Einstein gravity may suffice for the "quantization" to yield spin 2 gravitons livable in the almost Minkowski spacetime (though we bare in mind that spin 2 particles, not necessarily labeled as gravitons, have many issues that we still don't understand, like self coupling)! As for spin 2 "gravitons", many have expressed serious doubts (details omitted) that even a single graviton can ever be detected (recall a free spin 2 particle in flat $M^4$ can generally be formulated via the Fierz-Pauli action S involving the tensor field $h_{\mu\nu}$ subject to $\partial^\mu(\delta S/\delta h^{\mu\nu})=0$) and others have uttered doubts on any analogy between the graviton and the photon in the GR classical framework. And (b) option (2) is equally extendable to the bosonic SM fields, and that is important for wanting a unified view of gravity. But I see no serious objection in making variable G perhaps spin dependent in that fermions have one unified variable G while bosons another. Whether scenario (2) is reconcilable with at least one of the more prominent gravity theories around (like the string theory or perhaps the loop quantum gravity), provided some drastic changes are made to those theories without modifying their trademark character, remains to be seen. String theory, e.g., comes with a dilaton scalar field that regulates the string coupling strength and can be used, at least in concept, for deriving the asymptotically free variable G type gravity that we have in mind. But the lack of string background independent formulation, in the spirit of GR in generating spacetime dynamically, is problematic and still in the making, though it is not clear how its eventual background independent formulation can enhance the rudiments of scenario (2)! As for (3) it may be the most compelling of all proposed theories here and remains a strong contender for a gravity theory but its prime importance in tensorial type gravity, as opposed to scalar gravity studied here, is perhaps sturdily mitigated by other extra high up tensorial effects of geometrical and "energetic" character. Tampering with c, however, has too far reaching consequences beyond gravity. Though over here we by passed this issue by asserting that any tiny variation of c is the consequence of only gravity which remains, according to us, always weak at all energy and length scales. Still one must exercise caution while invoking VSL, especially in quantization matters related to quantum field and quantum gravity theories.

**Acknowledgment:** I wish to thank Drs. Bahram Mashhoon for providing ref [26], and informing me on his work on nonlocal Special Relativity and its possible extension to a nonlocal theory of gravity, and Tony Pitucco for various comments.